\documentclass[epj,nopacs,final]{svjour}
\usepackage{graphics}
\usepackage{latexsym,overpic,rotating}
\usepackage{subfigure,color,showkeys}
\usepackage{epsfig,color,rotating,amsmath,delarray,array,multirow}
\usepackage{epstopdf}
\usepackage{makeidx,pifont,float,amssymb}
\usepackage{ulem} 
\newcommand{\er}{$\pm$}
\newcommand{\be}{\begin{eqnarray}}
\newcommand{\ee}{\end{eqnarray}}
\newcommand{\bea}{\begin{eqnarray}}
\newcommand{\eea}{\end{eqnarray}}
\newcommand{\bc}{\begin{center}}
\newcommand{\ec}{\end{center}}

\newcommand{\nn}{{\nonumber}\\}
\newcommand{\beq}{\begin{equation}}
\newcommand{\eeq}{\end{equation}}
\newcommand{\ba}{\begin{eqnarray}}
\newcommand{\ea}{\nonumber \end{eqnarray}}

\newcommand{\bi}{\begin{enumerate}}
\newcommand{\ei}{\end{enumerate}}
\newcommand{\rd}{\color{darkred}}
\newcommand{\bl}{\color{darkblue}}

\newcommand{\gr}{\color{darkgreen}}

\newcommand{\yel}{\color{darkyellow}}
\newcommand{\vio}{\color{violett}}
\definecolor{darkyellow}{rgb}{1.0,0.5,0}
\definecolor{lightyellow}{cmyk}{0,0,0.5,0}
\definecolor{lightred}{rgb}{1,0.5,0.5}
\definecolor{lightgreen}{rgb}{0,0.4,0}
\definecolor{lightblue}{rgb}{0.5,0.5,1}
\definecolor{darkred}{rgb}{0.8,0,0}
\definecolor{darkgreen}{rgb}{0,0.4,0}
\definecolor{darkcyan}{cmyk}{1,0.3,0.3,0.3}
\definecolor{darkblue}{rgb}{0,0,0.6}
\definecolor{lightbrown}{rgb}{0.7,0.3,0.3}
\definecolor{darkbrown}{rgb}{0.5,0,0}
\definecolor{violett}{rgb}{0.6,0,0.8}
\begin{document}

\title{\boldmath Hyperon I: Partial wave amplitudes for $K^-p$ scattering}
\titlerunning{Hyperon I: Partial wave amplitudes for $K^-p$ scattering}

\author{M.~Matveev\inst{1,2}, A.V.~Sarantsev\inst{1,2}, V.A. Nikonov\inst{1,2}, A.V. Anisovich\inst{1,2}, U. Thoma\inst{1},
and E.~Klempt\inst{1}\thanks{email: klempt@hiskp.uni-bonn.de} }
\authorrunning{M.~Matveev {\it et al.}}

\institute{\inst{1}Helmholtz--Institut f\"ur Strahlen-- und Kernphysik,
Universit\"at Bonn, 53115 Bonn, Germany\\
\inst{2}National Research Centre ``Kurchatov Institute'', Petersburg Nuclear Physics Institute,
Gatchina, 188300 Russia}

\date{\today}

\abstract{Early data on $K^-$ induced reactions off protons are collected and used in a
coupled-channel partial wave analysis (PWA). Data which had been published in the form of Legendre
coefficients are included in the PWA. In a {\it primary} fit using 3* and 4* resonances only, we
observe some significant discrepancies with the data. In a systematic search for new $\Lambda$ and
$\Sigma$ hyperon resonances, additional candidates are found. The significance of the known
and of the additional resonances is evaluated. Seventeen resonances listed with 1* or 2* and one resonance
listed with 3* in the Review of Particle Properties cannot be confirmed, five new hyperons are
suggested. The partial-wave amplitudes deduced in this analysis are compared to those from other
analyses.}


\maketitle


\section{Introduction}

The spectrum of $N^*$'s and $\Delta^*$'s is presently studied vigorously in a number of
photoproduction experiments at ELSA, JLab, MAMI, and SPring-8 (see
Refs.~\cite{Klempt:2009pi,Crede:2013sze}). In the first excitation shell, seven negative-parity
resonances (five $N^*$'s and two $\Delta^*$'s) are expected in the quark model~\cite{Isgur:1977ef};
these are known since long. In the second excitation shell, 21 positive-parity states are
predicted~\cite{Isgur:1978wd}, 16 of them have been observed, even though only 11 of them are
considered to be established (with 3*'s or 4*'s in the notation of the Review of Particle Physics,
RPP)~\cite{Tanabashi:2018oca}. But there is the chance that existing or new data with further
analyses will add to our knowledge of the missing or not so well-known states.

In the hyperon spectrum, seven negative-parity $\Lambda^*$'s and seven negative-parity $\Sigma^*$'s
resonances are expected in the first excitation shell. SU(3) links the seven $\Sigma^*$'s and five
$\Lambda^*$'s to the spectrum of $N^*$'s and $\Delta^*$'s. Two further SU(3) singlet  $\Lambda^*$'s
are expected. Compared to this expectation, one $\Lambda^*$ resonance with spin-parity $J^P=3/2^-$
is missing but only four negative-parity $\Sigma^*$'s are established. Only a small fraction of the
predicted spectrum of positive-parity hyperon resonances is known. Clearly, our knowledge on the
hyperon spectrum needs to be improved.

In this paper we study the possibility to find missing hyperon ($\Lambda^*$ and $\Sigma^*$)
resonances in existing data. The paper is motivated  by recent advances of coupled-channel
partial-wave analyses of the existing data of $K^- p$ reactions
\cite{Zhang:2013cua,Zhang:2013sva,Kamano:2014zba,Kamano:2015hxa,Kamano:2016djv,Fernandez-Ramirez:2015tfa},
and by the prospects of new data on hyperon spectroscopy from J-PARC~\cite{Sako:2013prop},
JLAB~\cite{Adhikari:2017prop}, and the forthcoming PANDA experiment~\cite{Iazzi:2016fzb}.

The pioneering work of Ref.~\cite{Zhang:2013cua,Zhang:2013sva} reported the first coupled-channel
partial wave analysis of most available data on Kaon induced reactions. In a first
step~\cite{Zhang:2013cua}, the authors constructed the partial-wave amplitudes for the reaction
$\bar K N \to \bar K N, \pi\Sigma, \pi\Lambda$ in slices of the invariant mass in the range from
$W=\sqrt s= 1.48$\,GeV to 2.1\,GeV. In a second step~\cite{Zhang:2013sva}, these partial-wave
amplitudes were fit using a multichannel parametrization in the form $S=B^TRB=I+2iT$, where $T$ is
the partial-wave $T$-matrix, $R$ a generalized multichannel Breit-Wigner matrix, and $B$ and its
transpose $B^T$ are unitary matrices describing non-resonant background. In these energy-dependent
fits, the partial-wave amplitudes of two-body reactions in sliced bins in the invariant mass were
exploited as well as the results of partial wave analyses on the reactions
$K^-p\to\pi\Lambda(1520), \pi\Sigma(1385)$, $K^*N$, and $\bar K\Delta$.

The partial-wave amplitudes from Ref.~\cite{Zhang:2013cua} were also used by the authors of
Ref.~\cite{Fernandez-Ramirez:2015tfa} exploiting a $K$-matrix formalism where poles are described
as conventional pole terms and background contributions by poles at negative values of $s$. The
energy-dependent amplitudes from their fit
described reasonably well the energy-independent partial-wave amplitudes
from Ref.~\cite{Zhang:2013cua} even though some significant discrepancies can be seen.
When the observables were calculated from the energy-dependent
amplitudes~\cite{Fernandez-Ramirez:2015tfa}, severe discrepancies showed up.

Kamano {\it et al.} \cite{Kamano:2014zba,Kamano:2015hxa} fitted a similar set of data on the
reactions $\bar K N \to \bar K N, \pi\Sigma, \pi\Lambda$, $\eta\Lambda$, $K\Xi$ and quasi-two-body
final states as reported in~\cite{Zhang:2013cua}. They tried two different models $A$ and $B$
(containing different sets of resonances) and compared their amplitudes with the energy independent
amplitudes of~\cite{Zhang:2013cua}. The three sets of amplitudes (from \cite{Zhang:2013cua} and
from models $A$, $B$ of Ref.~\cite{Kamano:2014zba}), are consistent for the dominant partial waves
but show larger discrepancies for smaller partial-wave amplitudes. It is hence not surprising that
the spectrum of hyperon resonances obtained in the three analyses
\cite{Zhang:2013sva,Kamano:2015hxa,Fernandez-Ramirez:2015tfa} agree only in the leading
contributions. Technically, it often remains open why one resonance is included in a fit and
another one is not.

Already in 2000, the Gie\ss en group \cite{Keil:2000wr} applied modern analysis techniques to study
the reactions $K^-p\to  \pi \Sigma$ and to $\pi \Lambda$. The authors solved the Bethe-Salpeter equation
in an unitary K-matrix approximation and fitted the partial-wave amplitudes derived in~\cite{Gopal:1976gs}.
 The measurements of total cross sections compiled in~\cite{Baldini:1988lb} were imposed as constraints.
Masses, widths and partial decay widths of the leading resonances below 1700\,MeV were determined.

New data in the low-mass region stimulated further investigations. A new Crystal Ball collaboration
was formed at BNL -- exploiting a detector that had originally been built  to study the charmonium
spectrum~\cite{Gaiser:1982ua} and which was then transferred to DESY to study a wide range of
particle physics including two-photon collisions \cite{Bienlein:1992en}. Subsequently, the detector
was exploited for hadron spectroscopy at BNL \cite{Nefkens:2004ui} and is presently used at MAMI in
Mainz~\cite{Beck:2005aj} for photoproduction experiments.

At BNL, the Crystal Ball collaboration made significant contributions to the spectroscopy of
low-mass hyperons. Several reactions were studied at eight incident $K^-$ momenta between 514 and
750\,MeV. Differential and total cross sections and the hyperon polarization were reported for the
reaction $K^-p\to\eta\Lambda$ \cite{Starostin:2001zz} and $K^-p\to\pi^0\Lambda$,
$K^-p\to\pi^0\Sigma$, and  $K^-p\to \bar K^0 n$~\cite{Prakhov:2008dc} (see also
\cite{Manweiler:2008zz}), $K^-p\to\pi^0\pi^0\Lambda$~\cite{Prakhov:2004ri},
$K^-p\to\pi^0\pi^0\Sigma$~\cite{Prakhov:2004an}. A new path to hyperon spectroscopy was opened in
Refs.~\cite{Moriya:2013eb,Moriya:2013hwg,Moriya:2014kpv}.

\begin{table}[pt]
\caption{\label{list}List of reactions used in the partial wave analysis. $\Delta$ denotes the
$\Delta(1232)3/2^+$, $\Lambda^*$ the $\Lambda(1520)3/2^-$, $\Sigma^*$ the $\Sigma(1385)3/2^+$. }
\renewcommand{\arraystretch}{1.3}
\begin{tabular}{lll}
\hline\hline
 $K^-p\to K^- p$                    & $K^-p\to \bar K^0 n$              & $K^-p\to \pi^0 \Lambda$   \\
$K^-p\to\omega\Lambda$           & $K^-p\to \eta \Lambda$          & $K^-p\to \pi^0 \Sigma^0, \eta\Sigma^0$  \\
$K^-p\to \pi^{\mp} \Sigma^{\pm}$ & $K^-p\to K^{+/0} \Xi^{-/0}$   & $K^-p\to K^{-/0} \Delta^{+/0}$ \\
$K^-p\to \pi^\pm \Sigma^{*\mp}$ & $K^-p\to \pi^0 \Lambda^*$        &$K^-p\to K^{*-} p$  \\
$K^-p\to K^{*0} n$              & $K^-p\to \pi^0\pi^0 \Lambda$ & $K^-p\to \pi^0\pi^0 \Sigma$\\
\hline\hline
\end{tabular}
\renewcommand{\arraystretch}{1.0}
\end{table}

This paper is part of a comprehensive study of the hyperon spectrum. In this paper,
a fit is presented to (nearly) all available data on $K^-p$ induced reactions. The reactions used
in the fit are shown in Table~\ref{list}. The emphasis of this paper lies on two-body final states
and on a determination of the hyperon resonances needed to achieve a good fit to the data. In
\cite{Hyperon-III}, data on three-body and quasi-two-body final states are discussed. In that
paper, we present properties of hyperon resonances in detail and compare the resulting spectrum
with the Bonn quark model \cite{Loring:2001ky}. In \cite{Hyperon-I}, we present a fit of
low-energy data on $K^-$ induced reactions, including data on $K^-p$ at rest, and found that only
one pole is required to describe the $\Lambda(1405)1/2^-$ region. For the threshold region, data on
$K^-p$ properties at rest are also very important: The decay ratios ${\Gamma_{K^-p\rightarrow
 \pi^+\Sigma^-}}/{\Gamma_{K^-p\rightarrow \pi^-\Sigma^+}}$, ${\Gamma_{K^-p\rightarrow
 \pi^0\Lambda}}/{\Gamma_{K^-p\rightarrow
 \text{neutral}}}$, and ${\Gamma_{K^-p\rightarrow \pi^\pm\Sigma^\mp}}/{\Gamma_{K^-p\rightarrow
\text{inelastic}}}$ were taken from Refs.~\cite{Tovee:1971ga,Nowak:1978au}, The SIDDHARTA
experiment at DA$\Phi$NE determined the energy shift and width of the 1S level of the kaonic
hydrogen atom~\cite{Bazzi:2011zj,Bazzi:2012eq}. These data proved to be very important for the
study of the $\Lambda(1405)$
region~\cite{Oller:2000fj,Oset:2001cn,Jido:2003cb,Cieply:2009ea,Ikeda:2012au,Guo:2012vv,Mai:2012dt,Mai:2014xna,Roca:2013av,Roca:2013cca,Cieply:2016jby,Miyahara:2018onh} which suggested the existence of two isoscalar
poles in the $\Lambda(1405)$ region. In~\cite{Hyperon-IV} we present a coupled-channel
partial wave analysis of CLAS data on $\gamma p\to K^+ (\pi\Sigma)$~\cite{Moriya:2013hwg} and
$K^+(\bar K N)$ \cite{Dey:2014tfa} and argue that photoproduction may offer new chances to study
hyperon resonances.

The paper is organized as follows: First, in Section~\ref{pwa-bnga}, a short outline of the BnGa
partial-wave-analysis method is presented. In Section~\ref{data}, we list the data used in this
analysis. The search for new or less established resonances is described in Section~\ref{scan}. The
results of the final fit are presented in Section~\ref{results}. In Section~\ref{star}, we suggest
a possible star rating for the resonances used in the final fit. Our partial wave amplitudes for $\bar K
N\to \bar K N$ and $\bar K N \to \pi\Sigma$ scattering in isospin $I=0$ and $I=1$ and for $\bar K N\to
\pi^0\Lambda$ scattering are compared to those from other analyses in Section \ref{amplitudes}. The
paper concludes in Section~\ref{summary} with a short summary.

\section{\label{pwa-bnga}The BnGa partial wave analysis}
\subsection{The scattering amplitude}

The general form of the amplitude for meson-baryon scattering can be written as
\be
A(s,t)=\sum\limits_{IJN}C_IQ_{JN}(s,t)A_{IJN}(s)\,,
\label{ast}
\ee
where $IJN$ are isospin, total spin and ``naturality''. $C_I$ are the Clebsch-Gordan coefficients
which depend on the isospin of all particles in the process (including intermediate states).
$A_{IJN}(s)$ are partial wave amplitudes, and the $Q_{JN}(s,t)$ tensors describe the angular dependent
part of the partial wave amplitudes. The naturality is an alternative way to describe the parity;
it is given by $N=(-1)^{(n+1)}P$ where $n$ corresponds to the rank of the partial wave, $J=n+1/2$.
The angular dependent part is constructed from the decay vertices, polarization vectors and tensors
which describe the structure of the particle propagators (see \cite{Anisovich:2004zz}). For
example, the angular part for the scattering of a pseudoscalar meson and a $J^P=1/2^+$ baryon is
given by
\be
Q_{J\pm}(s,t)&=&\bar u(q_1)\tilde N^{(\pm)}_{\alpha_1\ldots\alpha_n}
(q^\perp) F^{\alpha_1\ldots\alpha_n}_{\xi_1\ldots\xi_n}(P)\times\nn
&&N^{(\pm)}_{\xi_1\ldots\xi_n}(k^\perp) u(k_1)\,.~~~~~
\ee
Here, the $k_1$ and $q_1$ are the momenta of the initial and final-state baryons, and $k_2$ and
$q_2$  are the momenta of the initial and final-state mesons; the relation
$P=(k_1+k_2)=(q_1+q_2),\,s=P^2$ holds true. The momenta $k^\perp$ and $q^\perp$ are relative
momenta in the initial and final states orthogonal to the total momentum $P$:
\be
k^\perp_\mu =\frac 12(k_1-k_2)_\nu g_{\mu\nu}^\perp,
&\qquad q^\perp_\mu =\frac 12(q_1-q_2)_\nu g^\perp_{\mu\nu},\nn
 g^\perp_{\mu\nu}=g_{\mu\nu}-\frac{P_\mu P_\nu}{P^2}\,. &
\label{gperp}
\ee
The baryons are described with bispinors:
\be
u(p)&=&\frac{1}{\sqrt{2m(p_0+m)}} \left (
\begin{array}{c}
(p_0+m)\omega \\
(\vec p\vec \sigma)\omega
\end{array}
\right ), \nonumber \\
\bar u(p)&=
&\frac{1}{\sqrt{2m(p_0+m)}} \left (
\begin{array}{c}
\omega^* (p_0+m) \\
-\omega^*(\vec p\vec \sigma)
\end{array}
\right )\,.
\ee
Here, $\omega$ represents a 2-dimensional spinor and $\omega^*$ the conjugated and transposed
spinor, and we use the normalization condition
\be
\bar u(p) u(p)=1\,,  \qquad \sum\limits_{polarizations}\!\!\!\!\! u(p)
\bar u(p)=\frac{m+\hat p}{2m}\,.
\label{bisp_norm}
\ee

The structure of the resonance propagator corresponds to a convolution of the polarization vectors
of the resonances and have the following covariant form:
\be
F^{\mu_1\ldots\mu_n}_{\nu_1\ldots\nu_n}\!=\!(-1)^n
\frac{\sqrt{s}\!+\!\hat P}{2\sqrt{s}}
O^{\mu_1\ldots\mu_n}_{\xi_1\ldots \xi_n}
T^{\xi_1\ldots\xi_n}_{\beta_1\ldots \beta_n} O^{\beta_1\ldots
\beta_n}_{\nu_1\ldots\nu_n}\,,
\label{fp}
\ee
where
\be T^{\xi_1\ldots\xi_n}_{\beta_1\ldots \beta_n}&=&
\frac{n+1}{2n\!+\!1} \big( g_{\xi_1\beta_1}\!-\!
\frac{n}{n\!+\!1}\sigma_{\xi_1\beta_1} \big)
\prod\limits_{i=2}^{n}g_{\xi_i\beta_i}, \nn
\sigma_{\alpha_i\alpha_j}&=&\frac 12
(\gamma_{\alpha_i}\gamma_{\alpha_j}-
\gamma_{\alpha_j}\gamma_{\alpha_i})\,.
\label{t1}
\ee
The $O^{\mu_1\ldots\mu_n}_{\xi_1\ldots \xi_n}$ describe the structure of the boson propagator for
the particle with spin $J=n$ and are constructed from the metrical tensors orthogonal to the
momentum of the resonance, see Eqn.~(\ref{gperp}). For the lowest spin states,
\be
O\!&=&\! 1\,,\qquad
O^\mu_\nu\!=\!g_{\mu\nu}^\perp=g_{\mu\nu}-\frac{P_\mu P_\nu}{s}\,, \nn
O^{\mu_1\mu_2}_{\nu_1\nu_2}\!&=&\! \frac 12 \hspace{-0.8mm}\left (
g_{\mu_1\nu_1}^\perp  g_{\mu_2\nu_2}^\perp \!+\!
g_{\mu_1\nu_2}^\perp  g_{\mu_2\nu_1}^\perp  \!- \!\frac 23
g_{\mu_1\mu_2}^\perp  g_{\nu_1\nu_2}^\perp \hspace{-0.8mm}\right ).~
\ee
For higher states, the operator can be calculated using the recurrent expression
\be &&O^{\mu_1\ldots\mu_L}_{\nu_1\ldots
\nu_L}=\frac{1}{L^2} \bigg (
\sum\limits_{i,j=1}^{L}g^\perp_{\mu_i\nu_j}
O^{\mu_1\ldots\mu_{i-1}\mu_{i+1}\ldots\mu_L}_{\nu_1\ldots
\nu_{j-1}\nu_{j+1}\ldots\nu_L}-
\nonumber \\
 &&  \frac{4}{(2L-1)(2L-3)} \times
\nn    && \sum\limits_{i<j\atop k<m}^{L}
g^\perp_{\mu_i\mu_j}g^\perp_{\nu_k\nu_m}
O^{\mu_1\ldots\mu_{i-1}\mu_{i+1}\ldots\mu_{j-1}\mu_{j+1}\ldots\mu_L}_
{\nu_1\ldots\nu_{k-1}\nu_{k+1}\ldots\nu_{m-1}\nu_{m+1}\ldots\nu_L}
\bigg )\,.
\ee
The operator $O^{\mu_1\ldots\mu_n}_{\nu_1\ldots \nu_n}$ provides the symmetry and traceless
condition for the indices within one group and the orthogonality to the particle momentum. The
structure of the $T^{\mu_1\ldots\mu_n}_{\nu_1\ldots \nu_n}$ operator is unique and is defined by
the orthogonality condition to the $\gamma$-matrix and the normalization condition
\be
F^{\mu_1\ldots\mu_n}_{\nu_1\ldots\nu_n}F^{\nu_1\ldots\nu_n}_{\xi_1\ldots\xi_n}=
(-1)^nF^{\mu_1\ldots\mu_n}_{\xi_1\ldots\xi_n}\,.
\ee

The vertex functions $N^{(\pm)}_{\alpha_1\ldots\alpha_n}$ describe the spin structure of a
resonance decaying into a baryon with $J^P=1/2^+$ and a pseudoscalar meson. The decay orbital
momentum is connected with the total spin as $J=L+\frac 12$ for the '+' states and as $J=L-\frac
12$ for the '-' naturality states.
\be
N^{(+)}_{\mu_1\ldots\mu_n}(k^\perp)\!&=\!
X^{(n)}_{\mu_1\ldots\mu_n}(k^\perp)&~~~L=n\,, \nn
N^{(-)}_{\mu_1\ldots\mu_{n}}(k^\perp)\!&=\! i\gamma_5 \gamma_\nu
X^{(n+1)}_{\nu\mu_1\ldots\mu_{n}}(k^\perp)&~~~L=n+1,
\label{vert_fun}
\ee
where $X^{(n)}_{\mu_1\ldots\mu_n}(k^\perp)$ is the orbital momentum operator which depends on the
relative momentum orthogonal to the momentum of the decaying particle:
\be
X^{(0)}&=&1\ , \qquad X^{(1)}_\mu=k^\perp_\mu=\frac{1}{2}(k_1-k_2)_\nu\,g^\perp_{\mu\nu}\ , \qquad\nonumber \\
X^{(2)}_{\mu_1 \mu_2}&=&\frac32\left(k^\perp_{\mu_1}
k^\perp_{\mu_2}-\frac13\, k^2_\perp g^\perp_{\mu_1\mu_2}\right), \nonumber  \\
X^{(3)}_{\mu_1\mu_2\mu_3}&=&\frac52\Big[k^\perp_{\mu_1}
k^\perp_{\mu_2 } k^\perp_{\mu_3} \nn
&-&\frac{k^2_\perp}5\left(g^\perp_{\mu_1\mu_2}k^\perp
_{\mu_3}+g^\perp_{\mu_1\mu_3}k^\perp_{\mu_2}+
g^\perp_{\mu_2\mu_3}k^\perp_{\mu_1} \right)\Big]\,.~~~
\ee
The operators $X^{(L)}_{\mu_1\ldots\mu_L}$ for $L\ge 1$ can be written in the form of the
recurrence expression
\be
X^{(L)}_{\mu_1\ldots\mu_L}&=&k^\perp_\alpha
Z^{\alpha}_{\mu_1\ldots\mu_L}\,,
\ee
where
\be
&&Z^{\alpha}_{\mu_1\ldots\mu_L}\!\!\!=\!\frac{2L-1}{L^2}
\sum^L_{i=1}X^{{(L-1)}}_{\mu_1\ldots\mu_{i-1}\mu_{i+1}\ldots\mu_L}
g^\perp_{\mu_i\alpha}-
\nonumber \\
&& \frac{2}{L^2}  \sum^L_{i,j=1 \atop i<j} g^\perp_{\mu_i\mu_j}
X^{{(L-1)}}_{\mu_1\ldots\mu_{i-1}\mu_{i+1}\ldots\mu_{j-1}\mu_{j+1}
\ldots\mu_L\alpha}\,.~~\,
\label{z}
\ee
In our calculations we use the $\gamma$-matrices in the standard representation
\be
\gamma_0=\left ( \begin{array}{cc} 1 & 0 \\ 0 & -1 \end{array}
\right ),\,\, \vec \gamma=\left (
\begin{array}{cc}
0 & \vec \sigma \\
-\vec \sigma & 0
\end{array}
\right ),\,\, \gamma_5=\left (
\begin{array}{cc}
0 & 1 \\
1 & 0
\end{array}
\right ).~
\ee

In the c.m.s. of the reaction the scattering amplitude~(\ref{ast}) can be rewritten as:
\be
&&A(s,t)=\omega^*\left [G(s,t)+H(s,t)i(\vec \sigma \vec n) \right
]\omega' \;,
\nonumber \\
&&G(s,t)=\sum\limits_L \big [(L\!+\!1)F_L^+(s)- L F_L^-(s)\big ]
P_L(z) \;, \nonumber \\
&&H(s,t)=\sum\limits_L \big [F_L^+(s)+ F_L^-(s)\big ] P'_L(z) \;,
\label{piN_others}
\ee
where $\omega$ and $\omega'$ are nonrelativistic spinors and $\vec n$ is a unit vector normal to
the decay plane. The $F$-functions are defined as follows:
\be
F^+_L&=&(|\vec k||\vec q|)^L \chi_i\chi_f\;\frac{\alpha_L}{2L\!+\!1}
\sum\limits_I A_{I(L+\frac 12)+}(s) \;,
\nonumber \\
F^-_L&=&(|\vec k||\vec q|)^L \chi_i\chi_f\;\frac{\alpha_L}{L}
\sum\limits_I A_{I(L-\frac 12)-}(s)\,, \\
\chi_i&=&\sqrt{\frac{m_N+k_{N0}}{2m_N}}\;, \qquad
\chi_f=\sqrt{\frac{m_N+q_{N0}}{2m_N}}\;,\nn
\alpha_L&=&\frac{(2L-1)!!}{L!}
\ee

The approach through the standard $G$ and $H$ functions is absolutely identical to the our
covariant approach, the covariant approach allows us to construct naturally the amplitudes with
multibody final states and to perform a combined analysis of all available data sets.

\subsection{t and u-channel exchange amplitudes}

\begin{figure}
\centerline{\includegraphics[width=0.30\textwidth]{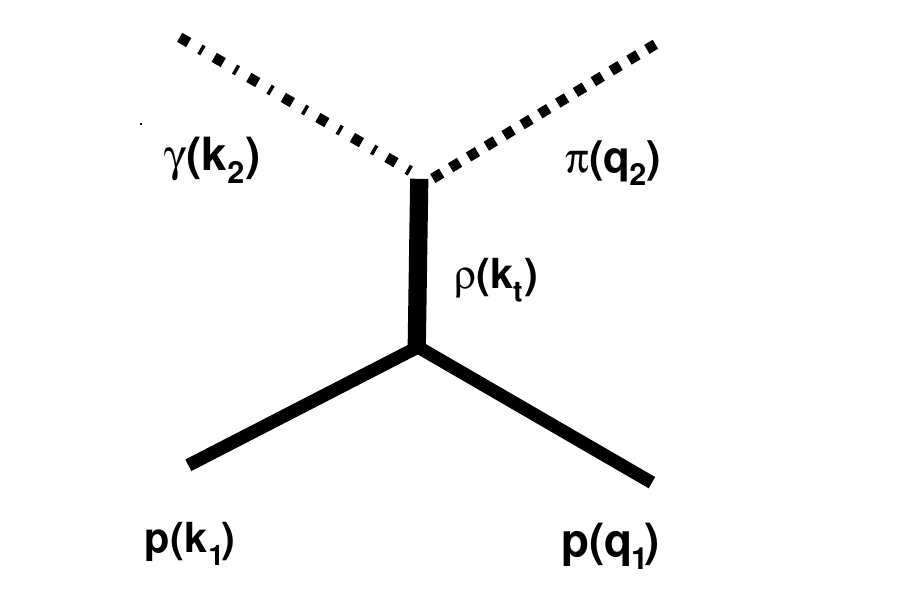}
\hspace{-12mm}\includegraphics[width=0.30\textwidth]{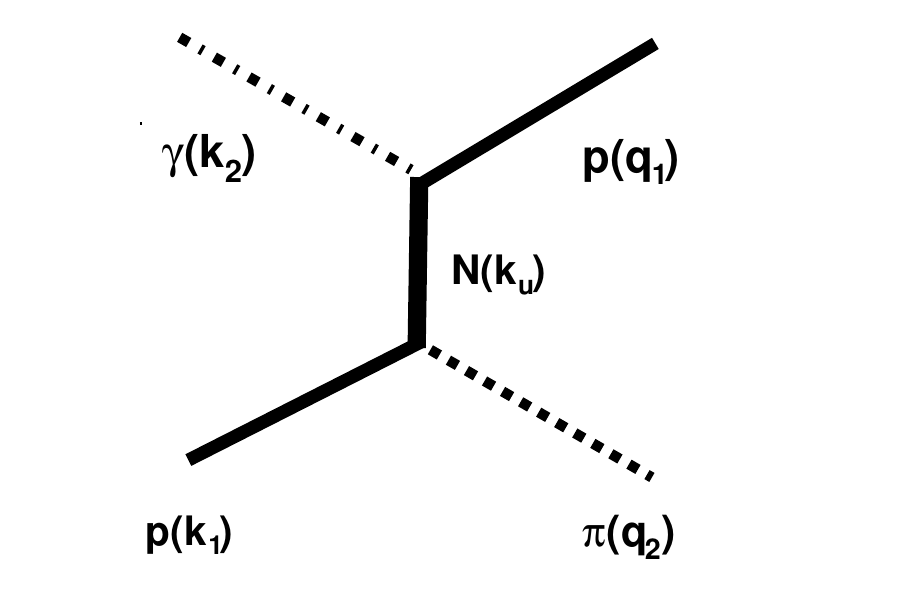}}
\caption{Diagrams for
the $t$- and $u$-channel exchange amplitudes.}
\label{tu}\vspace{-2mm}
\end{figure}

Non-resonance contributions to the reactions are described by constants in the
$K$-matrix (see below) and by amplitudes for $t$ and $u$-channel exchanges (see Fig.~\ref{tu}).

The Reggeized amplitudes for $t$-channel exchange are given by
\be
A(t)=g_1(t)g_2(t)\frac{1+\xi
exp(-i\pi\alpha(t))}{\sin(\pi\alpha(t))} \left (\frac{\nu}{\nu_0}
\right )^{\alpha(t)} .
\ee
Here $\nu=\frac 12 (s-u)$, $\alpha(t)$ is the intercept of the Regge trajectory, and $\xi$ is its
signature. We found a significant contribution only from the exchange by the $\rho$, $\omega$ and
$K^*$ vector-meson trajectories and the $f_0$, $a_0$ and $K^*_0$ trajectories of scalar mesons.

The $u$-channel amplitudes are described as the exchange of the corresponding baryon. We observe
significant contributions only from exchange of nucleons in the $Kp\to\Lambda\pi$ and
$Kp\to\Sigma\pi$ reactions. This exchange is represented by a spin-1/2 propagator\vspace{-2mm}
\be
1/(m_p-\hat k_u) \quad{\rm where}\quad u=k_u^2\,.
\ee
\subsection{\label{Structure}The structure of the partial wave amplitudes}

As the first step we parameterized the partial wave amplitudes as a sum of Breit-Wigner resonances:
\be
A(s)=\sum\limits_\beta\frac{g^{\beta}_{in}g^{\beta}_{out}}
{M_\beta^2-s-iM_\beta\Gamma^{\beta}_{tot}}
\ee
The total width of a resonance is equal to the sum of its partial widths:
\be
M_\beta\Gamma^{\beta}_{tot}=\sum\limits_j
M_\beta\Gamma^{\beta}_j=\sum\limits_j\rho_j^{n\pm}(s)
(g^\beta_j)^2\,,
\ee
where $J=n+\frac 12$; $\rho_j^{n\pm}(s)$ is the imaginary part of the loop diagram with vertices
given in Eq.~(\ref{vert_fun}). In the case of the pseudoscalar meson-baryon channel these functions
are given by
\be
\rho^{n+}_j(s)\!&=&\! \frac {\alpha_n}{2n+1}\frac{|\vec
k_j|^{2n}}{B(n,|\vec k_j|^2,R)}
 \frac{m\!+\!k_{0}}{2m} \frac{2|\vec k_j|}{16\pi \sqrt s}\,,~~
\nn \rho^{n-}_j(s)\!&=&\!\frac {\alpha_{n+1}}{n+1}\frac{|\vec
k_j|^{2n+2}}{B(n\!+\!1,|\vec k_j|^2,R)} \frac{m\!+\!k_{0}} {2m}
\frac{2|\vec k_j|}{16\pi \sqrt s}\,,~~
\label{rho_piN}
\ee
where $k_j=(k_0,\vec k)$ is momentum of the final-state baryon in
the channel $j$ calculated in the resonance rest system. The
Blatt-Weiskopf form factor ${B(n,|\vec k_j|^2,R)}$ (with $R=0.8$ fm)
is introduced to provide the correct asymptotic behavior of the
phase volume at large energies.

The phase volumes for the three particle final states are described
by the spectral integral which has all corresponding cuts and
branching points in the complex plane of the total energy. The
formulae for these functions are given in \cite{Anisovich:2006bc}.

\begin{table*}
\caption{\label{data-diff1}Differential cross sections for $K^-p\to K^-p$ elastic  and  $K^-p\to \bar K^0n$ charge exchange scattering
used in this analysis. Listed are the reaction, the momenta at which the data are given with a reference, the number
of data points, and the $\chi^2$ from the primary ($\chi_1^2$) and from the final fit ($\chi_2^2$).\vspace{2mm}
}
\centering\scriptsize
\renewcommand{\arraystretch}{1.35}
\begin{tabular}{|c|lc|ccc|}
\hline\hline
Reaction      & Mass & Ref. & $N_{\rm data}$ & $\chi^2_1$& $\chi^2_2$\\\hline
$K^-p\to K^-p$
& 1464 1466 1469 1472 1475 1478 1481 1484 1488 1491 1494 1498 1501 1505 1509 &&&&\\[-0.5ex]
& 1512 1516 1520 1524 1528 1532 1536 1540 1544 1548                                              & \cite{Mast:1975pv}             & 482  & 1313 &  990    \\\cline{2-6}
& 1732 1749 1758 1763 1768 1772 1777 1789                                                        & \cite{Cameron:1981qi}          & 320  &  760 &  694    \\\cline{2-6}
& 1775 1796 1815 1833 1852 1870 1889 1907 1925 1941 1957                                         & \cite{Conforto:1975nw}         & 451  & 1216 &  852    \\\cline{2-6}
& 1858 1869 1877 1887 1902 1911 1921 1930                                                        & \cite{Griselin:1975pa}         & 311  &  406 &  408    \\\cline{2-6}
& 2207 2246 2288 2328 2365 2397 2436                                                             & \cite{deBellefon:1976qr}       &  75  &  189 &  124    \\\cline{2-6}
& 1696 1687 1681 1671 1662 1652 1642 1633 1624 1615 1606 1595 1586 1578 1569 &&&&\\[-0.5ex]
& 1561 1552 1544 1536                                                                            & \cite{Armenteros:1970eg}       & 752  & 1054 &  827    \\\cline{2-6}
& 1611 1626 1640 1654 1667 1680 1692 1704 1715 1726 1735 1746 1757 1767                          & \cite{Adams:1975mg}            & 491  &  810 &  675    \\\cline{2-6}
& 1689 1702 1717 1724 1734 1744 1748 1754 1763 1772 1779 1789 1804 1814 1822 &&&&\\[-0.5ex]
& 1831 1841 1848 1856 1865 1875 1879 1898                                                        & \cite{Conforto:1971vy}         & 896  &  969 &  907    \\\cline{2-6}
& 1735 1763 1798 1810 1819 1876 1909 1946                                                        & \cite{Albrow:1971yu}           & 219  &  306 &  259    \\\cline{2-6}
& 1970 1992 2014 2037 2059 2080 2102 2123  2144 2186 2207 2227 2248 2268 2288                    & \cite{Daum:1968jey}            & 327  &  833 &  656    \\\cline{2-6}
& 1785 1832 1860 1892 1924 2122                                                                  & \cite{AnderssonAlmehed:1970nb} & 162  &  465 &  431    \\\cline{2-6}
& 2122 2137 2151 2166 2181 2196 2212 2229 2244 2260 2276 2292 2331 2348 2365 &&&&\\[-0.5ex]
& 2382 2400 2417                                                                                 & \cite{Barber:1975ci}           & 268  & 1202 &  811    \\\cline{2-6}
& 1837 1849 1859 1869 1879 1889 1854 1911 1922 1933 1944 1956 1967                               & \cite{Barber:1975rg}           & 416  & 1868 & 1687    \\\cline{2-6}
\hline
 $K^-p\to \bar K^0n$
& 1466 1469 1472 1475 1478 1481 1484 1488                                                        & \cite{Mast:1975pv}             &  64  &  383 &  182  \\\cline{2-6}
& 1491 1494 1498 1501 1505 1509 1512 1516 1520 1524 1528 1532 1536 1540 1544                     & \cite{Mast:1975pv}             & 297  &  696 &  475  \\\cline{2-6}
& 1732 1749 1758 1763 1768 1772 1777 1789                                                        & \cite{Cameron:1981qi}          & 160  &  276 &  267  \\\cline{2-6}
& 1775 1796 1815 1833 1852 1870 1889 1907 1925 1941 1957                                         & \cite{Conforto:1975nw}         & 220  &  744 &  558  \\\cline{2-6}
& 1858 1869 1877 1887 1902 1911 1921 1930 1935 1949 1963 1976 1992                               & \cite{Griselin:1975pa}         & 260  &  414 &  387  \\\cline{2-6}
& 2207 2246 2288 2328 2365 2397 2436                                                             & \cite{deBellefon:1976qr}       &  75  &  177 &  102  \\\cline{2-6}
& 1696 1687 1681 1671 1662 1652 1642 1633 1624 1615 1606 1595 1586 1578 1569 &&&&\\[-0.5ex]
& 1561 1552 1544 1536                                                                            & \cite{Armenteros:1970eg}       & 380  &  618 &  572  \\\cline{2-6}
& 1569 1589 1598 1620 1634 1647 1659 1676                                                        & \cite{Prakhov:2008dc}          & 128  &  325 &  216  \\\cline{2-6}
& 1729 1739 1741 1747 1755 1763 1775 1780 1794                                                   & \cite{Jones:1974at}            & 333  &  400 &  398  \\\cline{2-6}
& 1915 1939 1963 1984 2006 2027 2042 2068 2088 2111 2125 2151 2169                               & \cite{Litchfield:1971ri}       & 260  &  515 &  394  \\\cline{2-6}
& 1569 1578 1587 1597 1606 1615 1624 1633 1643 1652 1661 1670 1680 1689 1708 &&&& \\[-0.5ex]
& 1717 1726 1736 1745 1755 1763 1773                                                             & \cite{AlstonGarnjost:1977ct}   & 808  & 1837 & 1237  \\\cline{2-6}
& 1689 1702 1717 1724 1734 1744 1748 1754 1763 1772 1779 1789 1804 1814 1822 &&&&\\[-0.5ex]
& 1831 1841 1848 1856 1865 1875 1879 1898                                                        & \cite{Armenteros:1969kn}       & 460  &  498 &  534  \\\cline{2-6}
\hline
\end{tabular}
\end{table*}

In the case of overlapping resonances the sum of Breit-Wigner amplitudes can violate the unitarity
condition. In this case one can use the $K$-matrix approach which satisfies unitarity. Here the
partial wave amplitudes are represented by a matrix with elements which describe the transition
between channels $i$ and $j$ ($i,j=pK,\pi\Lambda,\pi\Sigma\ldots$) in the form
\be
A_{ij}(s)=K_{im}(I-i\hat \rho \hat K)_{mj}^{-1}\,,
\ee
where $\hat \rho$ is the diagonal matrix of the phase volumes. The $K$-matrix is parameterized as a
sum of $K$-matrix poles and nonresonant contributions,
\be
K_{ij}=\sum\limits_\beta\frac{g^{\beta}_{i}g^{\beta}_{j}}
{M_\beta^2-s}+f_{ij}\,.
\label{km_ampl}
\ee
This parameterization does not take into account rescattering effects described, e.g., by triangle
diagrams. For baryons this effects exists even for two-particle final states since the pseudoscalar
meson can be re-absorbed by the baryon. To take into account such effects, we substitute the
$K$-matrix approach by the so-called  $D$-matrix approach \cite{Anisovich:2011zz}. Here the
amplitude has the form:
\be
A_{ij}(s)=K_{im}(I-i\hat \rho \hat K)_{mk}^{-1}(i\hat \rho \hat D)_{kj}+D_{ij}\,,
\ee
where the $D$-matrix is given by
\be
D_{ij}=\sum\limits_\beta\frac{g^{\beta}_{i}G^{\beta}_{j}}
{M_\beta^2-s}+F_{ij}\,.
\label{d_ampl}
\ee
In the present analysis we describe the decay parameters of the
D-matrix by
\be
G^\beta_j=g^\beta_j\exp{(i\phi^\beta_j)},\quad{\rm and}\quad
F_{ij}=f_{ij}\exp{(i\phi_{ij})}\,.
\ee
The (small) phases take into account the contribution of multiparticle
scattering diagrams.

\begin{table*}
\caption{\label{data-diff2}Differential cross sections on $K^-p$ scattering into two-body final
states used in this analysis. Listed are the reaction, the momenta at which the data are given with
a reference, the number of data points, and the $\chi^2$ from the primary ($\chi_1^2$) and from the final
fit ($\chi_2^2$). (LC: data given as Legendre coefficients).  \vspace{2mm} } \centering\scriptsize
\renewcommand{\arraystretch}{1.32}
\begin{tabular}{|c|lc|ccc|}
\hline\hline
 $K^-p\to \pi^0\Lambda$
& 1732 1749 1758 1763 1768 1772 1777 1789                                                        & \cite{Cameron:1981qi}    & 160 & 467 & 362  \\\cline{2-6}
& 1775 1796 1815 1833 1852 1870 1889 1907  1925 1941 1957                                        & \cite{Conforto:1975nw}   & 220 & 468 & 362  \\\cline{2-6}
& 1858 1869 1877 1887 1902 1911 1921 1930 1935 1949 1963 1976 1992                               & \cite{Griselin:1975pa}   & 256 & 656 & 621  \\\cline{2-6}
& 1696 1687 1681 1671 1662 1652 1642 1633 1624 1615 1606 1595 1586 1578 1569 &&&&\\[-0.5ex]
& 1561 1552 1544 1536                                                                            & \cite{Armenteros:1970eg} & 380 & 743 & 610  \\\cline{2-6}
& 1569 1589 1598 1620 1634 1647 1659 1676                                                        & \cite{Prakhov:2008dc}    & 125 & 331 & 166  \\\cline{2-6}
& 1729 1739 1741 1747 1755 1763 1775 1780 1794                                                   & \cite{Jones:1974at}      & 341 & 488 & 427  \\\cline{2-6}
& 1689 1702 1717 1724 1734 1744 1748 1754 1763 1772 1779 1789 1804 1814 1822 &&&&\\[-0.5ex]
& 1831 1841 1848 1856 1865 1875 1879 1898                                                        & \cite{Armenteros:1969kn} & 460 & 756 & 701  \\\cline{2-6}
& 1915 1939 1963 1984 2006 2027 2042 2068 2088 2111 2125 2151 2169                               & \cite{Berthon:1970fd}    & 260 & 842 & 351  \\\cline{2-6}
& 1648 1657 1666 1675 1684 1692 1702 1711 1719 1728 1737 1746 1754 1763                          & \cite{Baxter:1974zs}     & 126 & 230 & 235  \\\cline{2-6}
& 1600 1630 1648 1663 1678 1693 1708 1722 1740                                                   & \cite{London:1975zz}     &  90 & 158 & 133  \\\cline{2-6}
& 2207 2246 2288 2328                                                                            & \cite{deBellefon:1975lp} &  60 & 152 & 141  \\\cline{2-6}
 \hline
 $K^-p\to \pi^0\Sigma^0$
& 1696 1687 1681 1671 1662 1652 1642 1633 1624 1615 1606 1595 1586 1578 1569 &&&&\\[-0.5ex]
& 1561 1552 1544 1536                                                                            & \cite{Armenteros:1970eg}  & 190 & 338 & 383  \\\cline{2-6}
& 1569 1589 1598 1620 1634 1647 1659 1676                                                        & \cite{Prakhov:2008dc}     & 125 & 361 & 281  \\\cline{2-6}
& 1648 1657 1666 1675 1684 1692 1702 1711 1719 1728 1737 1746 1754 1763                          & \cite{Baxter:1974zs}      & 140 & 235 & 229  \\\cline{2-6}
& 1605 1640 1660  1680 1700 1730                                                                 & \cite{London:1975zz}      &  54 &  88 &  77  \\\cline{2-6}
& 1569 1589 1598 1620 1634 1647 1659 1676                                                        & \cite{Manweiler:2008zz}   &  72 & 238 & 169  \\\cline{2-6}
 \hline
 $K^-p\to \pi^-\Sigma^+$
& 1732 1749 1758 1763 1768 1772 1777 1789                                                        & \cite{Cameron:1981qi}    & 160  &  460 &  298  \\\cline{2-6}
& 1775 1796 1815 1833 1852 1870 1889 1907                                                        & \cite{Conforto:1975nw}   & 220  &  490 &  400  \\\cline{2-6}
& 1858 1869 1877 1887 1902 1911 1921 1930 1935 1949 1963 1976 1992                               & \cite{Griselin:1975pa}   & 259  & 2838 & 2224  \\\cline{2-6}
& 1696 1687 1681 1671 1662 1652 1642 1633 1624 1615 1606 1595 1586 1578 1569 &&&&\\[-0.5ex]
& 1561 1552 1544 1536                                                                            & \cite{Armenteros:1970eg} & 380  &  551 &  439  \\\cline{2-6}
& 1729 1739 1741 1747 1755 1763 1775 1780 1794                                                   & \cite{Jones:1974at}      & 304  &  446 &  400  \\\cline{2-6}
& 1689 1702 1717 1724 1734 1744 1748 1754 1763 1772 1779 1789 1804 1814 1822 &&&&\\[-0.5ex]
& 1831 1841 1848 1856 1865 1875 1879 1898                                                        & \cite{Armenteros:1969kn} & 460  &  784 &  627  \\\cline{2-6}
& 1915 1939 1963 1984 2005 2028 2042 2068 2088 2111 2126 2151 2169                               & \cite{Berthon:1970sp}    & 252  &  866 &  577  \\\cline{2-6}
& 2207 2246 2288 2328 2365 2397 2436                                                             & \cite{deBellefon:1977sp} &  60  &  250 &  276  \\\cline{2-6}
 \hline
$K^-p\to \pi^+\Sigma^-$
& 1732 1749 1758 1763 1768 1772 1777 1789                                                        & \cite{Cameron:1981qi}    & 160 & 271 & 205  \\\cline{2-6}
& 1775 1796 1815 1833 1852 1870 1889 1907 1925 1941 1957                                         & \cite{Conforto:1975nw}   & 220 & 684 & 381  \\\cline{2-6}
& 1858 1869 1877 1887 1902 1911 1921 1930 1935 1949 1963 1976 1992                               & \cite{Griselin:1975pa}   & 259 & 383 & 234  \\\cline{2-6}
& 1696 1687 1681 1671 1662 1652 1642 1633 1624 1615 1606 1595 1586 1578 1569                 &&&&\\[-0.5ex]
& 1561 1552 1544 1536                                                                            & \cite{Armenteros:1970eg} & 380 & 863 & 889  \\\cline{2-6}
& 1729 1739 1741 1747 1755 1763 1775 1780 1794                                                   & \cite{Jones:1974at}      & 306 & 395 & 342  \\\cline{2-6}
& 1689 1702 1717 1724 1734 1744 1748 1754 1763 1772 1779 1789 1804 1814 1822                 &&&&\\[-0.5ex]
& 1831 1841 1848 1856 1865 1875 1879 1898                                                        & \cite{Armenteros:1969kn} & 460 & 666 & 597  \\\cline{2-6}
& 1915 1939 1963 1984 2005 2028 2042 2068 2088 2111 2126 2151 2169                               & \cite{Berthon:1970sp}    & 240 & 740 & 421  \\\cline{2-6}
& 2207 2246 2288 2328 2365 2397 2436                                                             & \cite{deBellefon:1977sp} &  57 & 173 & 210   \\\cline{2-6}
\hline
$K^-p\to \eta\Lambda$
& 1664 1665 1666 1667 1668 1669 1670 1672 1674 1676 1678 1680 1682 1683 1685 & \cite{Starostin:2001zz}    &  135  & 316 & 190 \\\cline{2-6}
& 1664 1671 1681 1687 1696                                                   & \cite{Armenteros:1970eg}   &   25  & 125 &  50 \\\cline{2-6}
\hline
$K^-p\to \eta\Sigma$
&  1750 1765 1780 1795                                                  & \cite{Jones:1974si}   &   16  & 12 &  11 \\\cline{2-6}
\hline
$K^-p\to K^0\Xi^0$
& 2020                                                                                          & \cite{Berge:1966jp}        & 17 & 15 & 14     \\\cline{2-6}
& 1970 2070 2140                                                                                & \cite{Burgun:1968g}        & 26 & 24 & 15     \\\cline{2-6}
& 2110 2280 2470                                                                                & \cite{Dauber:1969pm}       & 16 & 20 & 20     \\\cline{2-6}
& 2150                                                                                          & \cite{Carlson:1973jr}      & 8  & 24 & 22     \\\cline{2-6}
& 1263 1316 1368 1415 1462 1514 1546 1606 1653 1705 1741 1800 1843 1934   &&&&\\[-0.5ex]
& 2031 2135 2234 2331                                                                           & \cite{deBellefon:1977sw}   & 29 & 64 & 46   \\\cline{2-6}
\hline
$K^-p\to K^+\Xi^-$
& 1138 1161 1179 1201 1233 1253 1276 1296 1305 1336 1367 1396 1434  & \cite{Griselin:1975pa}      &      &      &    \\
& 1970 2070 2140                                                                                & \cite{Burgun:1968g}        & 101 & 182 & 168   \\\cline{2-6}
& 2110 2280 2420 2480                                                                           & \cite{Dauber:1969pm}       &  60 &  77 &  66   \\\cline{2-6}
& 2240                                                                                          & \cite{Trippe:1967tg}       &  20 &  25 &  18   \\\cline{2-6}
& 1950                                                                                          & \cite{Trower:1968wp}       &  12 &  10 &  10   \\\cline{2-6}
& 1263 1316 1368 1415 1462 1514 1546 1606 1653 1705 1741 1800 1843 1934   &&&&\\[-0.5ex]
& 2031 2135 2234 2331 2412 2516                                                                 & \cite{deBellefon:1977sw}   &  16 &  24 & 16   \\\cline{2-6}
\hline
$K^-p\to \bar K^0 n$
& 1709 1738 1758 1767 1782 1803 1821 1846 1865 1887 1919 1937 1953   &&&&\\[-0.5ex]
(LC)
& 1970 2001 2022 2051 2085 2106                                              & \cite{Horn:1972aj}      &  136 & 303 & 253   \\\cline{2-6}
\hline
$K^-p\to \pi^0\Lambda$
& 1709 1738 1758 1767 1782 1803 1821 1846 1865 1887 1919 1937 1953  &&&&\\[-0.5ex]
 (LC) & 1970 2001 2022 2051 2085 2106                                              & \cite{Horn:1972aj}      &  128 & 211 & 219   \\\cline{2-6}
& 1647 1656 1673 1677 1692 1698 1709 1715                                    & \cite{Ponte:1975ra}     &   40 & 138 &  97   \\\cline{2-6}
\hline
\end{tabular}\vspace{3mm}
\end{table*}
\begin{table*}
\caption{\label{data-pol}Data on the polarization observable $P$ in $K^-p$ scattering
used in this analysis. Listed are the reaction, the momenta at which the data are given with a reference, the number
of data points, and the $\chi^2$ from the primary ($\chi_1^2$) and from the final fit ($\chi_2^2$). \vspace{1mm}
}
\centering\scriptsize
\renewcommand{\arraystretch}{1.3}
\begin{tabular}{|c|lc|ccc|}
\hline
Reaction      & Mass & Ref. & $N_{\rm data}$ & $\chi^2_1$& $\chi^2_2$\\\hline
$K^-p\to K^-p$
& 1630 1652 1669 1687 1706 1720 1811 1827                                    & \cite{Ehrlich:1977bs}             & 134 & 245 & 285  \\\cline{2-6}
& 1735 1763 1798 1810 1819 1876 1909 1946                                    & \cite{Albrow:1971yu}              & 230 & 305 & 275  \\\cline{2-6}
& 1970 1992 2014 2037 2059 2080 2102 2123 2144 2186 2207 2227 2248 2268 2288 & \cite{Daum:1968jey}               & 327 & 801 & 485  \\\cline{2-6}
& 1785 1832 1860 1892 1924 2122                                              & \cite{AnderssonAlmehed:1970nb}    & 168 & 235 & 159  \\\cline{2-6}
& 1772 1791 1810 1828 1847 1865 1883 1902 1920                               & \cite{Bryant:1980yp}              & 321 & 617 & 455  \\\cline{2-6}
\hline
$K^-p\to \pi^0\Lambda$
& 1732 1749 1758 1763 1768 1772 1777 1789                                    & \cite{Cameron:1981qi}             & 160 & 236 & 162  \\\cline{2-6}
& 1775 1796 1815 1833 1852 1870 1889 1907  1925 1941 1957                    & \cite{Conforto:1975nw}            & 187 & 371 & 238  \\\cline{2-6}
& 1569 1589 1598 1620 1634 1647 1659 1676                                    & \cite{Prakhov:2008dc}             & 123 & 346 & 140  \\\cline{2-6}
& 1729 1739 1741 1747 1755 1763 1775 1780 1794                               & \cite{Jones:1974at}               & 153 & 183 & 155  \\\cline{2-6}
& 1689 1702 1717 1724 1734 1744 1748 1754 1763 1772 1779 1789 1804 1814                 &&&&\\[-0.5ex]
& 1822 1831 1841 1848 1856 1865 1875 1879                                    & \cite{Armenteros:1969kn}          & 160 & 182 & 167  \\\cline{2-6}
& 1915 1939 1963 1984 2006 2027 2042 2068 2088 2111 2125 2151 2169           & \cite{Berthon:1970fd}             &  88 & 309 & 210  \\\cline{2-6}
& 2207 2246 2288 2328                                                        & \cite{deBellefon:1975lp}          &  21 &  69 &  38  \\\cline{2-6}
\hline
$K^-p\to \pi^0\Sigma^0$
& 1569 1589 1598 1620 1634 1647 1659 1676                                    & \cite{Prakhov:2008dc}             & 124 & 572 & 299  \\\cline{2-6}
\hline
$K^-p\to \pi^-\Sigma^+$
& 1732 1749 1758 1763 1768 1772 1777 1789                                    & \cite{Cameron:1981qi}             & 160 & 663 & 672  \\\cline{2-6}
& 1775 1796 1815 1833 1852 1870 1889 1907 1925 1941 1957                     & \cite{Conforto:1975nw}            & 146 & 236 & 161  \\\cline{2-6}
& 1729 1739 1741 1747 1755 1763 1775 1780 1794                               & \cite{Jones:1974at}               &  92 & 114 & 116  \\\cline{2-6}
& 1689 1702 1717 1724 1734 1744 1748 1754 1763 1772 1779 1789 1804 1814                  &&&&\\[-0.5ex]
& 1822 1831 1841 1848 1856 1865 1874 1879 1898                               & \cite{Armenteros:1969kn}          & 180 & 236 & 232  \\\cline{2-6}
& 2207 2246 2288 2328                                                        & \cite{deBellefon:1977sp}          &  15 &  35 &  61  \\\cline{2-6}
\hline
$K^-p\to K^0\Xi^0$
& 1973 1973 2022 2065 2109 2138 2151 2274 2274 2468                          & \cite{Sharov:2011xq}              & 11  &  22 &  19  \\\cline{2-6}
\hline
$K^-p\to K^-\Xi^+$
& 1973 1973 2022 2065 2109 2138 2151 2274 2274 2468                          & \cite{Sharov:2011xq}              & 18  &  17 &  17  \\\cline{2-6}
\hline
\end{tabular}\vspace{3mm}
\end{table*}

\begin{table*}
\caption{\label{data-three}Data on the $K^-p$ induced reactions with three-body final states. For
all reactions, differential cross sections and three $\rho$-density matrix elements were expanded
into associated Legendre polynomials and Legendre coefficients for $l=1, \cdots 7$ were determined.
Listed are the reaction, the momenta at which the data are given with a reference, the number of
data points, and the $\chi^2$ from the primary ($\chi_1^2$) and from the final fit ($\chi_2^2$). Here, $\Delta$ denotes
$\Delta(1232)3/2^+$, $\Lambda^*$ is $\Lambda(1520)3/2^-$, and $\Sigma^*$ denotes
$\Sigma(1385)3/2^+$.  (LC: data given as Legendre coefficients). \vspace{2mm} } \centering\scriptsize
\renewcommand{\arraystretch}{1.35}
\begin{tabular}{|c|lc|ccc|}
\hline
                 & Mass & Ref. & $N_{\rm data}$ & $\chi^2_1$& $\chi^2_2$\\\hline
$K^-p\to  \omega\Lambda$
& 1915 1940 1963 1984 2005 2028 2042 2068 2088 2111 2126 2151 2169   \qquad\qquad        & \cite{Brandstetter:1972xp} & 130 & 242 & 147   \\\cline{2-6}
(LC)& 1915 1940 1963 1984 2005 2028 2042 2068 2088 2111 2126 2151 2169           & \cite{Brandstetter:1972xp} &  94 & 314 & 487   \\\cline{2-6}
& 1988 2012 2031 2051 2070 2088 2105 2120 2136 2151                          & \cite{Nakkasyan:1975yz}    & 100 &  91 &  72  \\\cline{2-6}
& 2207 2246 2288 2328 2365 2397 2436                                         & \cite{Baccari:1976ik}      &  70 & 116 &  89  \\\cline{2-6}
& 1988 2012 2031 2051 2070 2088 2105 2120 2136 2151                          & \cite{Nakkasyan:1975yz}    &  62 & 173 & 326   \\\cline{2-6}
\hline
$K^-p\to 2\pi^0\Lambda$
&  1571 1589 1598 1620 1632 1650 1659                       &\cite{Prakhov:2004ri} & 26513 &  \multicolumn{2}{c|}{$\delta\chi^2=-244^{a}$} \\\cline{2-6}
\hline
$K^-p\to 2\pi^0\Sigma$
&  1569 1589 1598 1620 1634 1647 1659 1676                   &\cite{Prakhov:2004an} & 3286 & \multicolumn{2}{c|}{$\delta\chi^2=-498^{a}$} \\\cline{2-6}
\hline
$K^-p\to \pi^0\Lambda^*$
& 1919 1942 1964 1986 2006 2027 2042 2068 2088 2111 2126 2151 2168  &\cite{Litchfield:1973ap} & 377 &  766 & 444 \\\cline{2-6}

(LC)& 1988 2011 2030 2052 2070 2088 2104 2120 2134 2150                 &\cite{Litchfield:1973ap} & 290 & 1169 & 372 \\\cline{2-6}
\hline
$K^-p\to \pi^0\Lambda^*$
&  1775 1796 1815 1833 1852 1870 1889 1907 1925 1941 1957  &\cite{Cameron:1977jr} & 319 & 606 & 523 \\\cline{2-6}
(LC)&  1710 1728 1747 1759 1775                                &\cite{Cameron:1977jr} &  25 & 193 &  66 \\\cline{2-6}
\hline
$K^-p\to \pi^-\Sigma^{*+}$
& 1775 1796 1815 1833 1852 1870 1889 1907 1926 1957  &\cite{Cameron:1978en} & 319 & 547 & 440 \\\cline{2-6}
(LC)& 2005 2028 2042 2068 2088 2110 2126 2151 2167       &\cite{Cameron:1978en} & 135 & 282 & 232 \\\cline{2-6}
\hline
$K^-p\to \pi^+\Sigma^{*-}$
& 1775 1796 1815 1833 1852 1870 1889 1907 1926 1957  &\cite{Cameron:1978en} & 319 & 524 & 538 \\\cline{2-6}
(LC)& 2005 2028 2042 2068 2088 2110 2126 2151 2167       &\cite{Cameron:1978en} & 126 & 266 & 185 \\\cline{2-6}
\hline
$K^-p\to K^{*-}p$
& 1815 1833 1852 1870 1889 1925 1941 1957                  &\cite{Cameron:1978qi} & 232 & 704 & 636 \\\cline{2-6}
(LC)& 1962 1985 2005 2025 2042 2067 2088 2111 2126 2151 2167   &\cite{Cameron:1978qi} & 231 & 323 & 266 \\\cline{2-6}
\hline
$K^-p\to K^{*0}n$
& 1815 1833 1852 1870 1889 1907 1925 1941 1957            &\cite{Cameron:1978qi} & 261 & 863 & 853 \\\cline{2-6}
(LC)& 1962 1985 2005 2025 2042 2067 2088 2111 2126 2151 2167  &\cite{Cameron:1978qi} & 110 & 167 & 129 \\\cline{2-6}
\hline
$K^-p\to \bar K\Delta$
& 1919 1942 1964 1986 2006 2027 2042 2068 2088 2111 2126 2151 2168 &\cite{Litchfield:1973ey}  & 377 & 547 & 484 \\\cline{2-6}
(LC)& 1988 2011 2030 2052 2070 2088 2104 2120 2134 2150                &\cite{Litchfield:1973ey}  & 290 & 499 & 384 \\\cline{2-6}
\hline
\end{tabular}\\[2ex]
$^{a}$: From the improvement of the likelihood when primary and final fits are compared.
\end{table*}

\section{\label{data}The data and the {\it primary} fit}

A detailed survey of the data that are used here in a coupled-channel analysis is presented in
Tables~\ref{data-diff1} to~\ref{data-three}. The tables list the reactions, the invariant masses at
which measurements have been performed, and the references to the data. $N_{\rm data}$ gives the
total number of data points for a reaction.  In the last line of Table~\ref{data-diff1}, e.g.,
there are differential cross sections at 20 angles and for 23 momenta giving $N_{\rm data}$=460. The
$\chi^2$ for these data achieved in the {\it primary} fit (using only established hyperons
and a relativistic Breit-Wigner description of the contributing resonances) and in the final fit are given in the last two columns.

Some data are slightly incompatible with other data in the normalization. The following scaling
factors were applied in the fits:
\bc
\renewcommand{\arraystretch}{1.4}
\begin{tabular}{lccccc}
$K^-p$:  & $d\sigma/d\Omega$              & 1.04   &\hspace{-3mm}\cite{AnderssonAlmehed:1970nb} & 1.10 &\hspace{-3mm}\cite{Barber:1975ci,Barber:1975rg}  \\
             &    $P$                                   & 0.95   &\hspace{-3mm}\cite{Albrow:1971yu}  &\\
$\bar K^0n$: & $d\sigma/d\Omega$               & 1.025 &\hspace{-3mm}\cite{Prakhov:2008dc} & \\
$\pi^-\Sigma^+$: & $d\sigma/d\Omega$ & 1.03  &\hspace{-3mm}\cite{Griselin:1975pa}   & 0.95 &\hspace{-3mm}\cite{deBellefon:1977sp}\\
$\pi^0\Sigma^0$: & $d\sigma/d\Omega$ & 1.06  &\hspace{-3mm}\cite{Prakhov:2008dc}   &  &\\
\end{tabular}
\renewcommand{\arraystretch}{1.}
\ec
Polarization data are important for partial wave analyses but often limited in statistics. Also
other low-statistics data like $K^-p\to K\Xi$ can be described rather badly without significantly affecting
the total $\chi^2_{\rm tot}$. To avoid this, data are given a weight. The differential cross
section for the final states $\eta\Lambda$, $\omega\Lambda$, and $K\Xi$ get a weight of 3, the
corresponding polarization data a weight of 6. Also the data on the quasi-two-body reactions get a
weight: 2 for $K^*p$; 3 for $\bar K\Delta(1232)$; 10 for $\pi\Lambda(1520)$. The polarization data for the
$\pi\Lambda$ final state get a weight factor 2, those for $\pi\Sigma$ of 4. The weights are chosen
to get a reasonable fit to these data without significantly distorting the fit to the other data
sets.

Most data are limited to the region below $W=2.0$\,GeV in the invariant mass; a few data extend the
mass region up to $W=2.4$\,GeV. We limit our study to resonances below 2.25\,GeV. The data of
Tables~\ref{data-diff1} and~\ref{data-diff2}  include elastic and charge-exchange scattering, the inelastic channels with a
$\Lambda$ hyperon in the final state produced with a $\pi^0$ or an $\eta$ meson, inelastic
scattering into the three $\pi\Sigma$ charge states or into $\eta\Sigma$, and the production of
cascade hyperons. Table~\ref{data-pol} gives references to publications on the polarization
observable $P$. For $K^-p\to K^-p$ elastic scattering, $P$ was measured by scattering off a
polarized target. Hyperons in the final state reveal their polarization $P$ via the asymmetry of
their decay.

Table~\ref{data-three} lists the data on  $K^-p\to \Lambda\omega$ and on reactions with three particles
in the final state,
including quasi-two-body final states. In the low-mass region, we include the data on $K^-p\to
2\pi^0\Lambda$~\cite{Prakhov:2004ri} and $K^-p\to 2\pi^0\Sigma$~\cite{Prakhov:2004an}. For these
data, the individual events are available, and we include the data event-by-event in an event-based
likelihood fit. We also fit data on the quasi-two-body reactions $K^- p \to \omega \Lambda$ $K^- p
\to \pi^0 \Lambda (1520)$~\cite{Litchfield:1973ap,Cameron:1977jr}, $K^-
p\to\pi^\mp\Sigma^\pm(1385)$ and $\rho\Lambda$~\cite{Cameron:1978en}, $\bar{K} N \to \bar{K}^*
N$~\cite{Cameron:1978qi}, and $K^-p\to\bar K\Delta(1232)$~\cite{Litchfield:1973ey}. Details on
these reactions are shown in an accompanying paper~\cite{Hyperon-III}.

\begin{table}[pt]
\caption{\label{used-hyperons}Hyperons used in the primary fit to the data
in which the Breit-Wigner masses and widths from the RPP~\cite{Tanabashi:2018oca}
are imposed.}
\renewcommand{\arraystretch}{1.4}
\bc
\begin{tabular}{lllcc}
\hline\hline
                & $J^P$   & Status & Mass                 & Width         \\
\hline
$\Lambda(1405)$ & $1/2^-$ & **** & $1405^{+1.3}_{-1.0}$   & $50.5\pm 2$  \\
$\Lambda(1670)$ & $1/2^-$ & **** & $1660 - 1680$                & $25 - 50$      \\
$\Lambda(1800)$ & $1/2^-$ & ***  & $1720 - 1850$                 & $200 - 400$    \\
$\Lambda(1520)$ & $3/2^-$ & **** & $1519.5 \pm 1.0$           & $15.6\pm 1.0$       \\
$\Lambda(1690)$ & $3/2^-$ & **** & $1685 - 1695$                & $50 - 70$      \\
$\Lambda(1830)$ & $5/2^-$ & **** & $1810 - 1830$                & $60 - 110$          \\
$\Lambda(2100)$ & $7/2^-$ & **** & $2090 - 2110$                & $100 - 250$         \\
\hline
$\Lambda(1600)$ & $1/2^+$ & ***  & $1560 - 1700$               & $50 - 250$     \\
$\Lambda(1810)$ & $1/2^+$ & ***  & $1750 - 1850$               & $50 - 250$     \\
$\Lambda(1890)$ & $3/2^+$ & **** & $1850 - 1910$              & $60 - 200$          \\
$\Lambda(1820)$ & $5/2^+$ & **** & $1815 - 1825$              & $70 - 90$      \\
$\Lambda(2110)$ & $5/2^+$ & ***  & $2090 - 2140$               & $150 - 250$    \\
\hline
                & $J^P$   & Status & Mass               & Width            \\
\hline
$\Sigma(1750)$  & $1/2^-$ & ***  & $1730 - 1800$                & $60-160$    \\
$\Sigma(1670)$  & $3/2^-$ & **** & $1665 - 1685$               & $40-80$          \\
$\Sigma(1940)$  & $3/2^-$ &  *** & $1900 - 1950$                & $150-300$   \\
$\Sigma(1775)$  & $5/2^-$ & **** & $1770 - 1780$               & $105-135$        \\
\hline
$\Sigma(1660)$  & $1/2^+$ & ***  & $1630 - 1690$              & $40-200$    \\
$\Sigma(1385)$  & $3/2^+$ & **** & $1382.80\pm 0.35$      & $36.0\pm 0.7$    \\
$\Sigma(1915)$  & $5/2^+$ & **** & $1900 - 1935$             & $80-160$    \\
$\Sigma(2030)$  & $7/2^+$ & **** & $2025 - 2040$             & $150-200$    \\
\hline\hline\vspace{-5mm}
\end{tabular}
\renewcommand{\arraystretch}{1.3}
\ec
\end{table}
We performed four types of fits: one {\it primary} fit, a series of exploratory fits called {\it
mass scans}, and the {\it final} fit. At the end, we performed fits in which
the significance of the contributing resonances is estimated, and a series of {\it error defining}
fits in which resonances are added in all contributing partial waves.

The {\it primary} fit to the data listed in Tables~\ref{data-diff1} to~\ref{data-three} used only
those hyperons which were listed in the Review of Particle Properties
(RPP)~\cite{Tanabashi:2018oca} with three or four stars. A list of hyperons used in
the {\it primary} fit and the ranges of masses and widths is given in Table~\ref{used-hyperons}.
The resonances are mostly well separated
and Breit-Wigner parametrisations were used for the resonances. Masses, widths and coupling
constants are allowed to vary within the limits quoted in the RPP.

\begin{figure}[pt]
\begin{center}
\begin{tabular}{ccc}
\hspace{-3mm}\includegraphics[width=0.16\textwidth,height=0.18\textwidth]{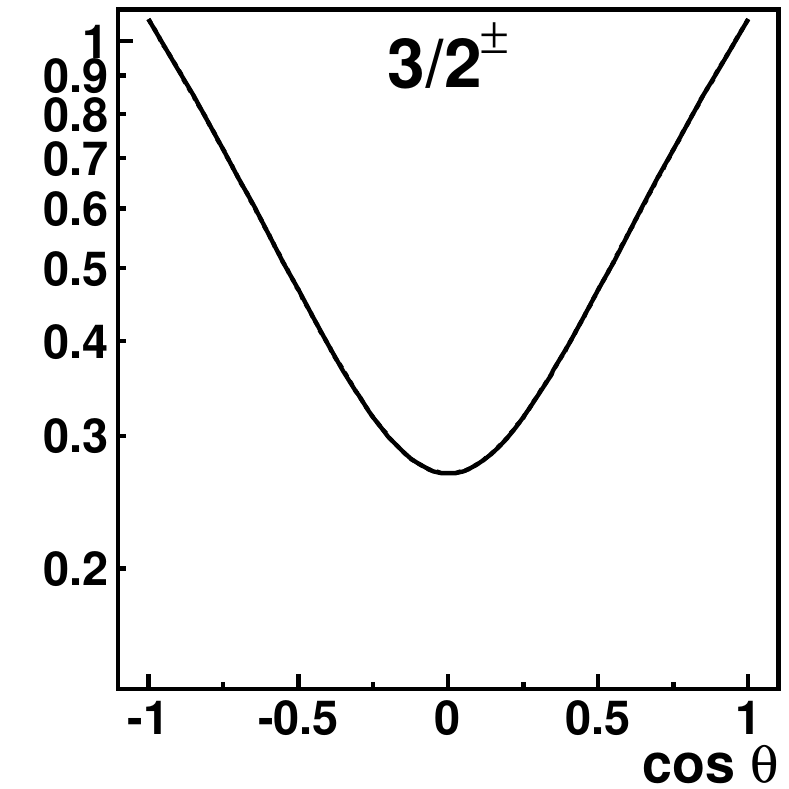}&
\hspace{-3mm}\includegraphics[width=0.16\textwidth,height=0.18\textwidth]{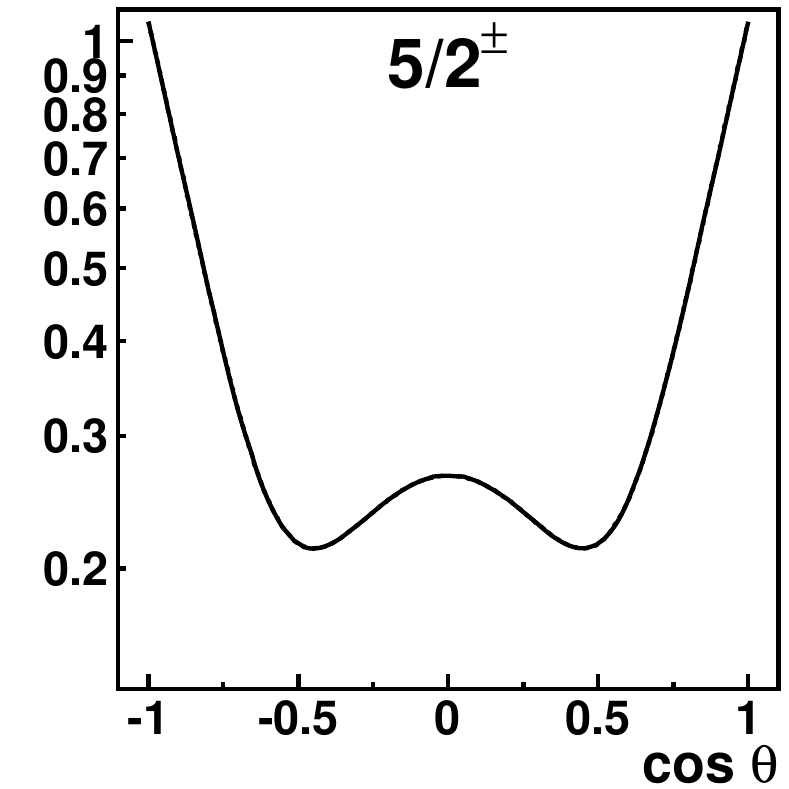}&
\hspace{-3mm}\includegraphics[width=0.16\textwidth,height=0.18\textwidth]{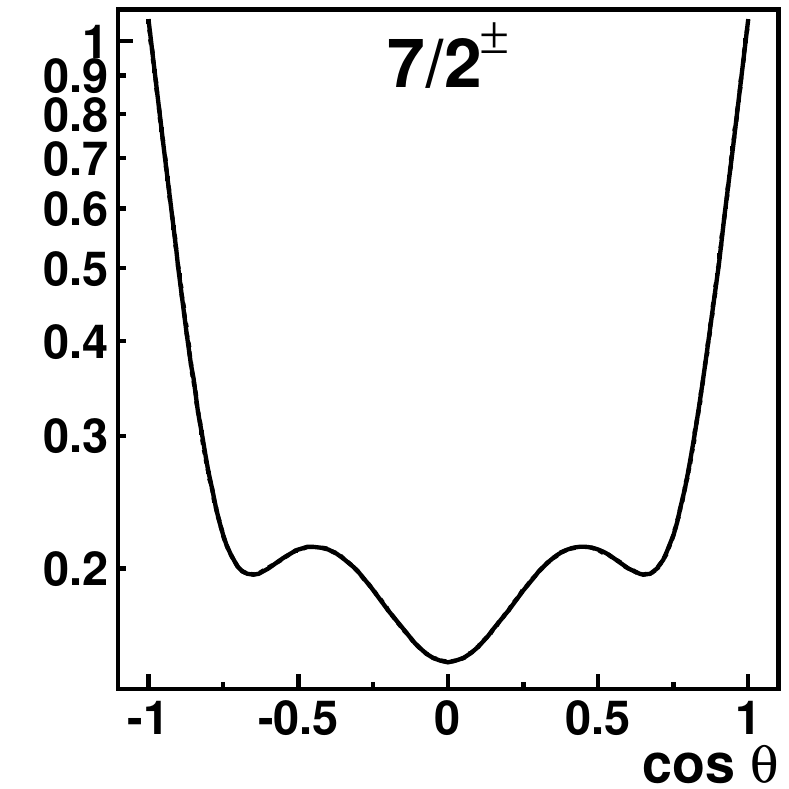}
\end{tabular}
\vspace{-3mm}\end{center}
\caption{\label{simulation}Shape of angular distributions for $K^-p$ into a $\Lambda^*$ or
$\Sigma^*$ resonance with given $J^P$ decaying into a $J^P=1/2^+$ baryon and a pseudoscalar meson.
Resonances with  $J^P=1/2^\pm$ yield a flat angular distribution, those with  $J^P=3/2^\pm$:
$3\cos^2\theta+1$,  $J^P=5/2^\pm$: $9/4\cdot$($5\cos^4\theta-2\cos^2\theta +1$),  $J^P=7/2^\pm$:
1/4$\cdot$($175\cos^6\theta-165\cos^4\theta+45\cos\theta^2+9$).
}
\end{figure}

Figure~\ref{simulation} shows the angular distributions expected for $J=3/2, 5/2$, and $7/2$. They
are identical for both parities. Forward-backward asymmetries come from the interference of even
and odd waves. This figure can serve as a guide when the data are interpreted.

\begin{figure*}
\centerline{\includegraphics[width=0.85\textwidth]{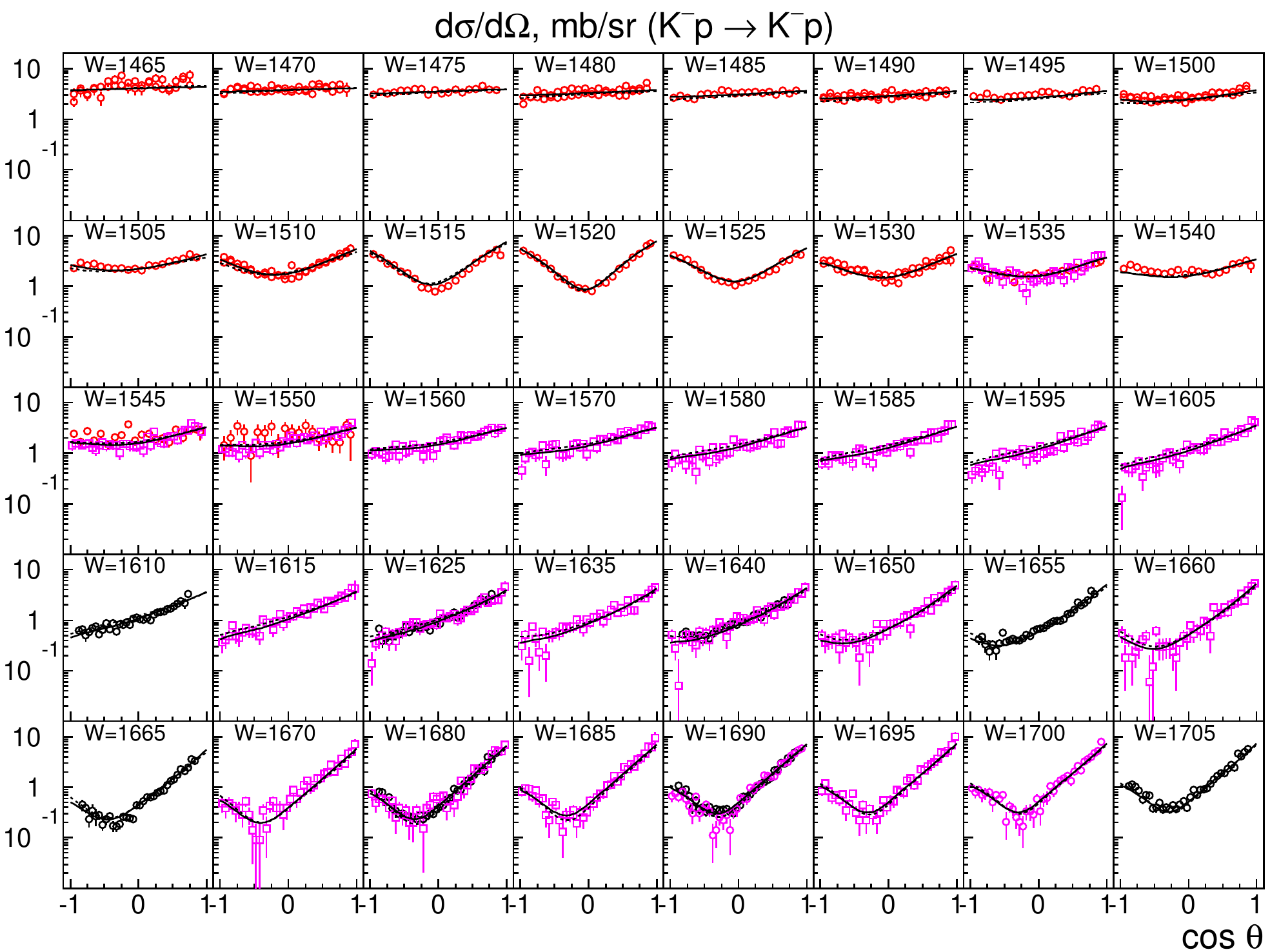}\vspace{2mm}}
\centerline{\includegraphics[width=0.85\textwidth]{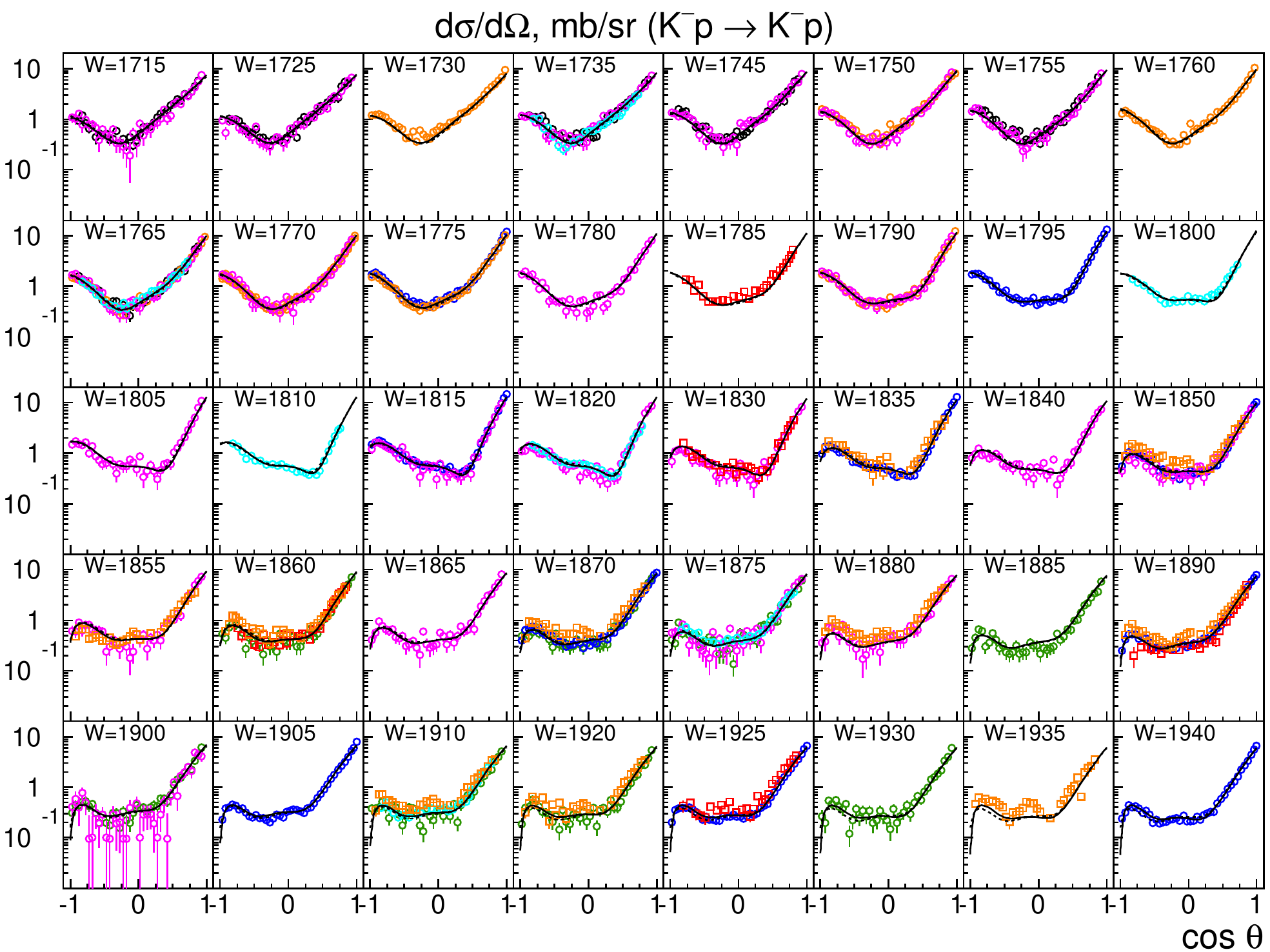}}
\caption{\label{fig:elast1}Differential cross section for the elastic $K^-p$ scattering. The data
are from \boldmath
  $\Box$~\cite{Daum:1968jey},
  {\rd$\Box$}~\cite{AnderssonAlmehed:1970nb},
 {\vio$\Box$}~\cite{Armenteros:1970eg},
 {\color{cyan}~$\circ$}~\cite{Albrow:1971yu},
 {\vio$\circ$}~\cite{Conforto:1971vy},
 {\bl$\Box$}~\cite{Barber:1975rg},
  {\gr$\circ$}~\cite{Griselin:1975pa},
 {\rd$\circ$}~\cite{Mast:1975pv},
 {\yel$\Box$}~\cite{Barber:1975ci},
 {\bl$\circ$}~\cite{Conforto:1975nw},
  {\gr$\Box$}~\cite{deBellefon:1976qr},
{\yel$\circ$}~\cite{Cameron:1981qi}.
 The solid line represent our final fit. }
\end{figure*}

\begin{figure*}
\centerline{\includegraphics[width=0.85\textwidth]{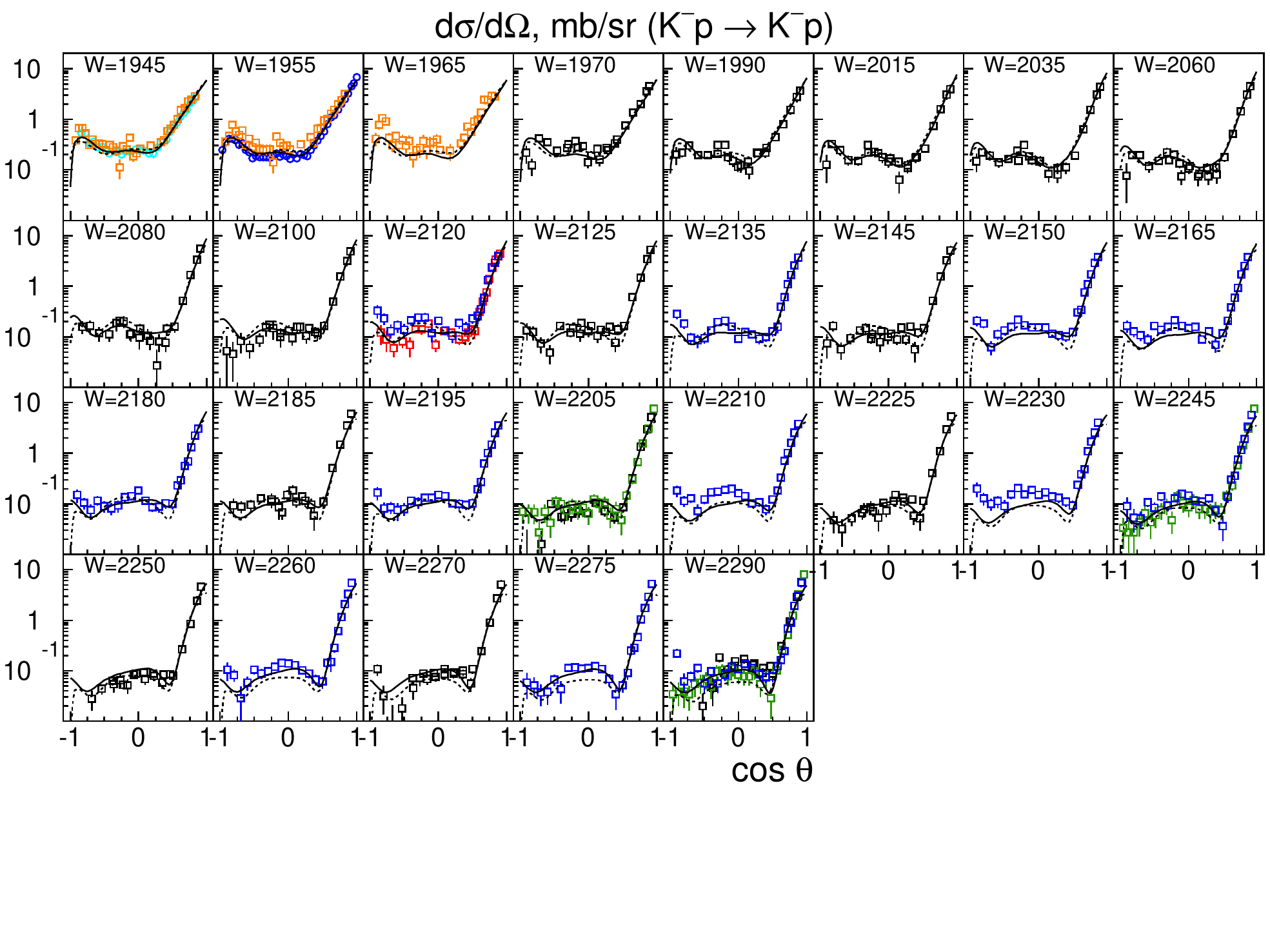}
\vspace{-20mm}}
\caption{\label{fig:elast2}Differential cross section for the
elastic $K^-p$ scattering. See Fig.~\ref{fig:elast1} for the color code. \vspace{4mm}}
\centerline{\includegraphics[width=0.85\textwidth]{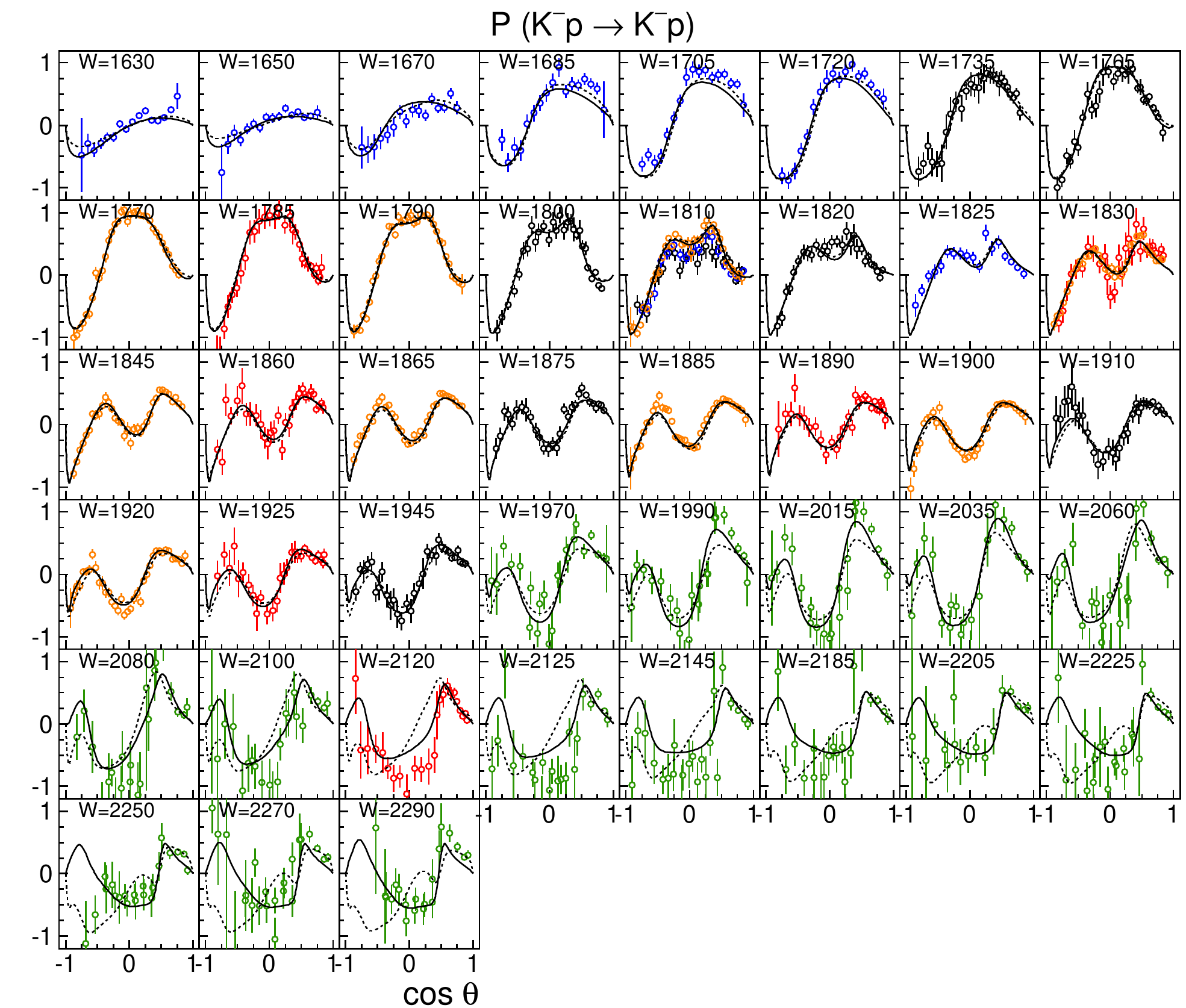}}
\caption{\label{fig:elast} The polarization observable $P$ for elastic $K^-p$ scattering. The data
are from \boldmath
 {\gr$\circ$}~\cite{Daum:1968jey},
  {\rd$\circ$}~\cite{AnderssonAlmehed:1970nb},
  $\circ$\cite{Albrow:1971yu},
  {\bl$\circ$}~\cite{Ehrlich:1977bs},
 {\yel$\circ$}~\cite{Bryant:1980yp}.
}
\end{figure*}

\begin{figure*}
\centerline{\includegraphics[width=0.85\textwidth]{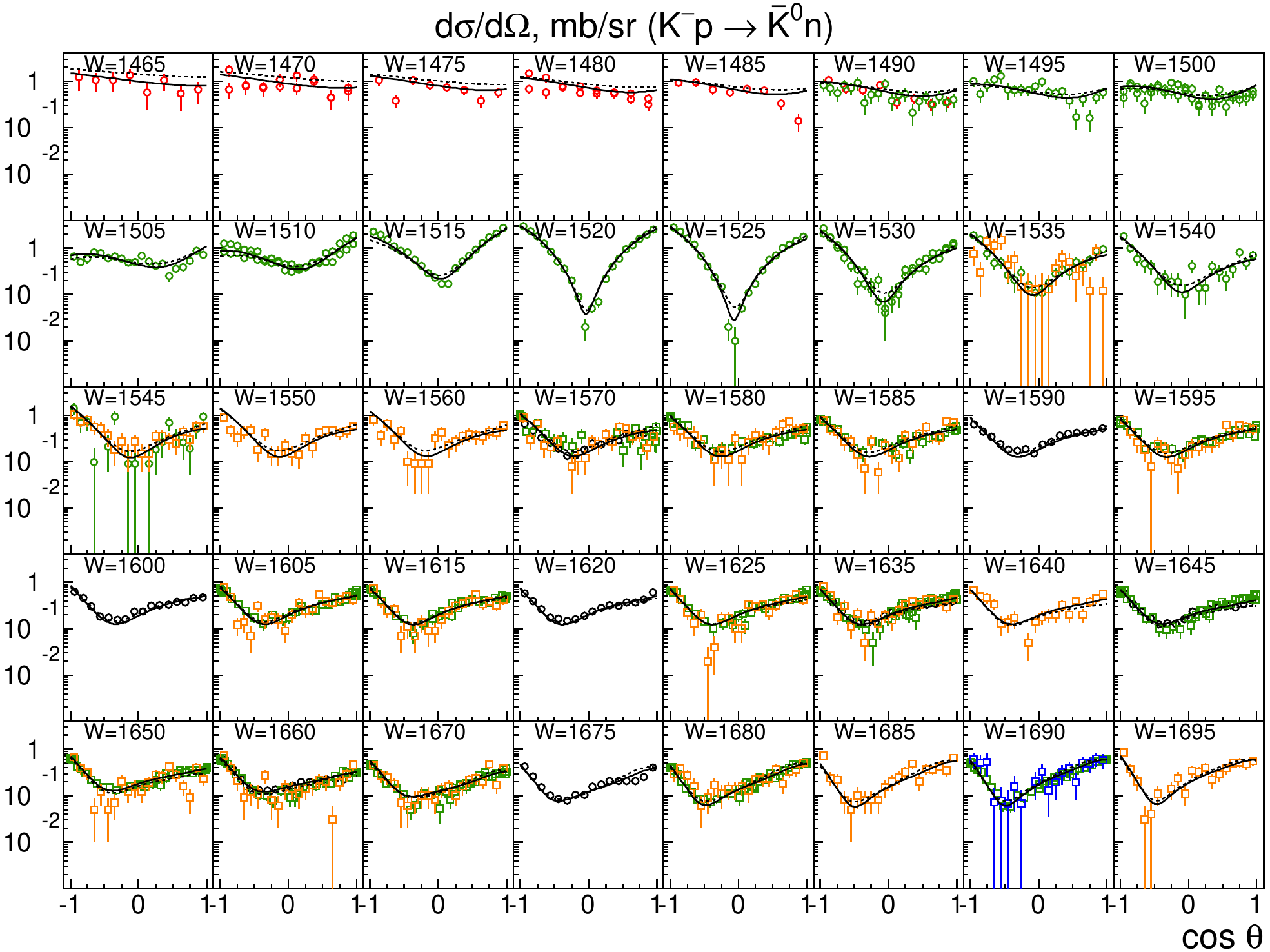}\vspace{2mm}}
\centerline{\includegraphics[width=0.85\textwidth]{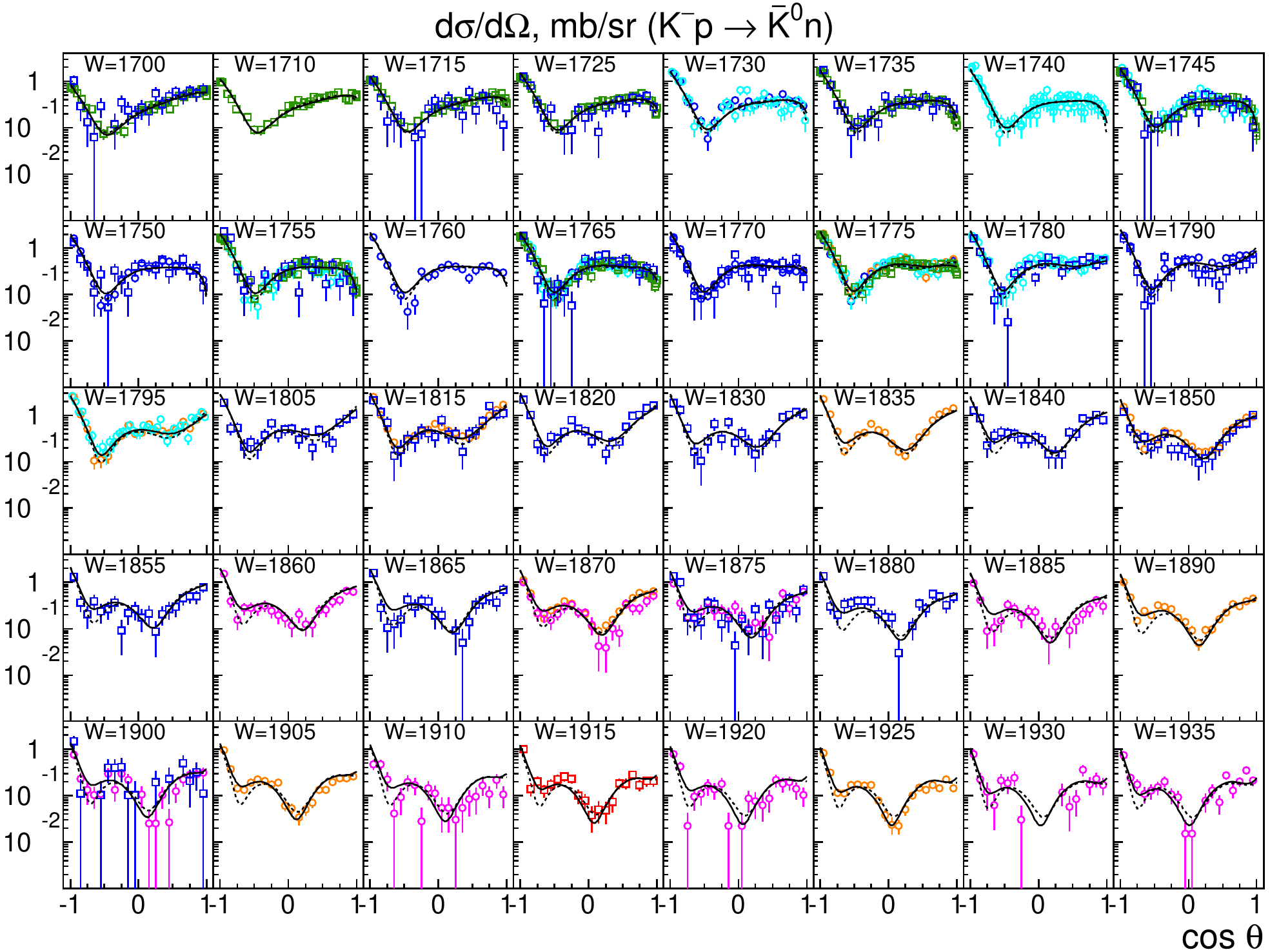}}
\caption{\label{fig:CEX1}Differential cross section for the charge-exchange reaction. The data
are from
\boldmath
 {\yel$\Box$}~\cite{Armenteros:1970eg},
 {\rd$\Box$}~\cite{Litchfield:1971ri},
 {\vio$\circ$}~\cite{Griselin:1975pa},
 {\rd$\circ$~}\cite{Mast:1975pv}a,
 {\gr$\circ$}~\cite{Mast:1975pv}b,
 {\yel$\circ$}~\cite{Conforto:1975nw},
 $\Box$~\cite{deBellefon:1976qr},
 {\bl$\circ$}~\cite{Cameron:1981qi},
 {\bl$\Box$}~\cite{Armenteros:1969kn},
 {\color{cyan}$\circ$}~\cite{Jones:1974at},
 {\gr$\Box$}~\cite{AlstonGarnjost:1977ct},
 $\circ$~\cite{Prakhov:2008dc}.
The solid line represent our final fit. }
\end{figure*}

\begin{figure*}
\centerline{\includegraphics[width=0.85\textwidth]{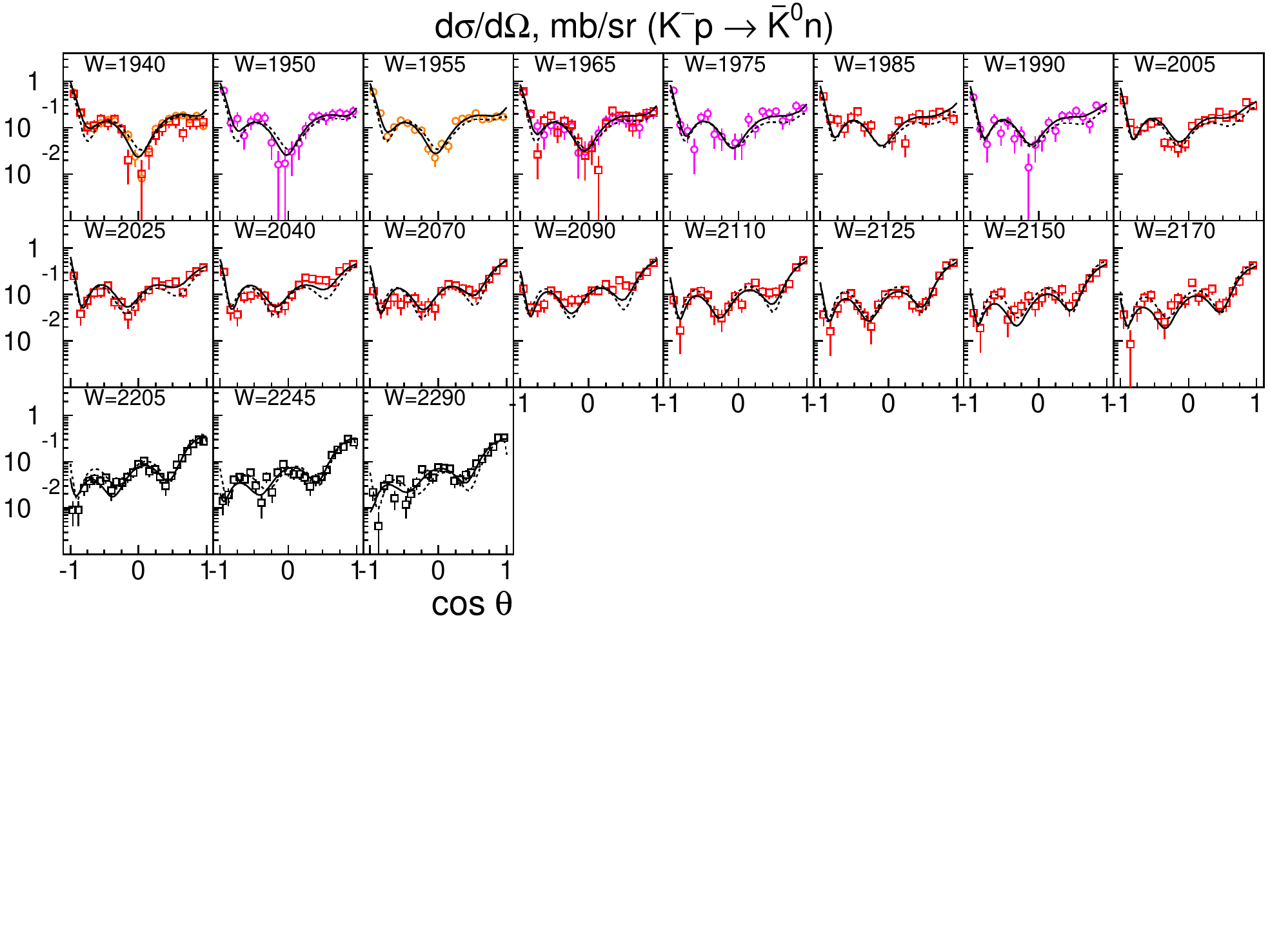}
\vspace{-40mm}}
\caption{\label{fig:CEX2}Differential cross section for the
charge-exchange reaction. See Fig.~\ref{fig:CEX1} for the color code.
\vspace{4mm}}
\centerline{\includegraphics[width=0.85\textwidth]{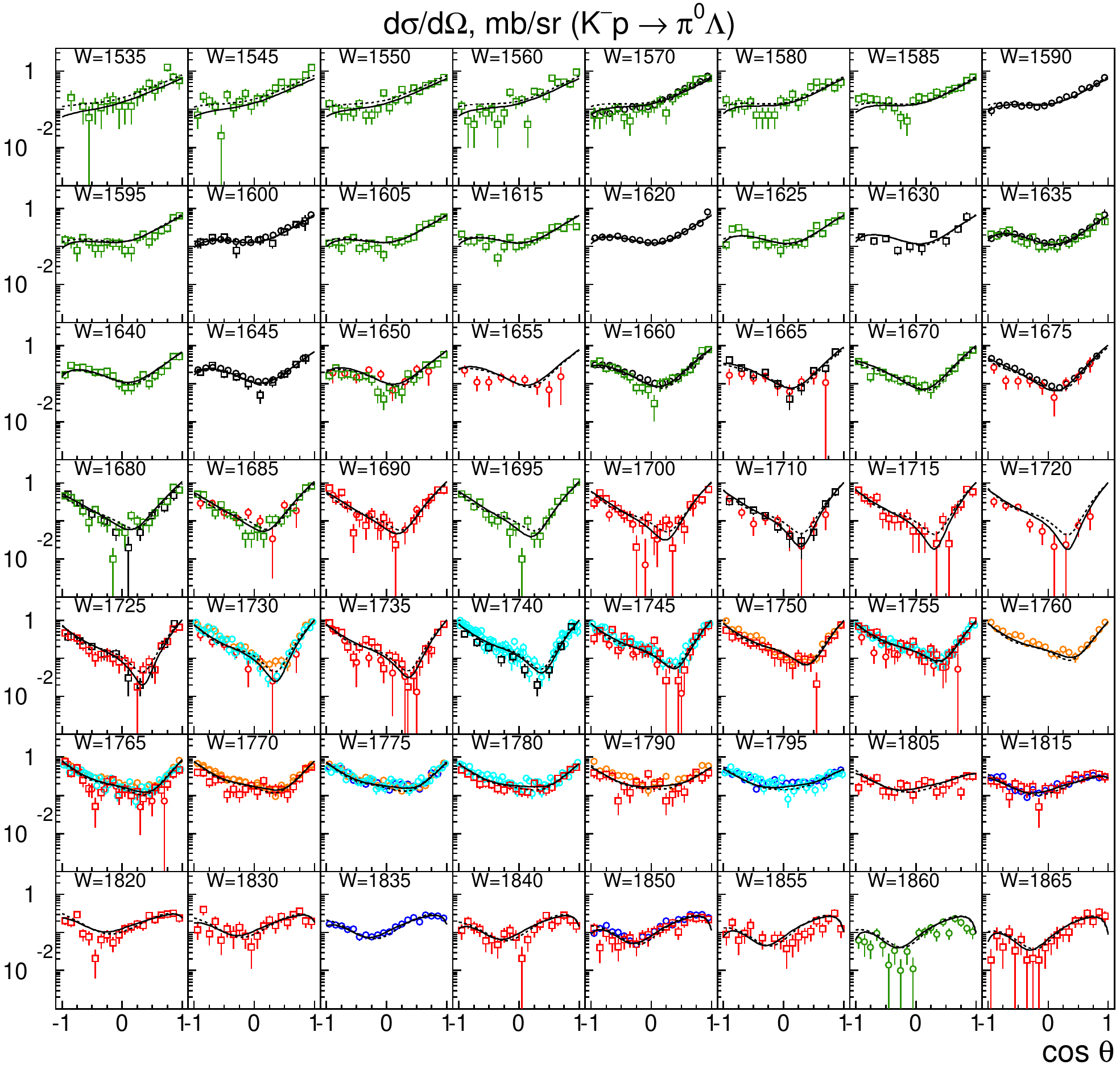}}
\caption{\label{fig:SpiL1}Differential cross sections for $K^-p\to\pi^0\Lambda$. The data are from
\boldmath
  {\gr$\Box$}~\cite{Armenteros:1970eg},
  {\gr$\circ$}~\cite{Griselin:1975pa},
 {\bl$\circ$}~\cite{Conforto:1975nw},
 {\yel$\circ$}~\cite{Cameron:1981qi},
  {\rd$\Box$}~\cite{Armenteros:1969kn},
 {\color{cyan}$\circ$}~\cite{Jones:1974at},
 {$\circ$}~\cite{Prakhov:2008dc},
  {\bl$\Box$}~\cite{Berthon:1970fd},
 {\rd$\circ$}~\cite{Baxter:1974zs},
  {$\Box$}~\cite{London:1975zz},
 {\vio$\circ$}~\cite{deBellefon:1975lp}. }
\end{figure*}

\begin{figure*}
\centerline{\includegraphics[width=0.85\textwidth]{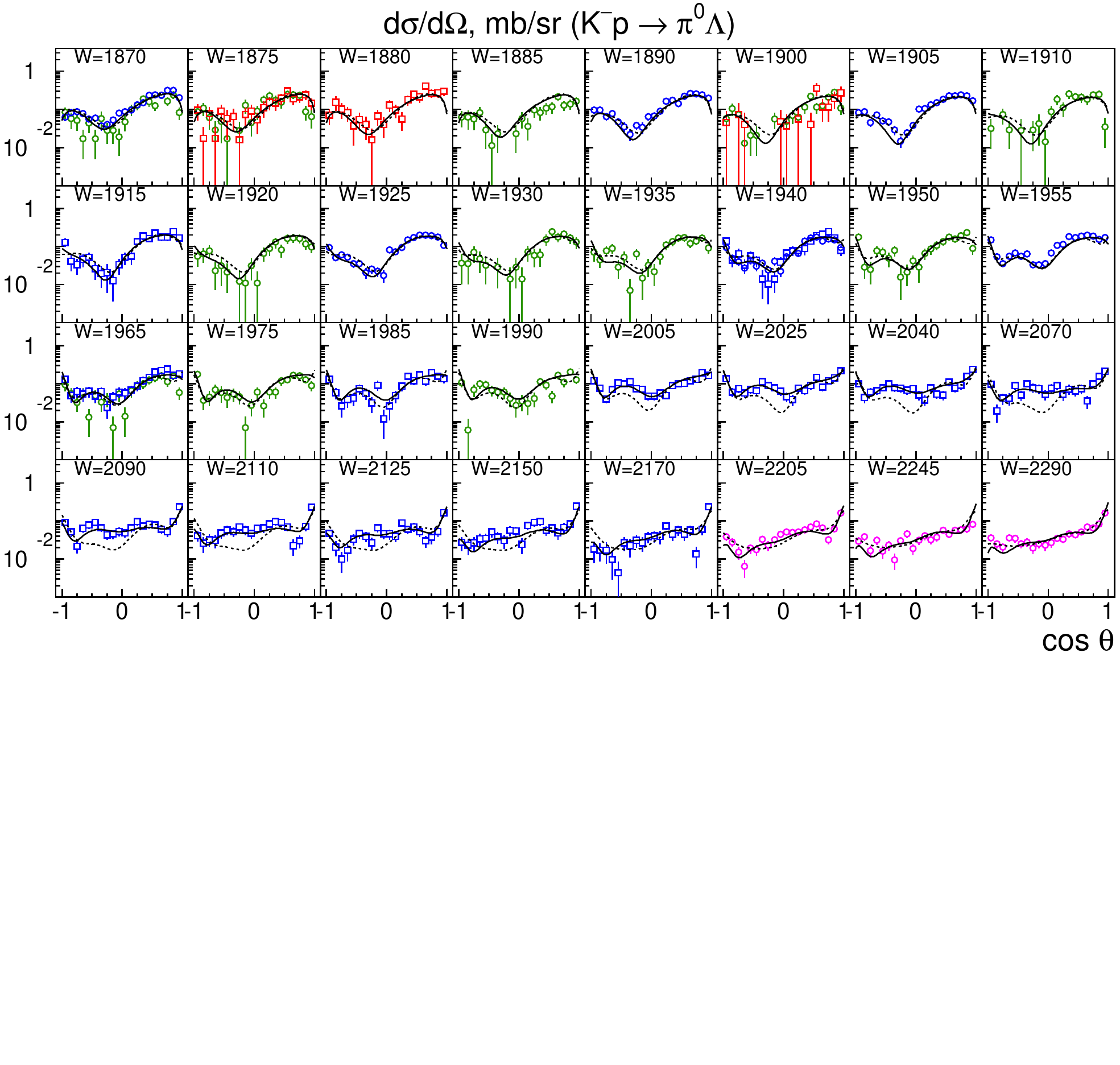}\vspace{-54mm}}
\caption{\label{fig:SpiL2}Differential cross sections for $K^-p\to\pi^0\Lambda$.
See Fig.~\ref{fig:SpiL1} for the color code. \vspace{2mm}}
\centerline{\includegraphics[width=0.85\textwidth]{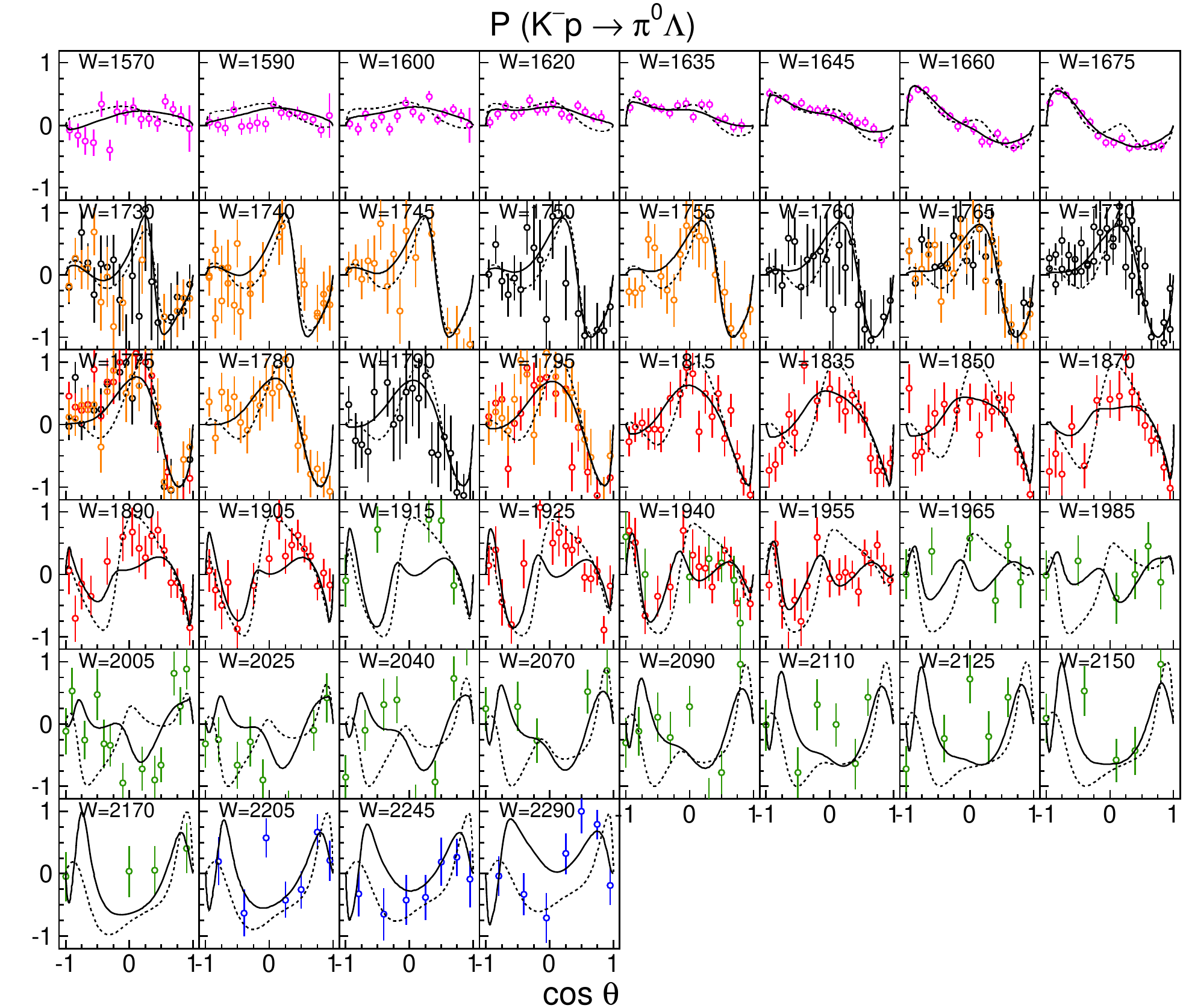}}
\caption{\label{fig:PpiL}$P$ for $K^-p\to\pi^0\Lambda$. The data are from \boldmath
   {\rd$\circ$}~\cite{Conforto:1975nw},
  {$\circ$}~\cite{Cameron:1981qi},
   {\yel$\circ$}~\cite{Jones:1974at},
  {\vio$\circ$}~\cite{Prakhov:2008dc},
  {\bl$\circ$}~\cite{deBellefon:1975lp}. }
\end{figure*}

\begin{figure*}
\centerline{\includegraphics[width=0.85\textwidth]{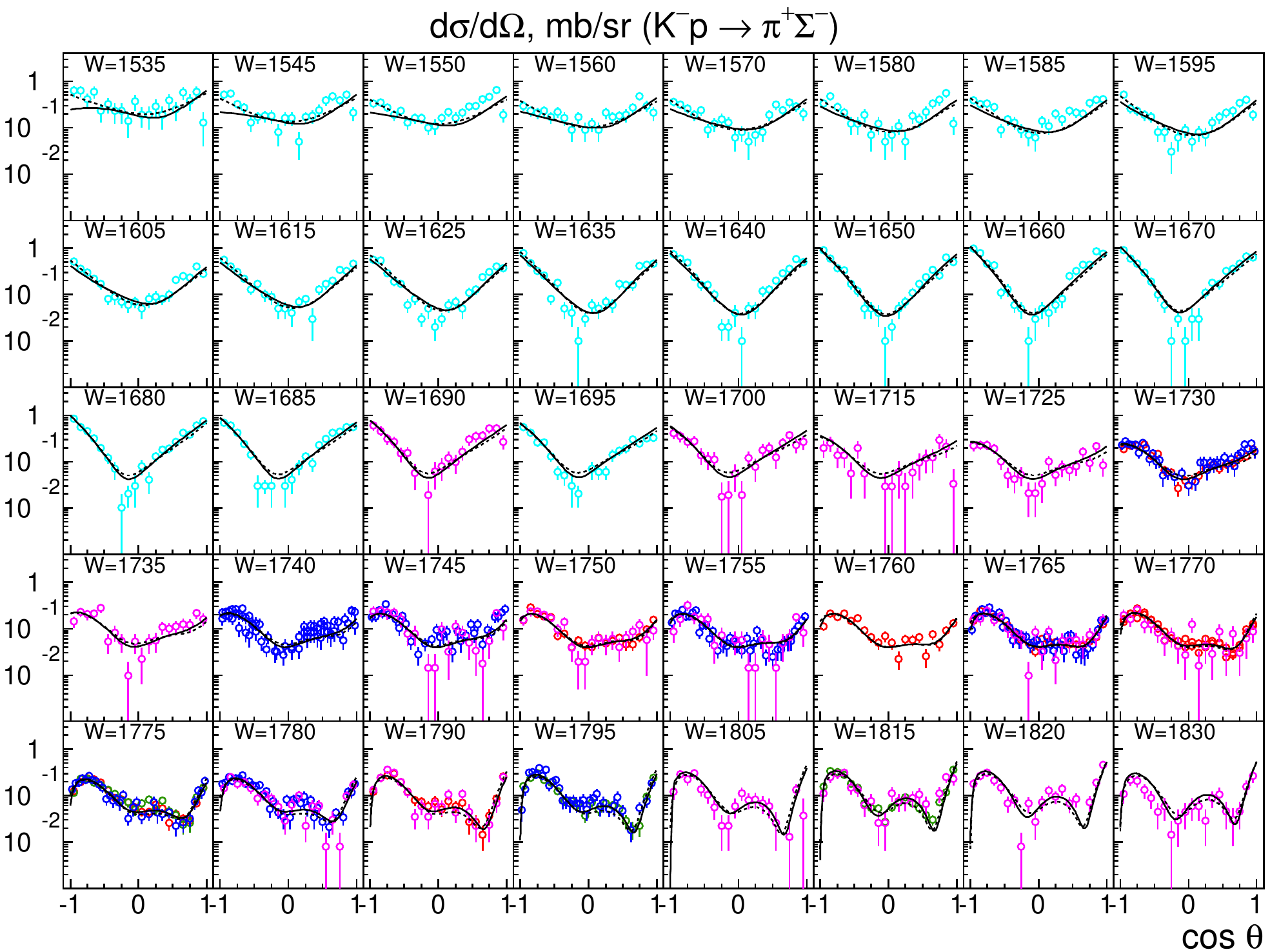}}
\centerline{\includegraphics[width=0.85\textwidth]{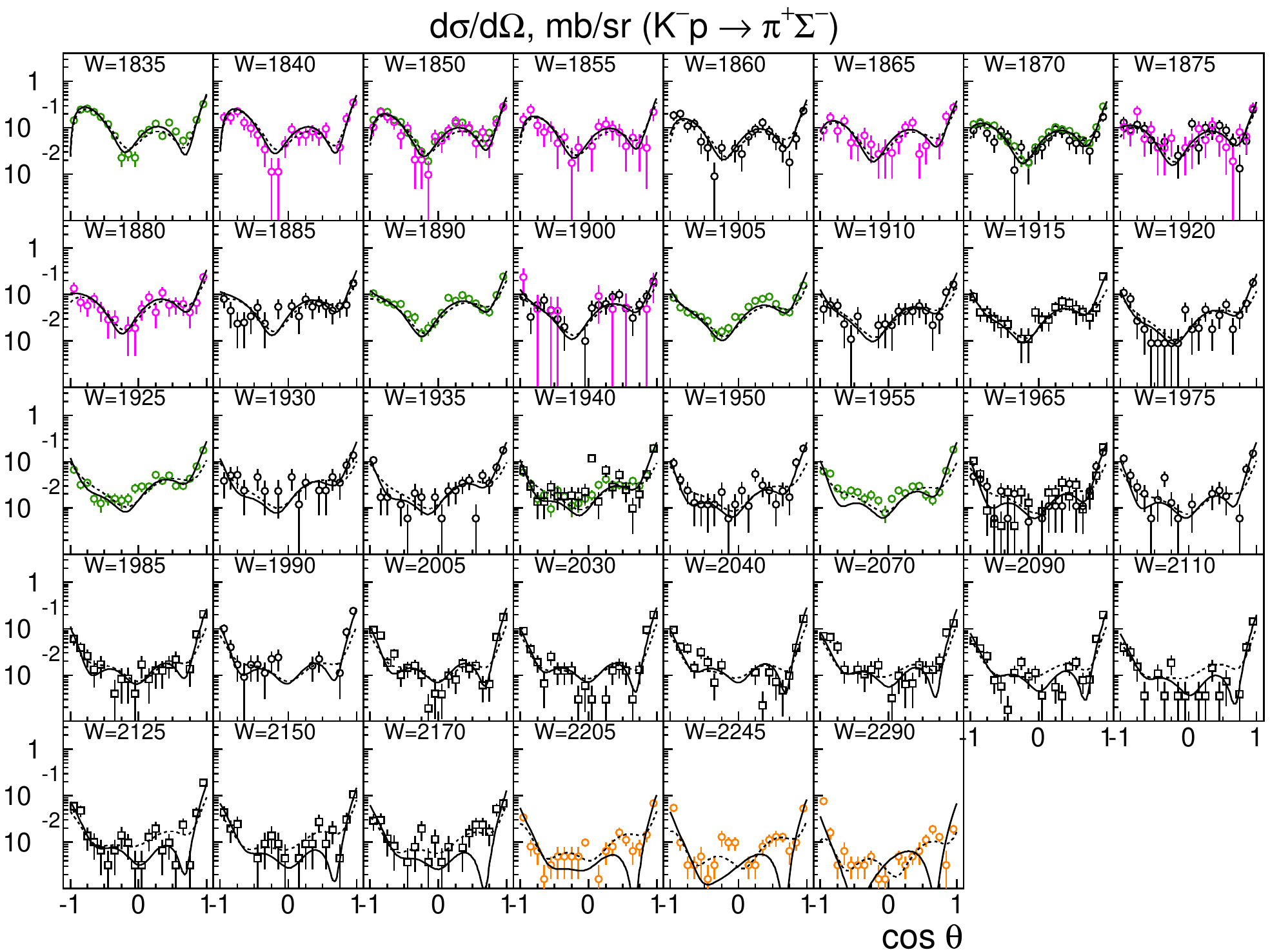}}
\caption{\label{fig:SpiSp1}Differential cross sections for $K^-p\to\pi^+\Sigma^-$. The data are
from \boldmath
 {\color{cyan}$\circ$}~\cite{Armenteros:1970eg},
{$\circ$}~\cite{Griselin:1975pa},
 {\gr$\circ$}~\cite{Conforto:1975nw},
  {\rd$\circ$}~\cite{Cameron:1981qi},
  {\vio$\circ$}~\cite{Armenteros:1969kn},
  {\bl$\circ$}~\cite{Jones:1974at},
{$\Box$}~\cite{Berthon:1970sp},
{\yel$\circ$}~\cite{deBellefon:1977sp}.}
\end{figure*}

\begin{figure*}
\centerline{\includegraphics[width=0.85\textwidth]{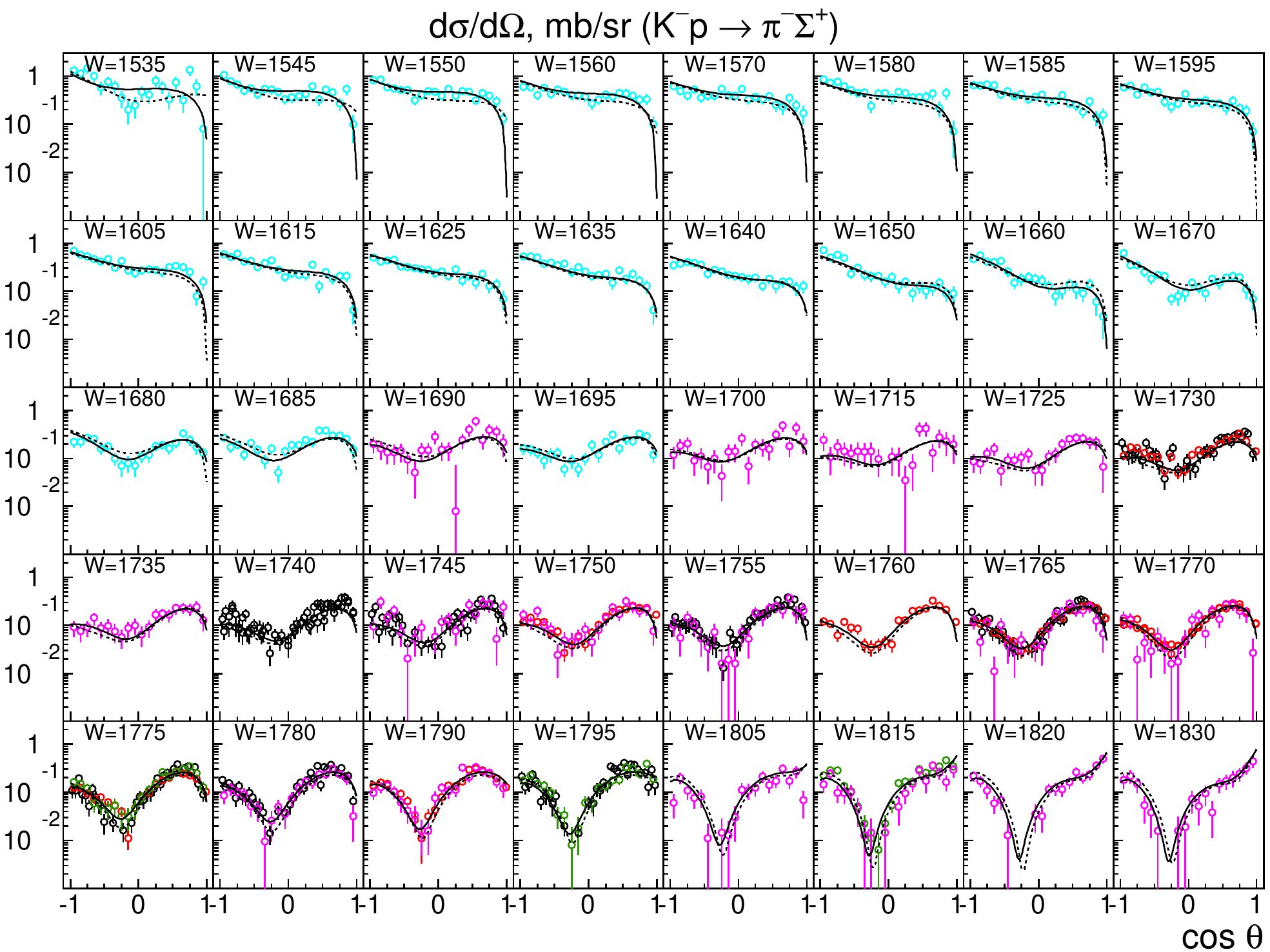}}
\centerline{\includegraphics[width=0.85\textwidth]{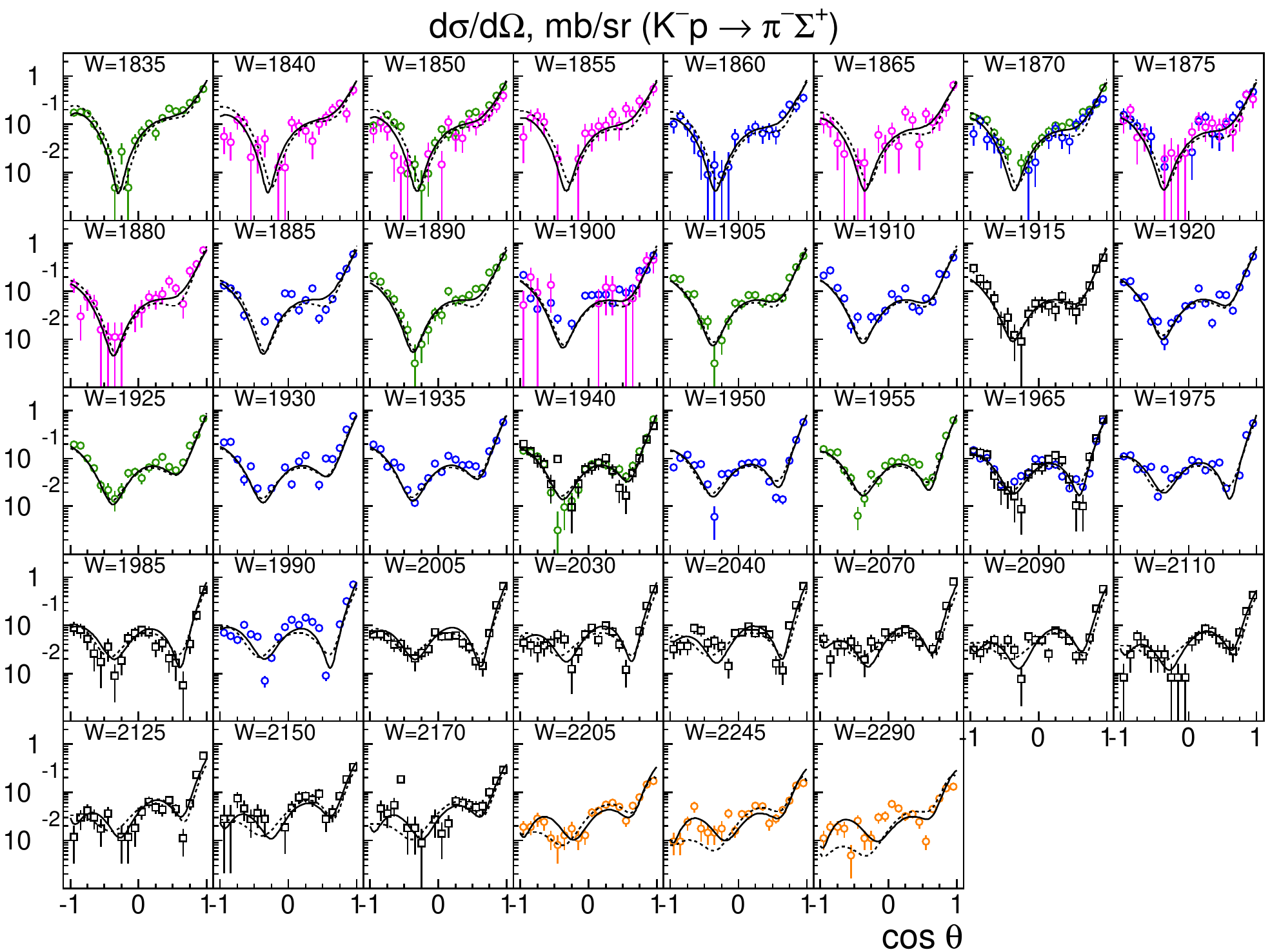}}
\caption{\label{fig:SpiSp2}Differential cross sections for $K^-p\to\pi^-\Sigma^+$. The data are
from \boldmath
 {\bl$\circ$}~\cite{Griselin:1975pa},
 {\gr$\circ$}~\cite{Conforto:1975nw},
   {\rd$\circ$}~\cite{Cameron:1981qi},
{\vio$\circ$}~\cite{Armenteros:1969kn},
 {\color{cyan}$\circ$}~\cite{Armenteros:1969kn},
{$\circ$}~\cite{Jones:1974at},
 {$\Box$}~\cite{Berthon:1970sp},
 {\yel$\circ$}~\cite{deBellefon:1977sp}.
}
\end{figure*}

\begin{figure*}
\centerline{\includegraphics[width=0.85\textwidth]{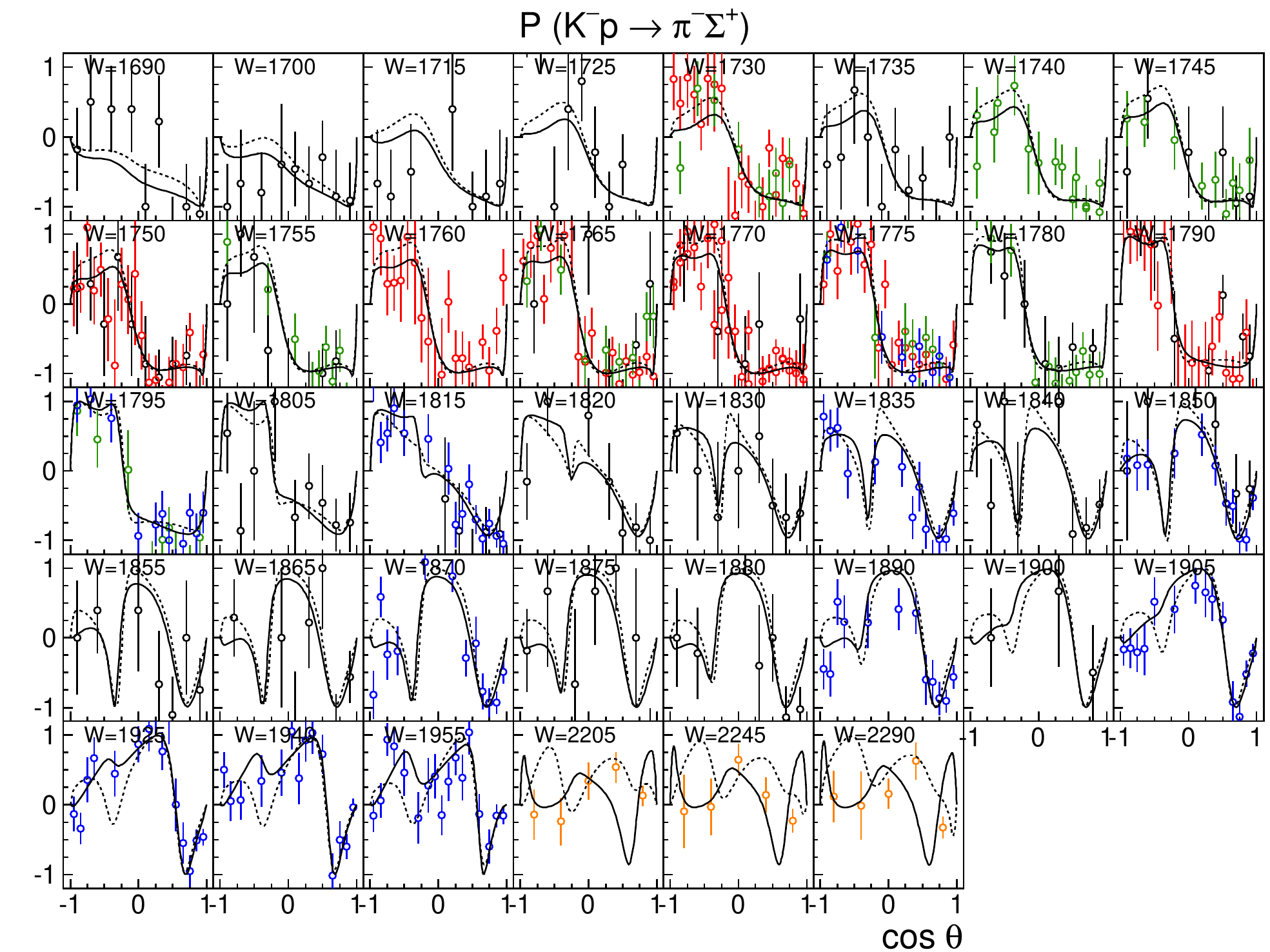}}
\caption{\label{fig:PpiSp}$P$ for $K^-p\to\pi^-\Sigma^+$. The data are from \boldmath
   {\bl$\circ$}~\cite{Conforto:1975nw},
 {\rd$\circ$}~\cite{Cameron:1981qi},
   {$\circ$}~\cite{Armenteros:1969kn},
  {\gr$\circ$}~\cite{Jones:1974at},
 {\yel$\circ$}~\cite{deBellefon:1977sp}. \vspace{2mm}}
\centerline{\includegraphics[width=0.85\textwidth]{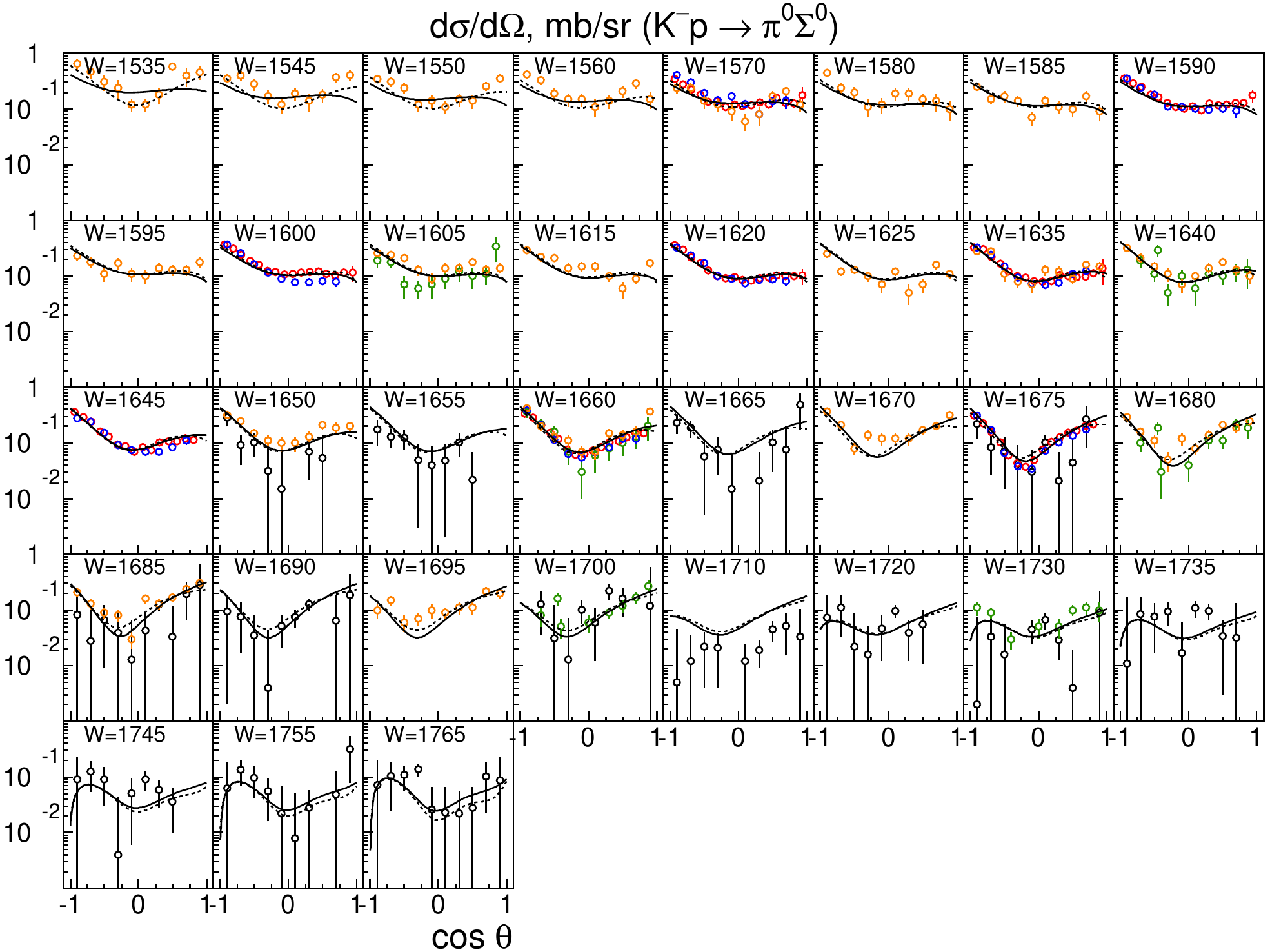}}
\caption{\label{fig:SpiS0}Differential cross sections for $K^-p\to\pi^0\Sigma^0$. Data: \boldmath
   {\yel$\circ$}~\cite{Armenteros:1970eg},
 {\rd$\circ$}~\cite{Prakhov:2008dc},
 {$\circ$}~\cite{Baxter:1974zs},
 {\gr$\circ$}~\cite{London:1975zz},
 {\bl$\circ$}~\cite{Manweiler:2008zz}. }
\end{figure*}

\begin{figure*}
\centerline{\includegraphics[width=0.85\textwidth]{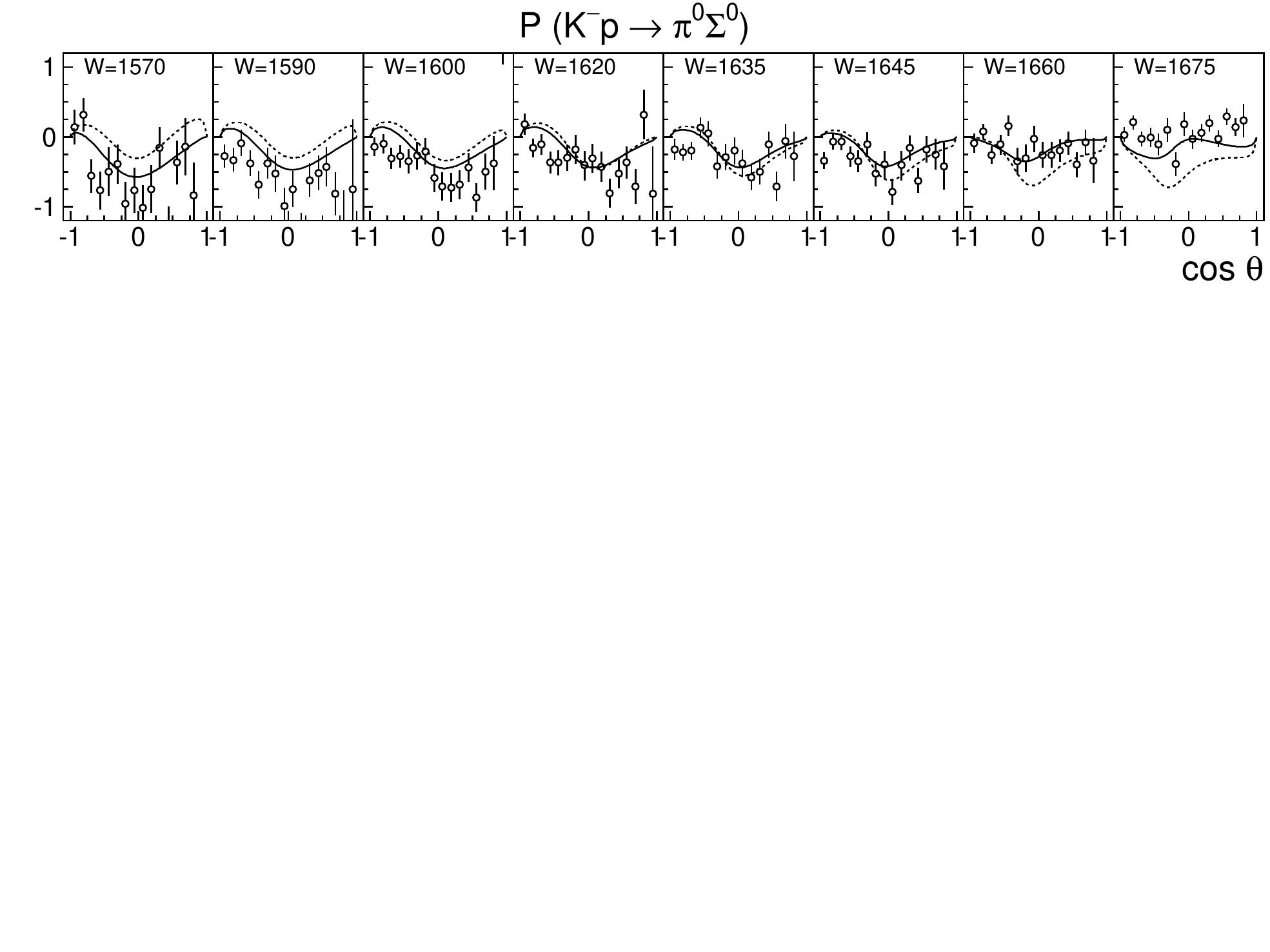}
\vspace{-82mm}}
\caption{\label{fig:PpiS0}$P$ for
$K^-p\to\pi^0\Sigma^0$. The data are from \boldmath
{$\circ$}~\cite{Prakhov:2008dc}.  \vspace{4mm}}
\centerline{\includegraphics[width=0.95\textwidth]{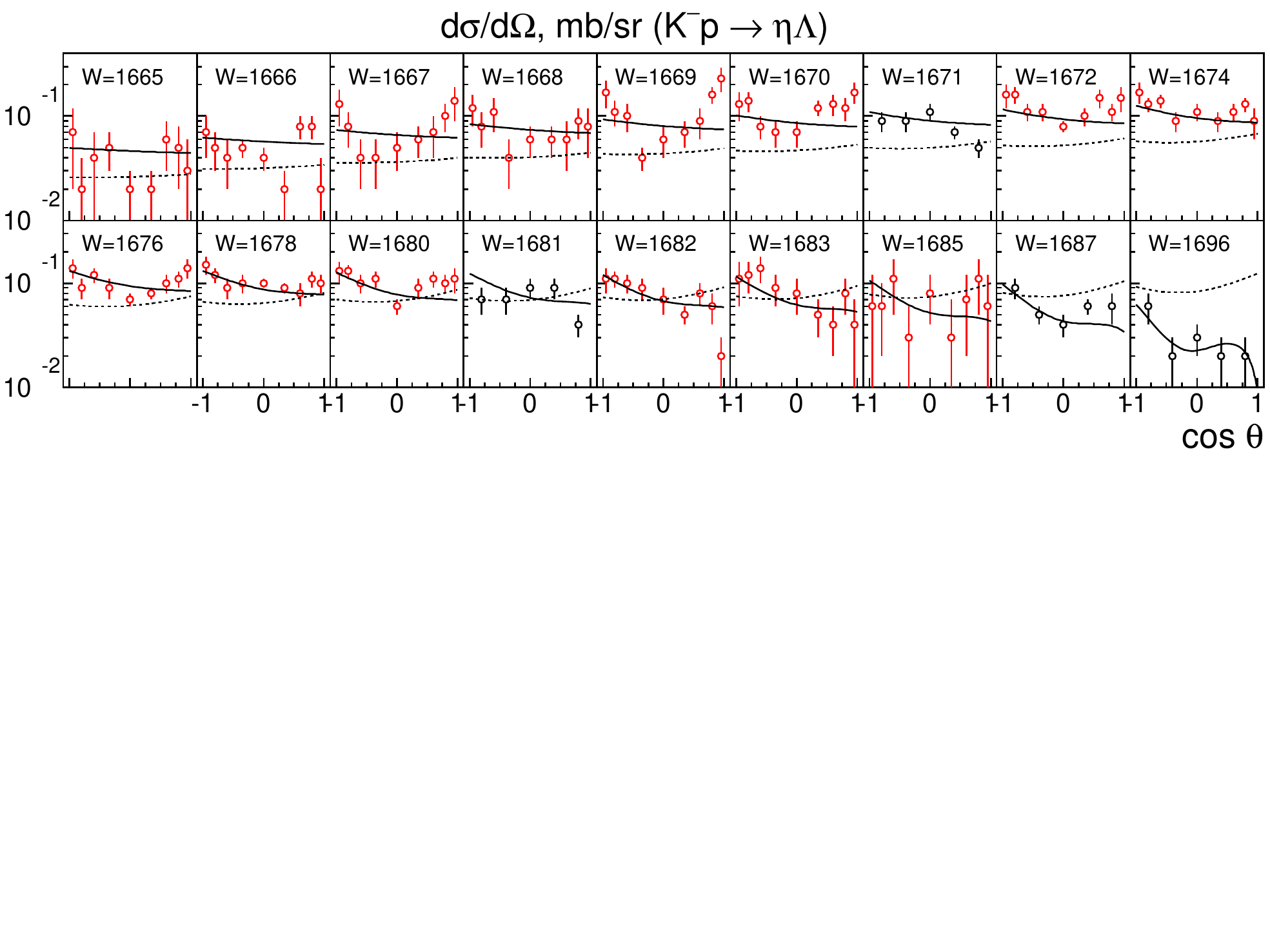}\vspace{-67mm}}
\caption{\label{fig:etalam}Differential cross sections for
$K^-p\to\eta\Lambda$ . The data are from \boldmath
  {$\circ$}~\cite{Armenteros:1970eg},
  {\rd$\circ$}~\cite{Starostin:2001zz}. \vspace{4mm}}
\bc
\includegraphics[width=0.85\textwidth]{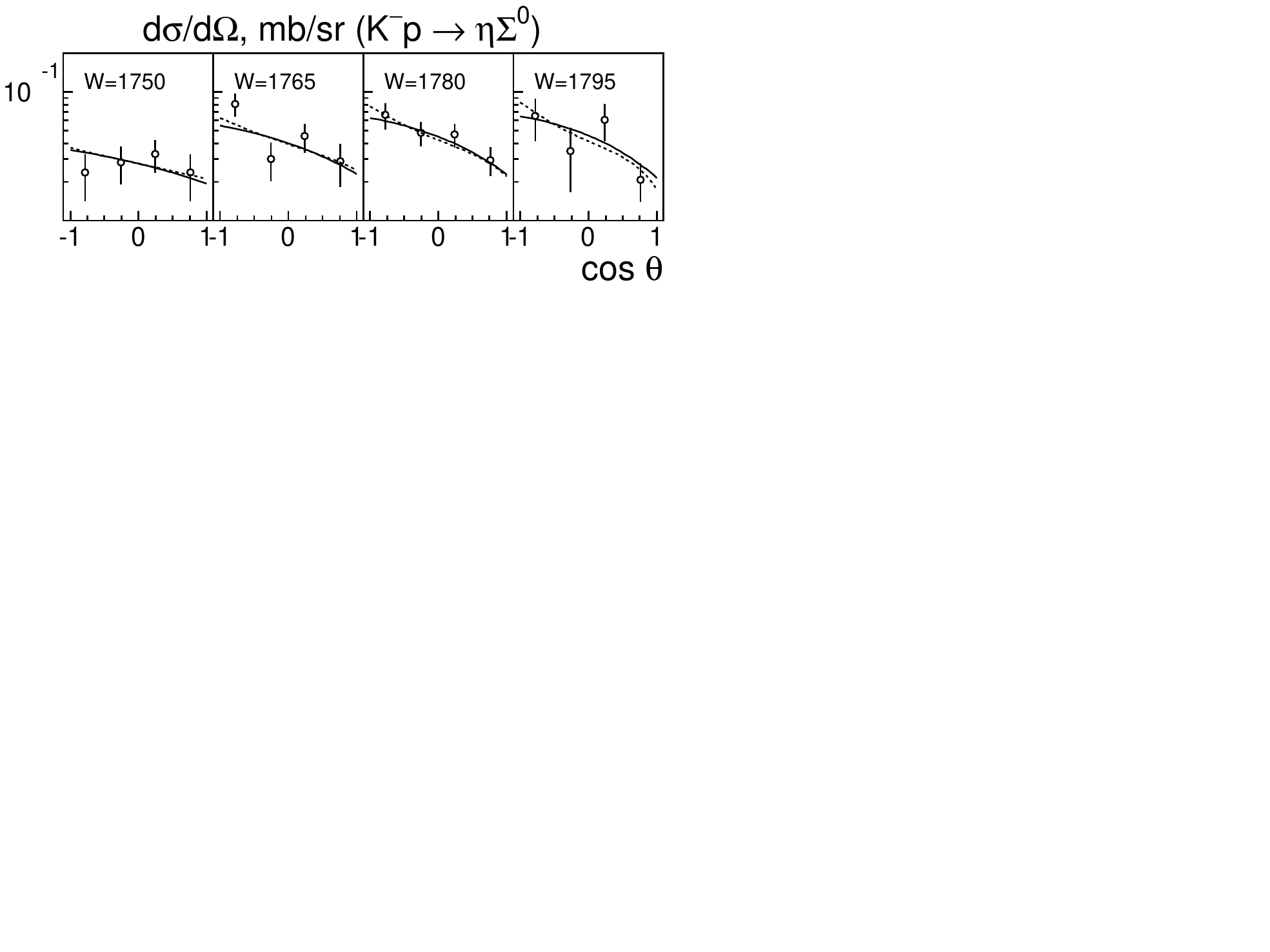}\vspace{-82mm}
\ec
\caption{\label{fig:etalam}Differential cross sections for
 $K^-p\to\eta\Sigma$. The data are from \boldmath
  {$\circ$}~\cite{Jones:1974si}  \vspace{2mm}}
\centerline{\includegraphics[width=0.95\textwidth]{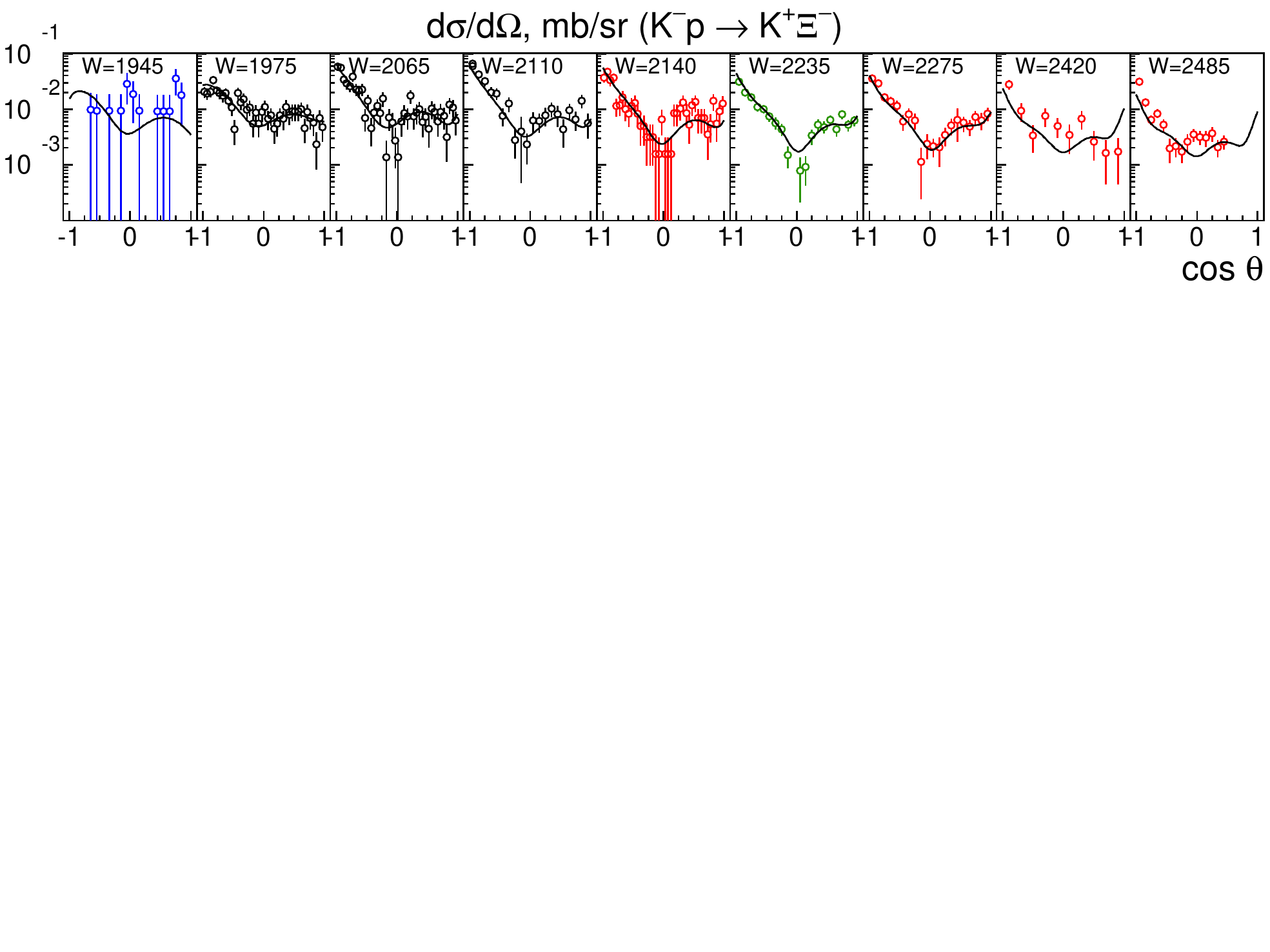}\vspace{-88mm}}
\caption{\label{fig:Xipm}Differential cross sections for
$K^-p\to K^+\Xi^-$. Data: \boldmath
   {$\circ$}~\cite{Burgun:1968g},
{\rd$\circ$}~\cite{Dauber:1969pm},
{\gr$\circ$}~\cite{Trippe:1967tg},
{\bl$\circ$}~\cite{Trower:1968wp}.}
\centerline{\includegraphics[width=0.85\textwidth]{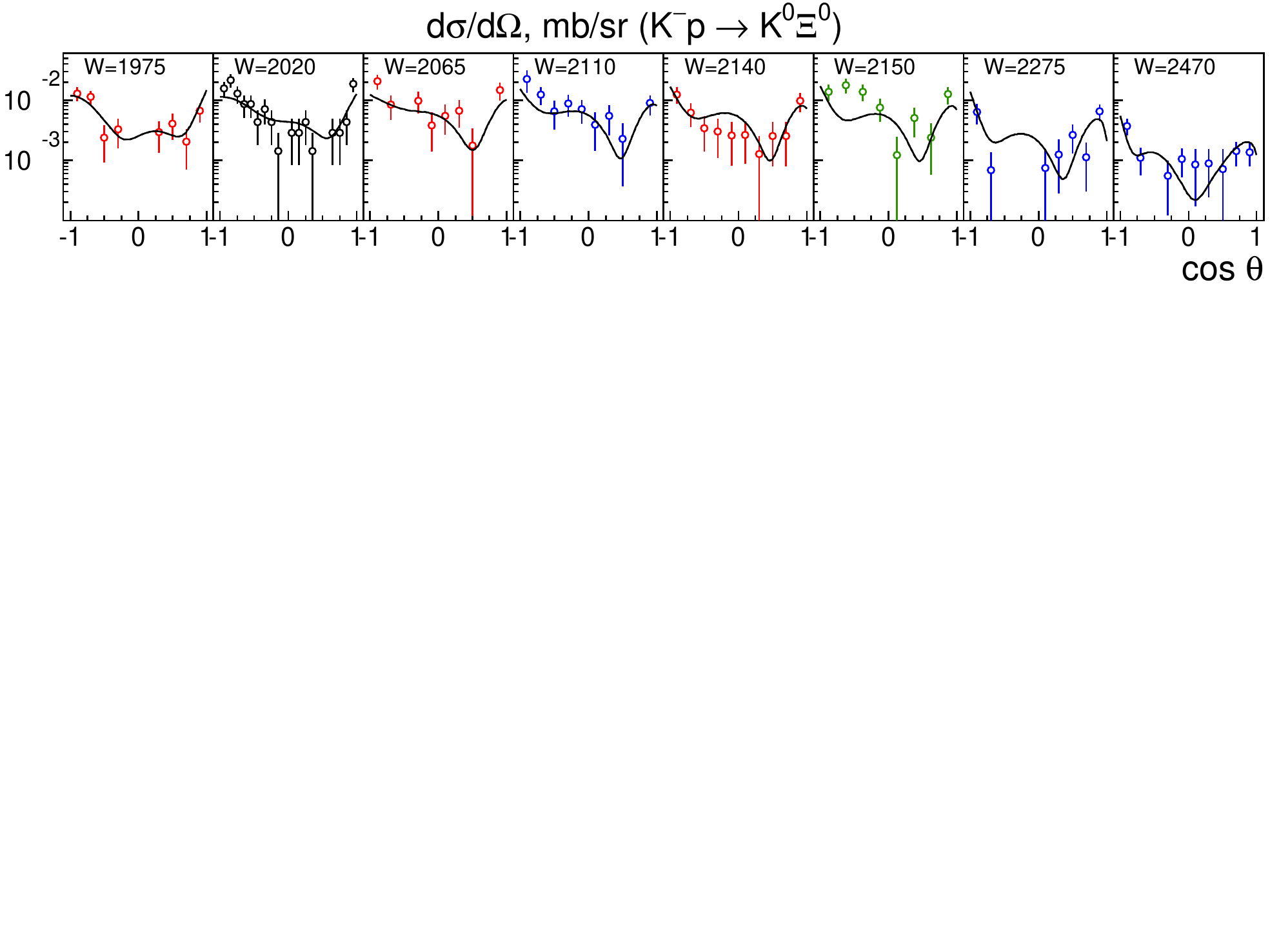}\vspace{-80mm}}
\caption{\label{fig:Xi00}Differential cross sections for $K^-p\to
K^0\Xi^0$. Data: \boldmath
{$\circ$}~\cite{Berge:1966jp}, {\rd$\circ$}~\cite{Burgun:1968g},
{\bl$\circ$}~\cite{Dauber:1969pm},
{\gr$\circ$}~\cite{Carlson:1973jr}. }
\end{figure*}

Figures~\ref{fig:elast1}-\ref{fig:Xipm} show a comparison of the data with our {\it primary}
(dashed curves) and our {\it final} (solid curves) fit. Mostly, the two fits show hardly any
difference: The established states serve as a rather good approximation. In these figures,
no scaling factors are applied, neither to the data nor to the fit.

First, we compare Figs.~\ref{fig:elast1}, \ref{fig:elast2} with Figs.~\ref{fig:CEX1},
\ref{fig:CEX2}. Note that both isospins can contribute to these reactions, with defined
Clebsch-Gordan coefficients. Thus, $\Lambda^*$ and $\Sigma^*$ resonances both contribute with
interfering amplitudes. The $K^-p$ and $\bar K^0 n$ decay amplitudes of $\Sigma^*$ resonances have the
same, of $\Lambda^*$ resonances opposite signs: this allows one to separate the $\Sigma^*$ and
$\Lambda^*$ contributions from the data shown in Figs.~\ref{fig:elast1} - \ref{fig:CEX2}. Similar
arguments hold true for measurements of $K^-p\to \pi^-\Sigma^+$ and $\pi^+\Sigma^-$ or $K^+\Xi^-$
and $K^0\Xi^0$.




\begin{figure*}
\centerline{\includegraphics[width=0.85\textwidth]{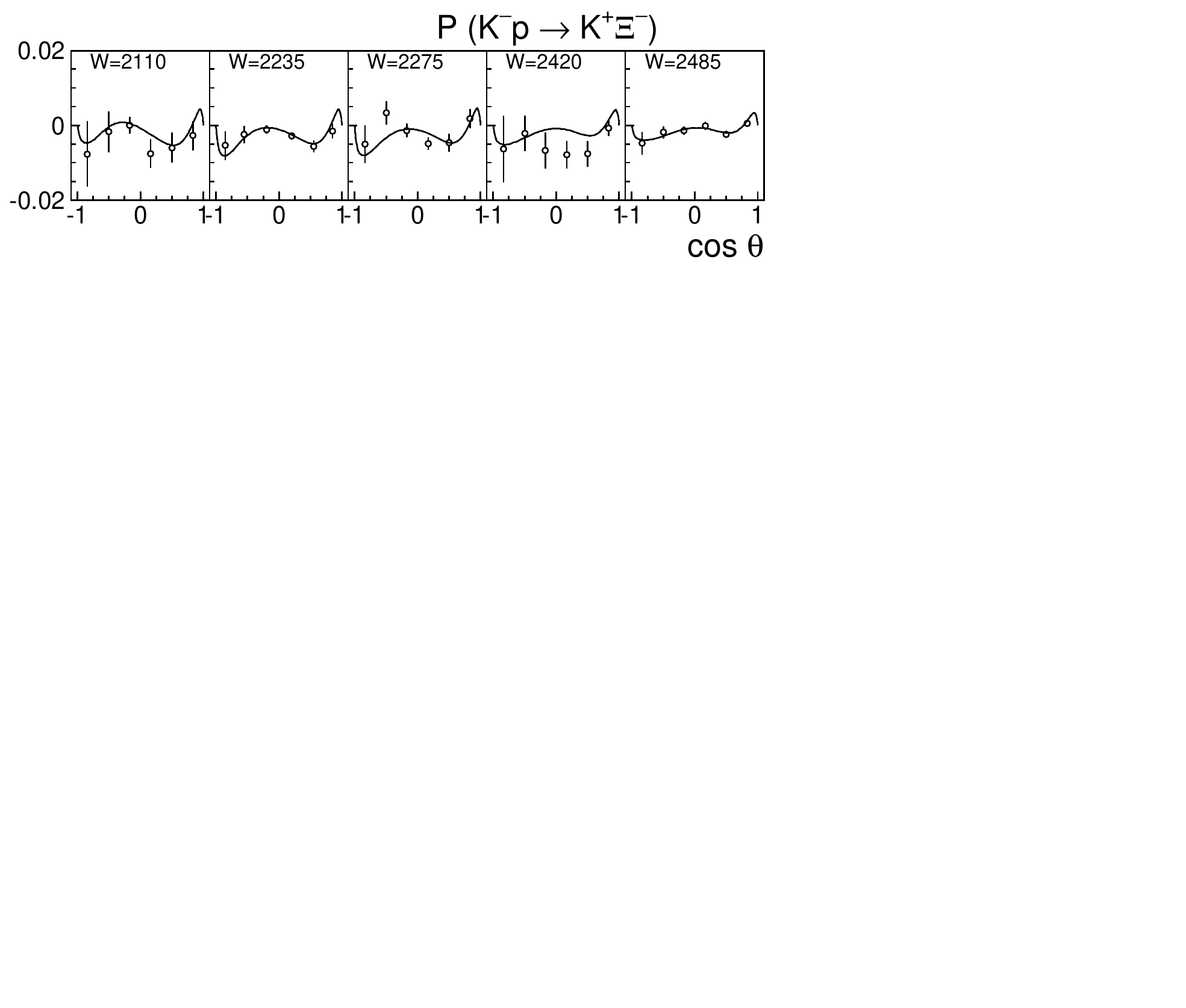}\hspace{-5cm}
\includegraphics[width=0.85\textwidth]{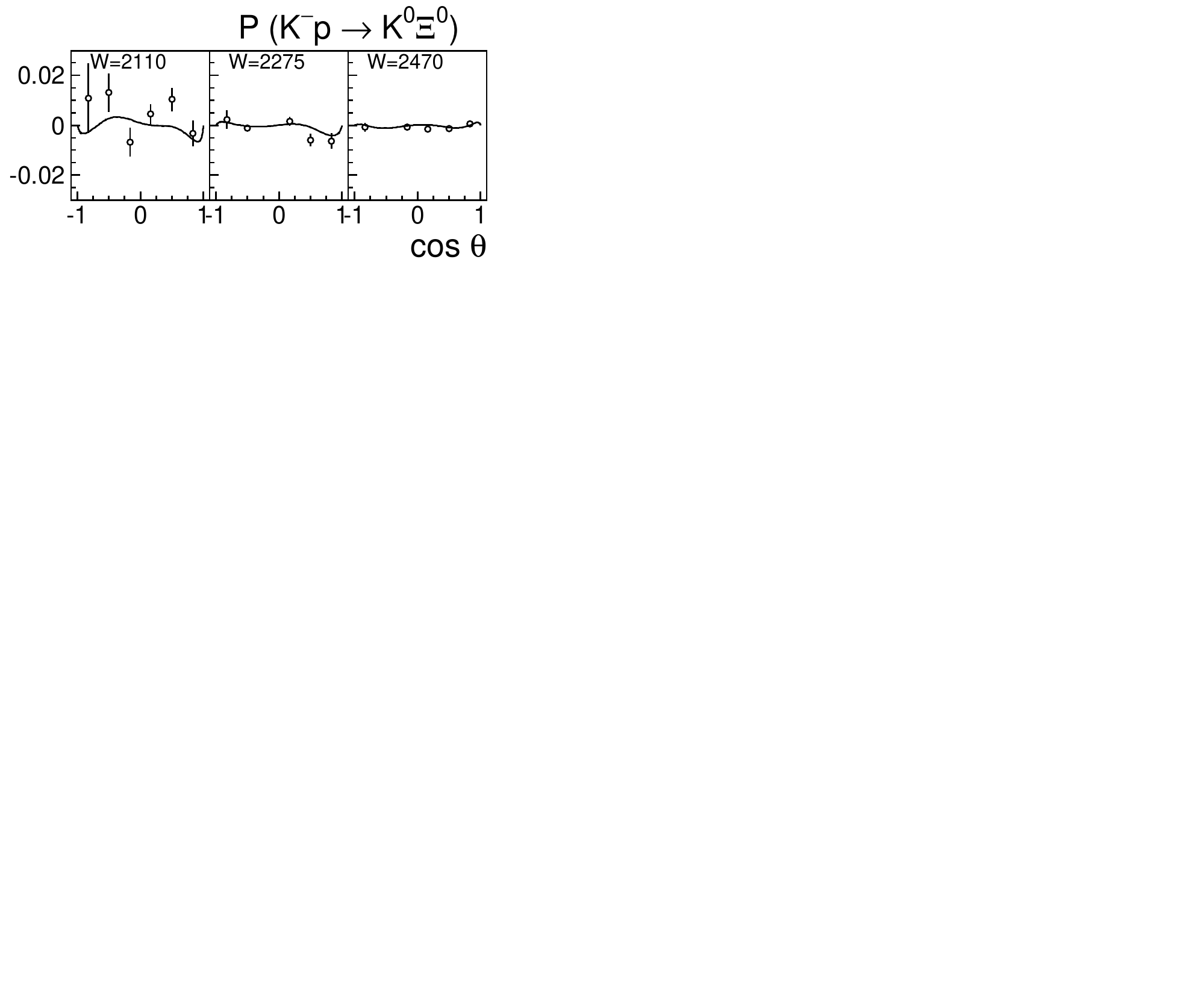}\hspace{-9cm}\vspace{-96mm}}
\caption{\label{fig:Xi_p}Recoil asymmetry for  $K^-p\to K^+\Xi^-$
and $K^-p\to K^0\Xi^0$. The data are from \boldmath
{$\circ$}~\cite{Sharov:2011xq}.}
\end{figure*}
\begin{figure*}[pt]
\begin{tabular}{cc}
\includegraphics[width=0.49\textwidth]{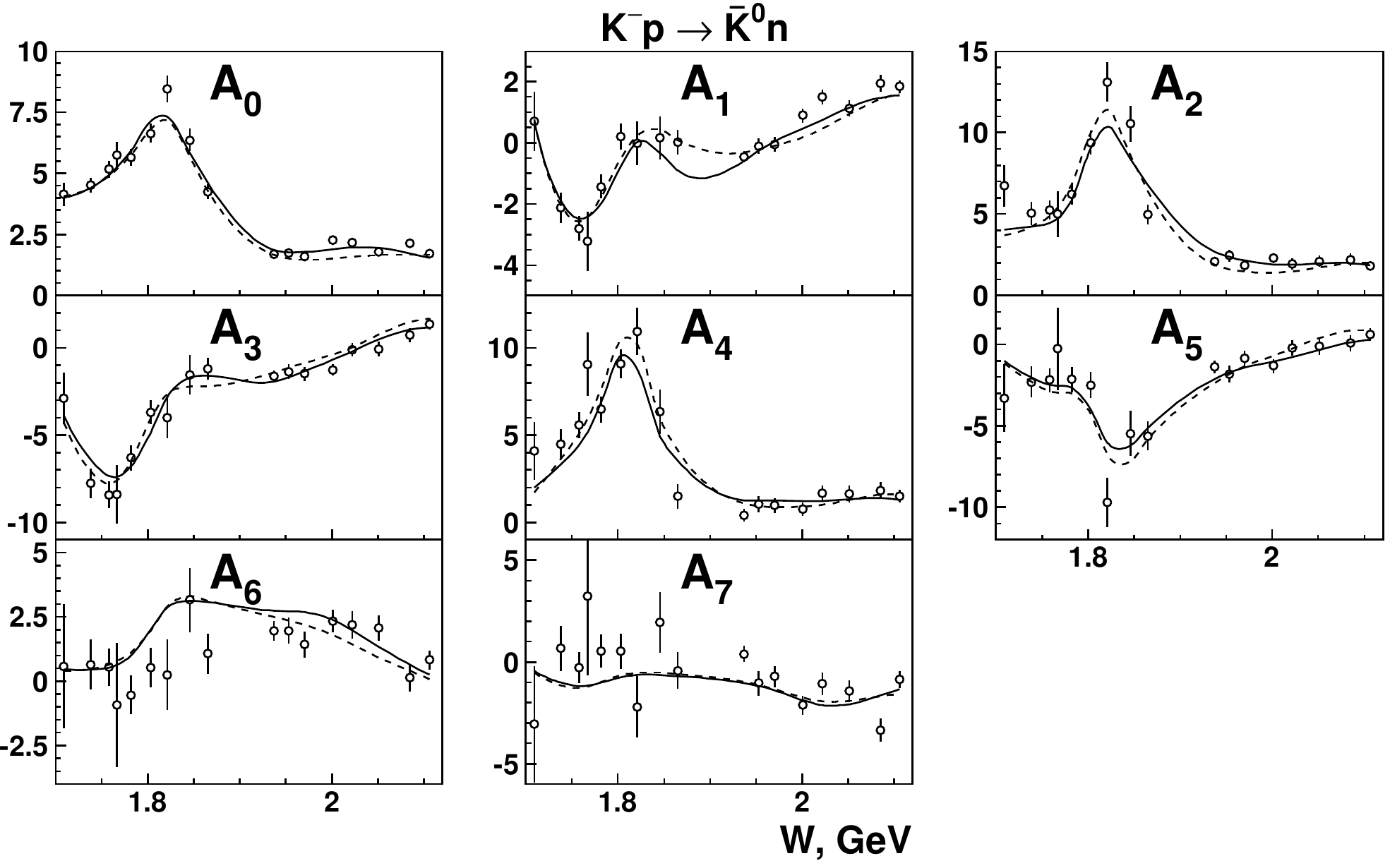}
&
\includegraphics[width=0.49\textwidth]{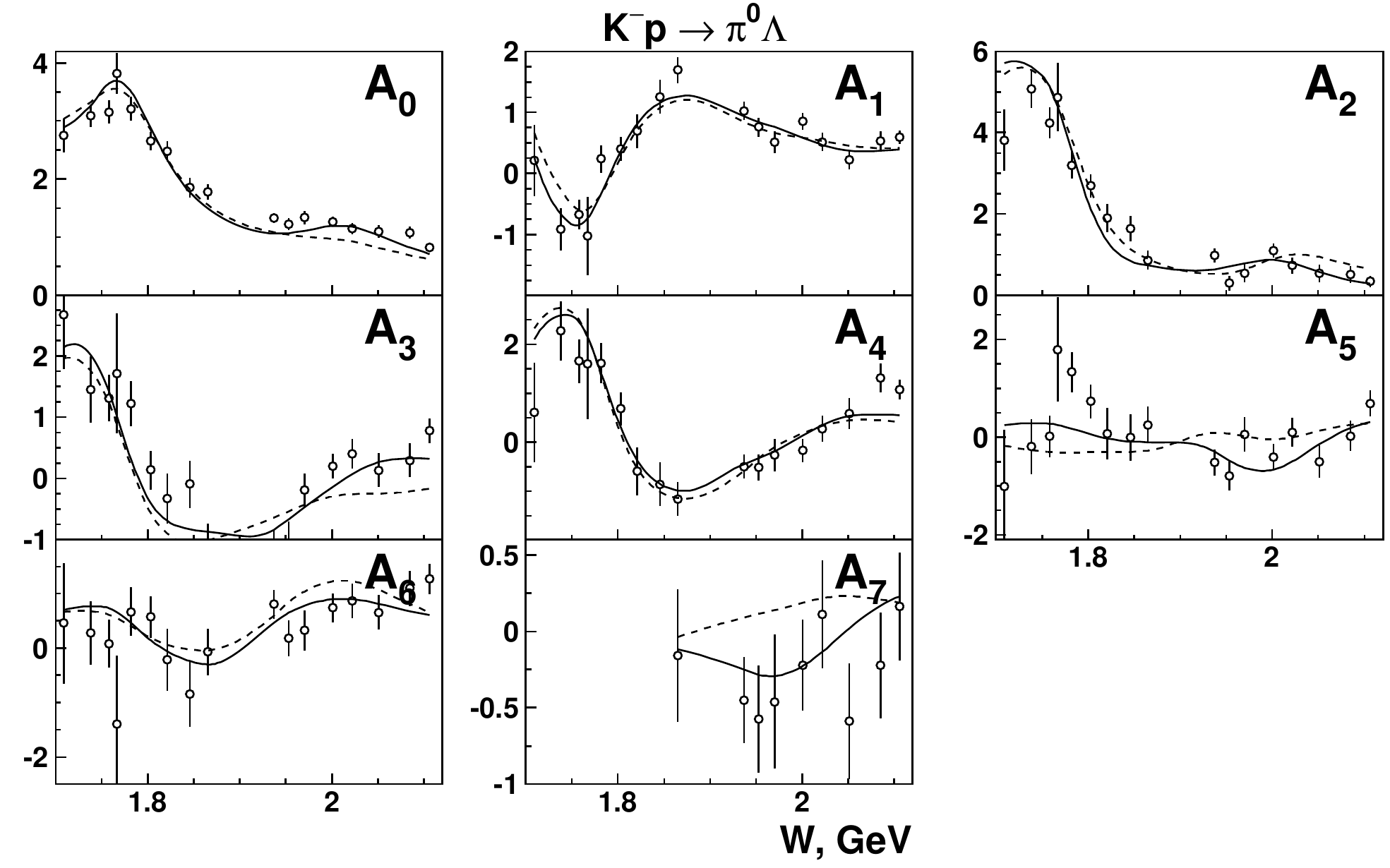}
\end{tabular}
\caption{\label{fig:Legendre}Legendre coefficients for the expansion of the differential cross sections for
for $K^-p\to \bar K^0n$ (left) and $K^-p\to \pi^0\Lambda$ (right). The data are from \cite{Horn:1972aj}.}
\end{figure*}
\begin{figure*}
\bc
\begin{tabular}{cccc}
\hspace{-3mm}\includegraphics[width=0.25\textwidth,height=0.23\textheight]{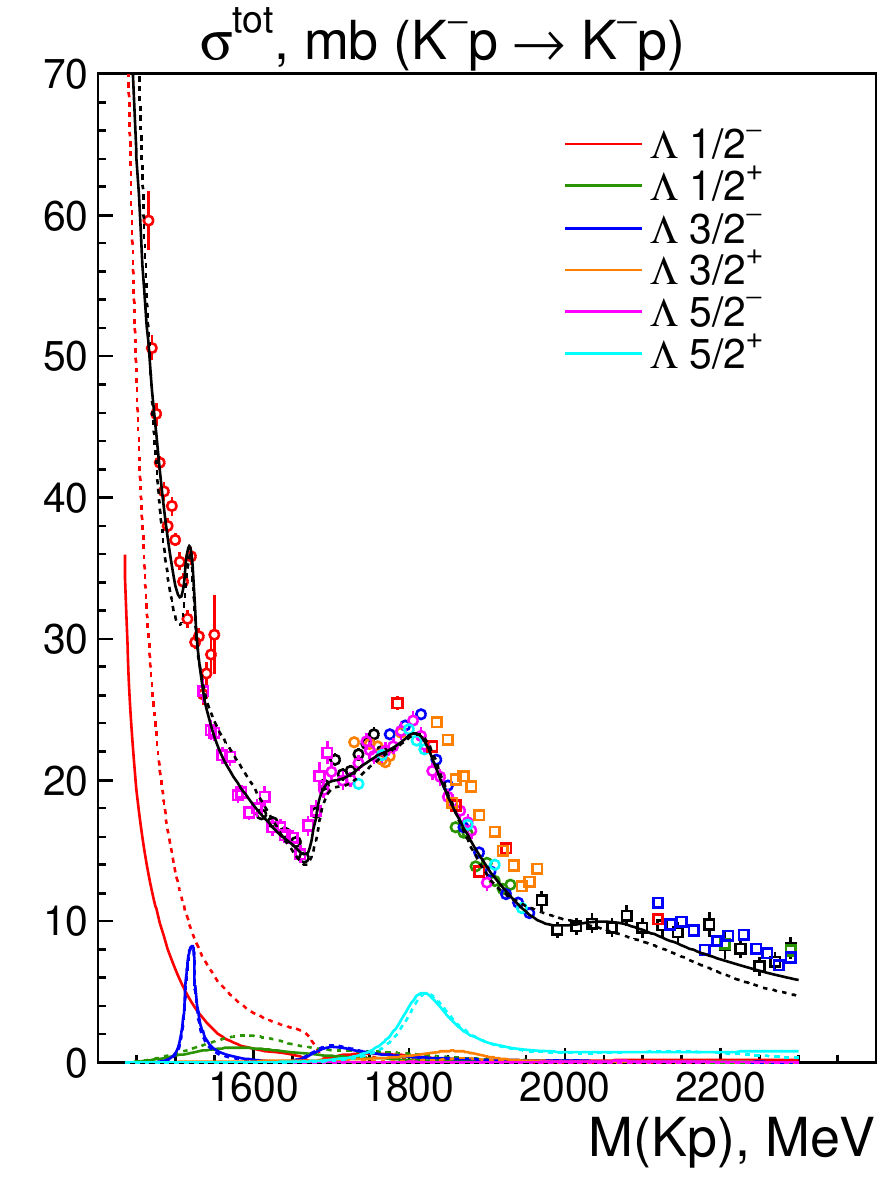}&
\hspace{-3mm}\includegraphics[width=0.25\textwidth,height=0.23\textheight]{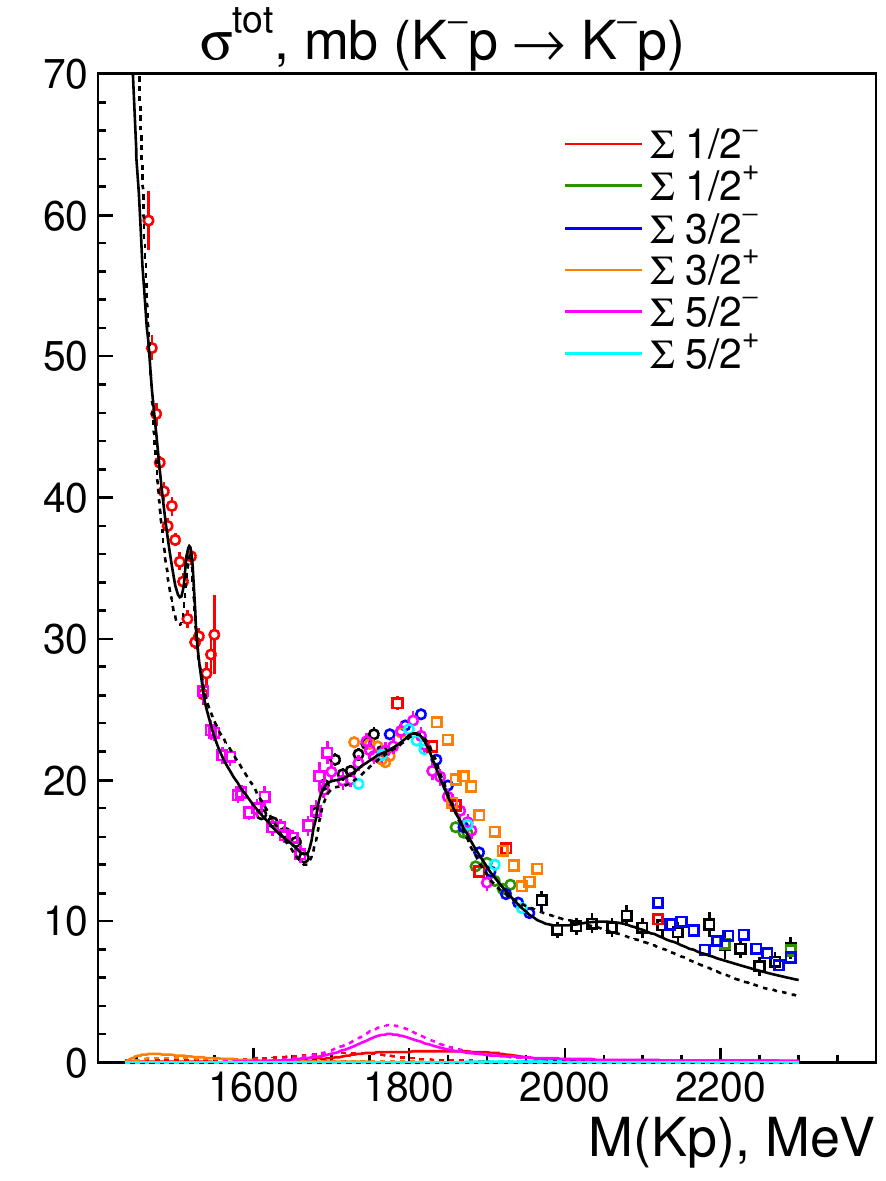}&
\hspace{-3mm}\includegraphics[width=0.25\textwidth,height=0.23\textheight]{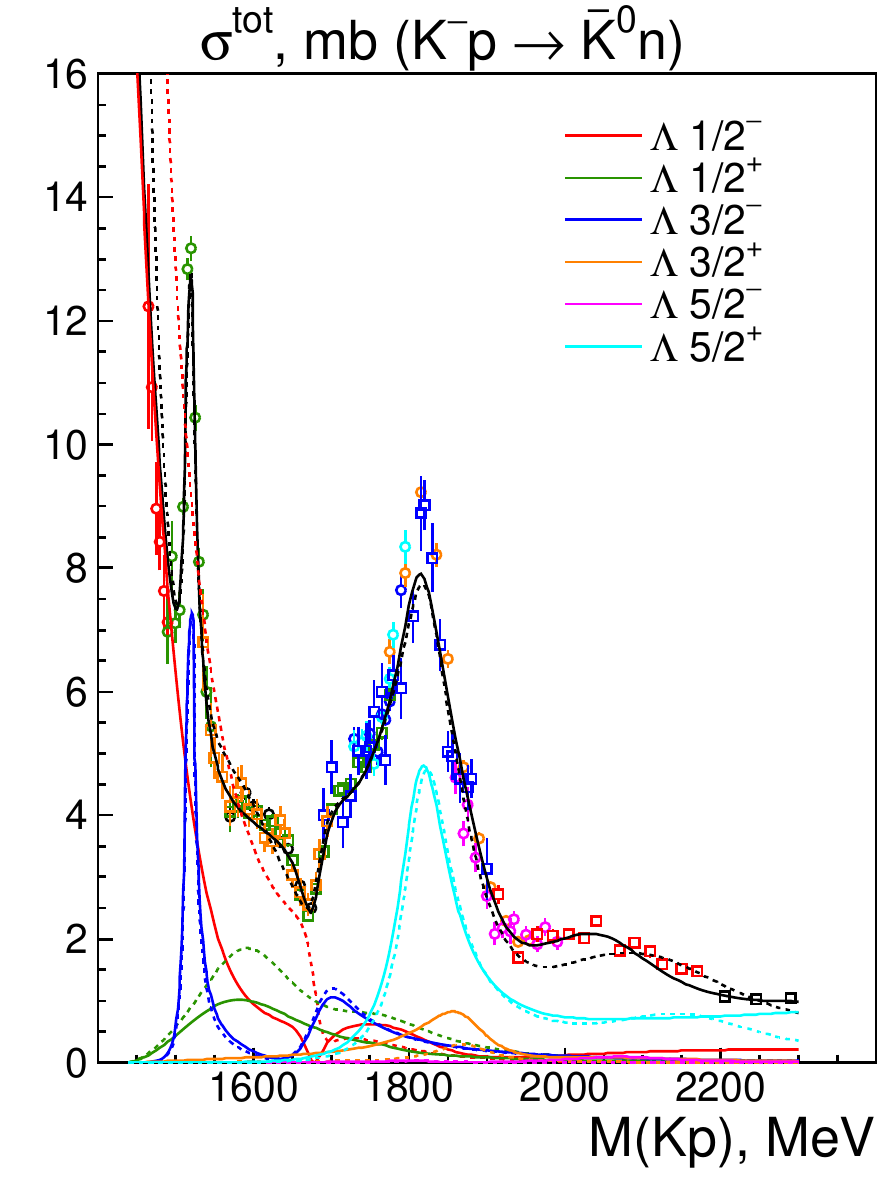}&
\hspace{-3mm}\includegraphics[width=0.25\textwidth,height=0.23\textheight]{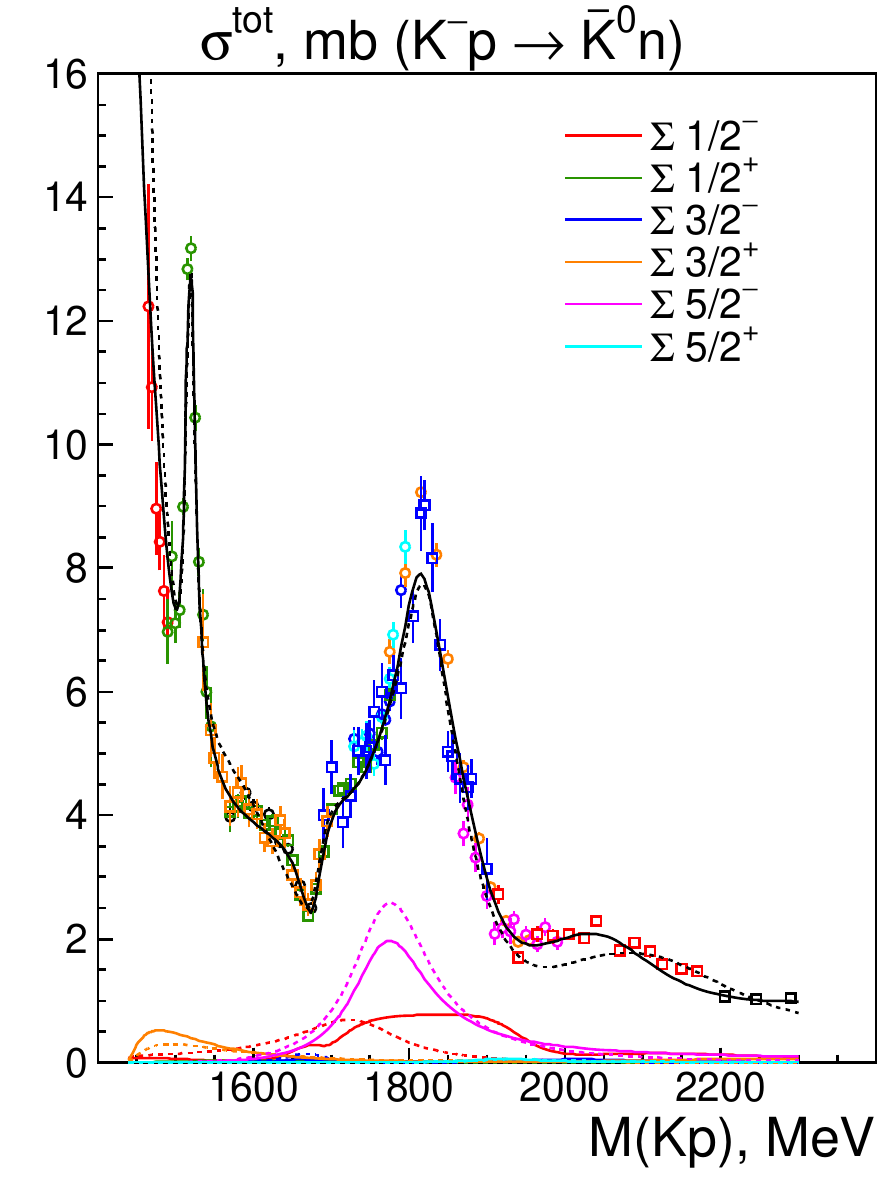}\\
\hspace{-3mm}\includegraphics[width=0.25\textwidth,height=0.23\textheight]{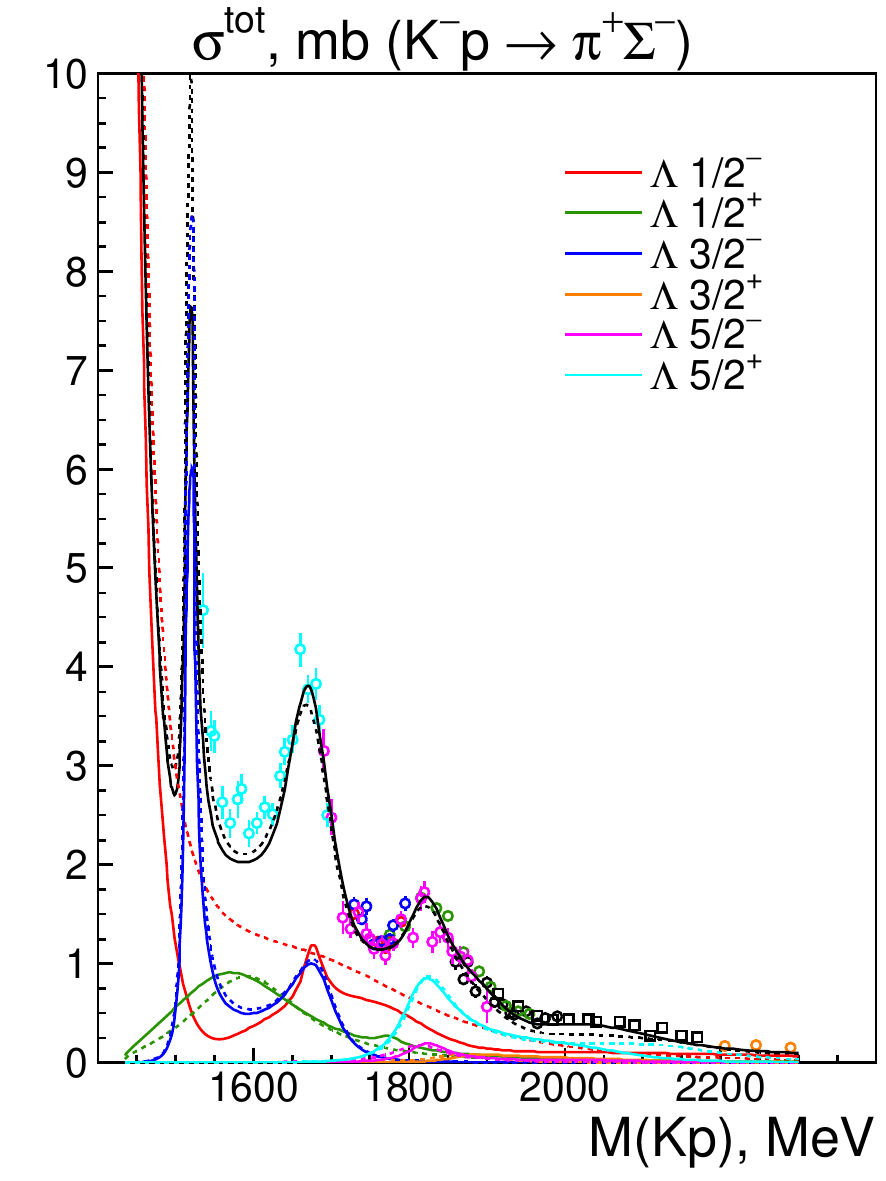}&
\hspace{-3mm}\includegraphics[width=0.25\textwidth,height=0.23\textheight]{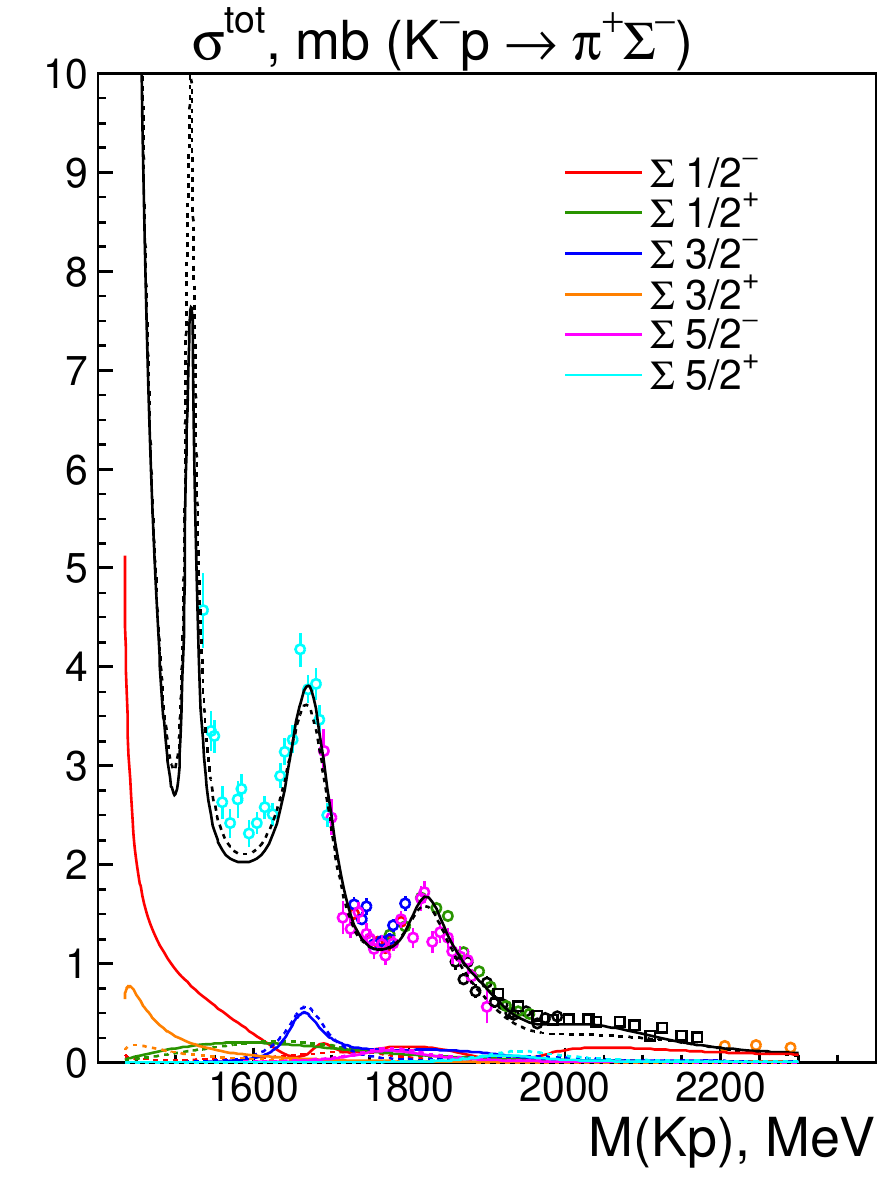}&
\hspace{-3mm}\includegraphics[width=0.25\textwidth,height=0.23\textheight]{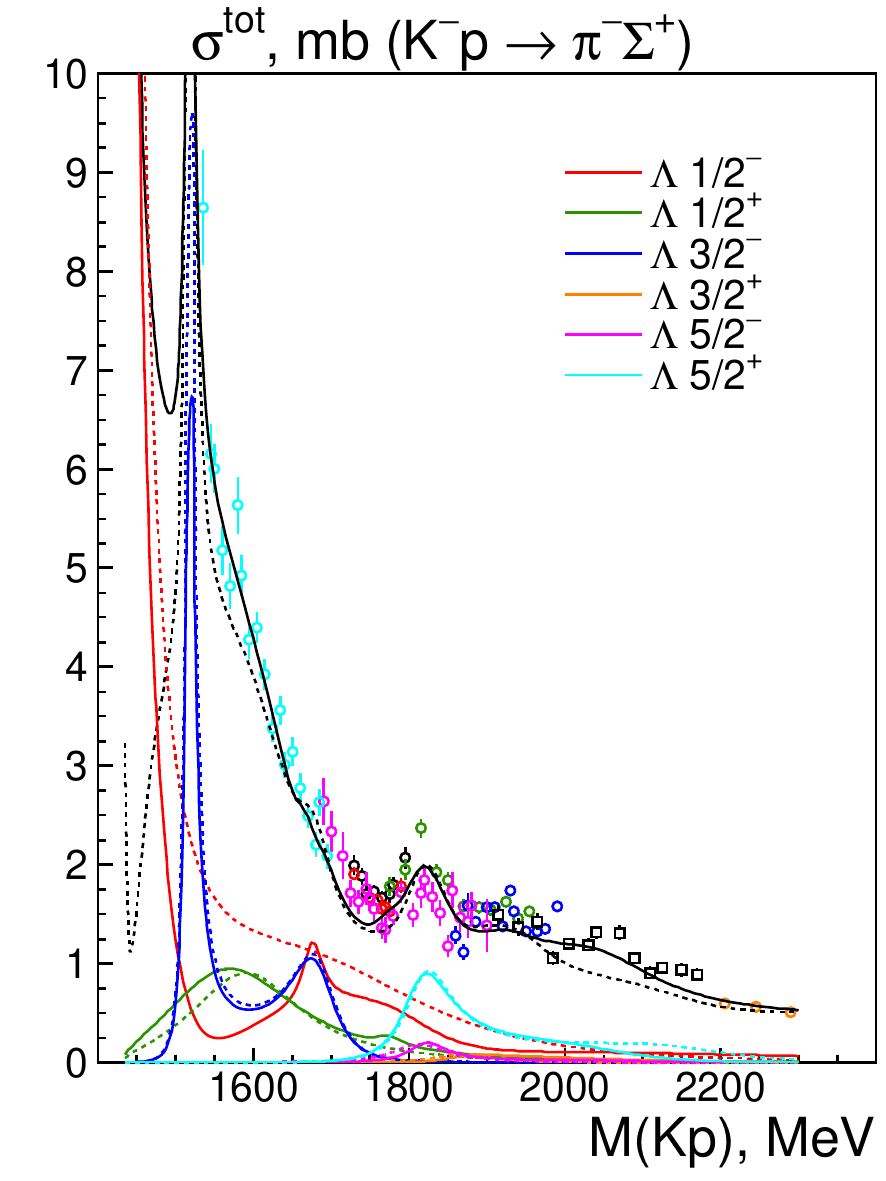}&
\hspace{-3mm}\includegraphics[width=0.25\textwidth,height=0.23\textheight]{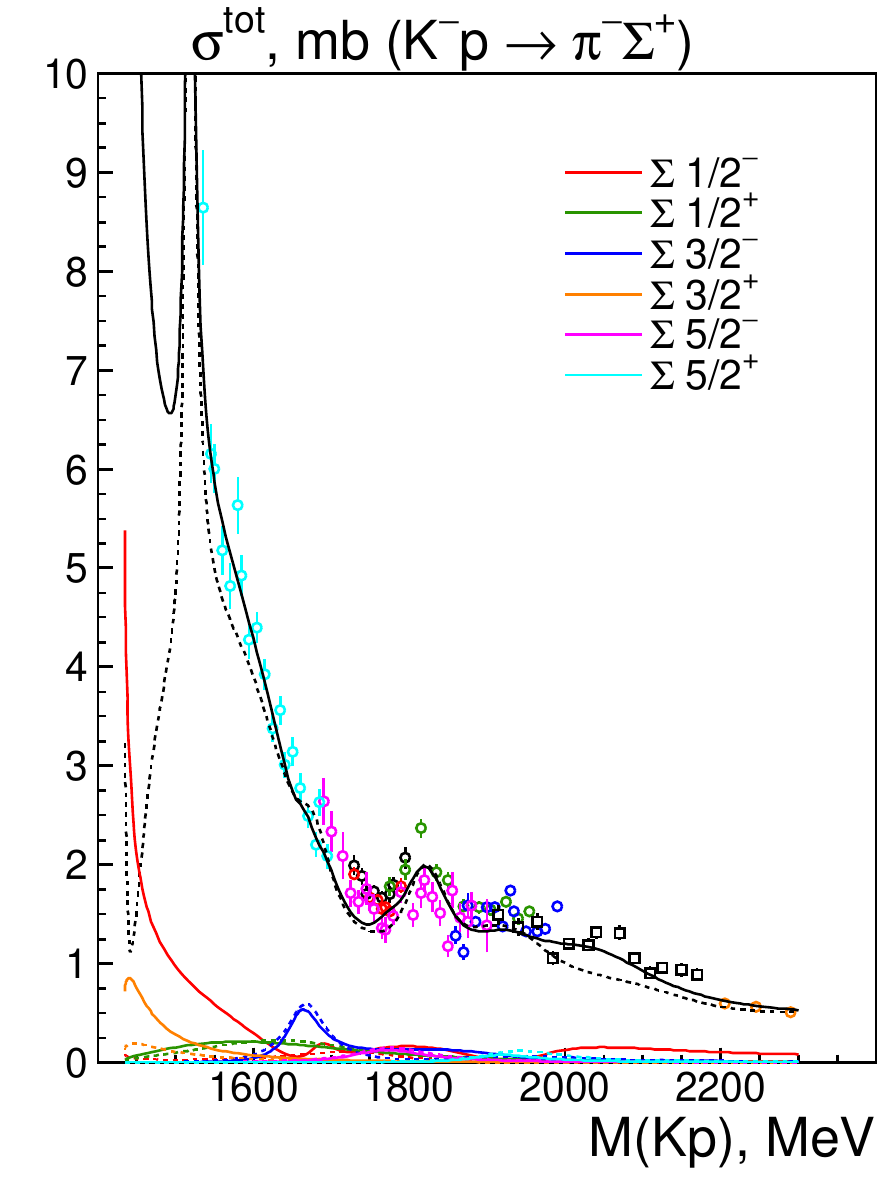}\\
\hspace{-3mm}\includegraphics[width=0.25\textwidth,height=0.23\textheight]{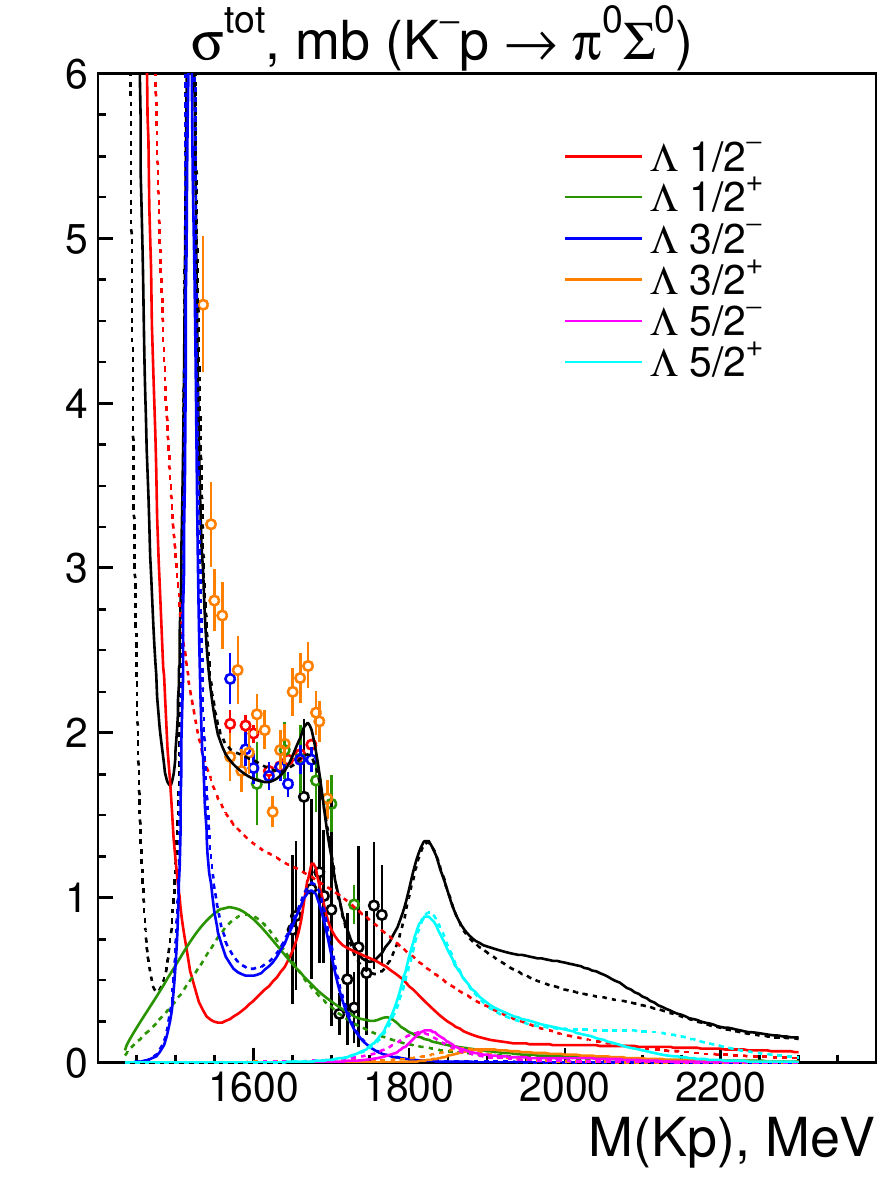}&
\hspace{-3mm}\includegraphics[width=0.25\textwidth,height=0.23\textheight]{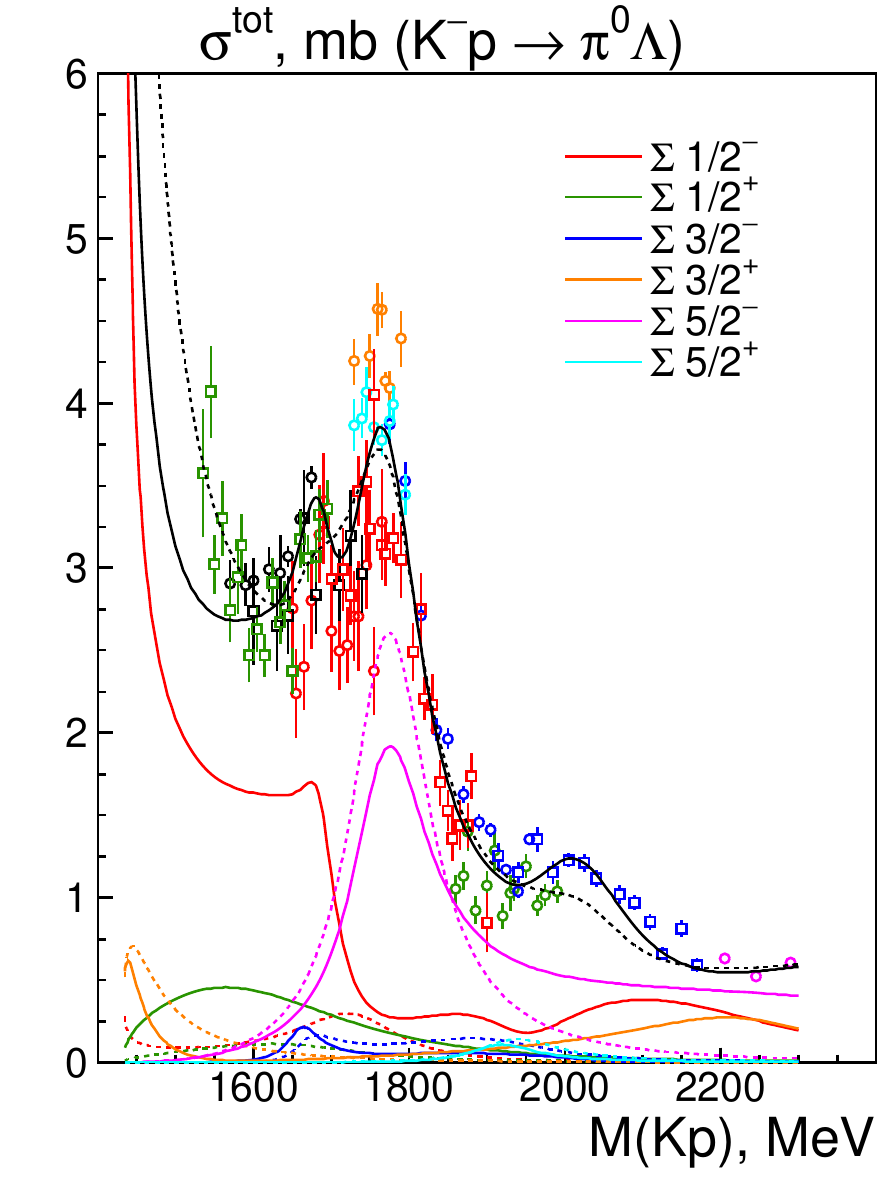}&
\hspace{-3mm}\includegraphics[width=0.25\textwidth,height=0.23\textheight]{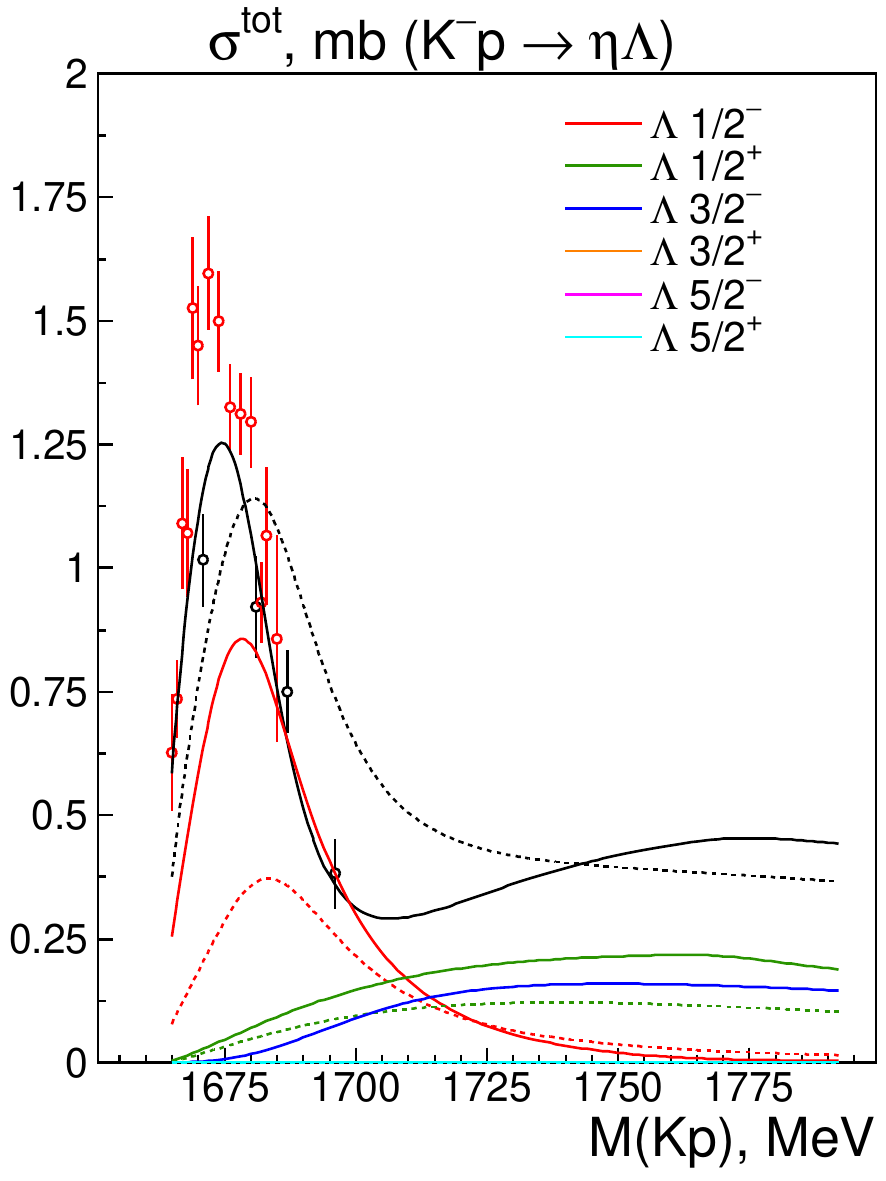}&
\hspace{-3mm}\includegraphics[width=0.25\textwidth,height=0.23\textheight]{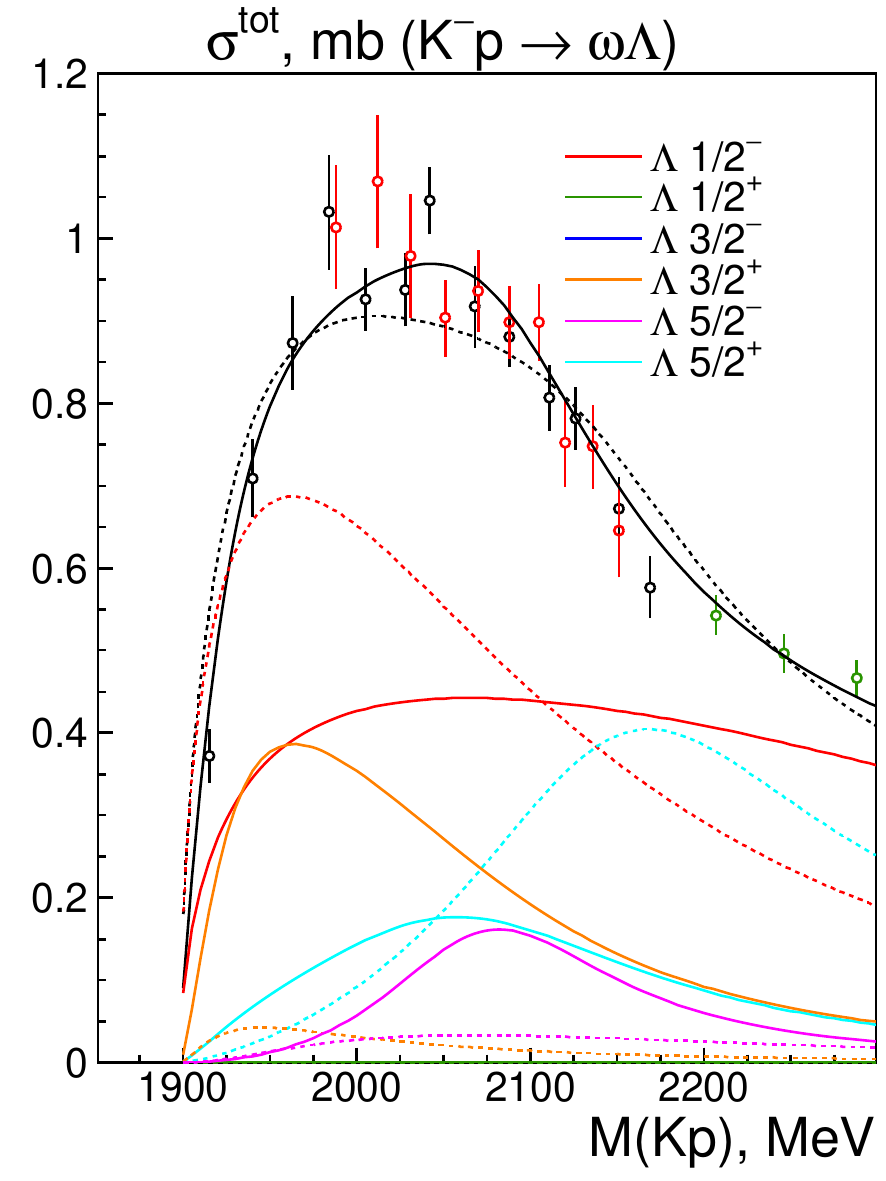}\\
\hspace{-3mm}\includegraphics[width=0.25\textwidth,height=0.23\textheight]{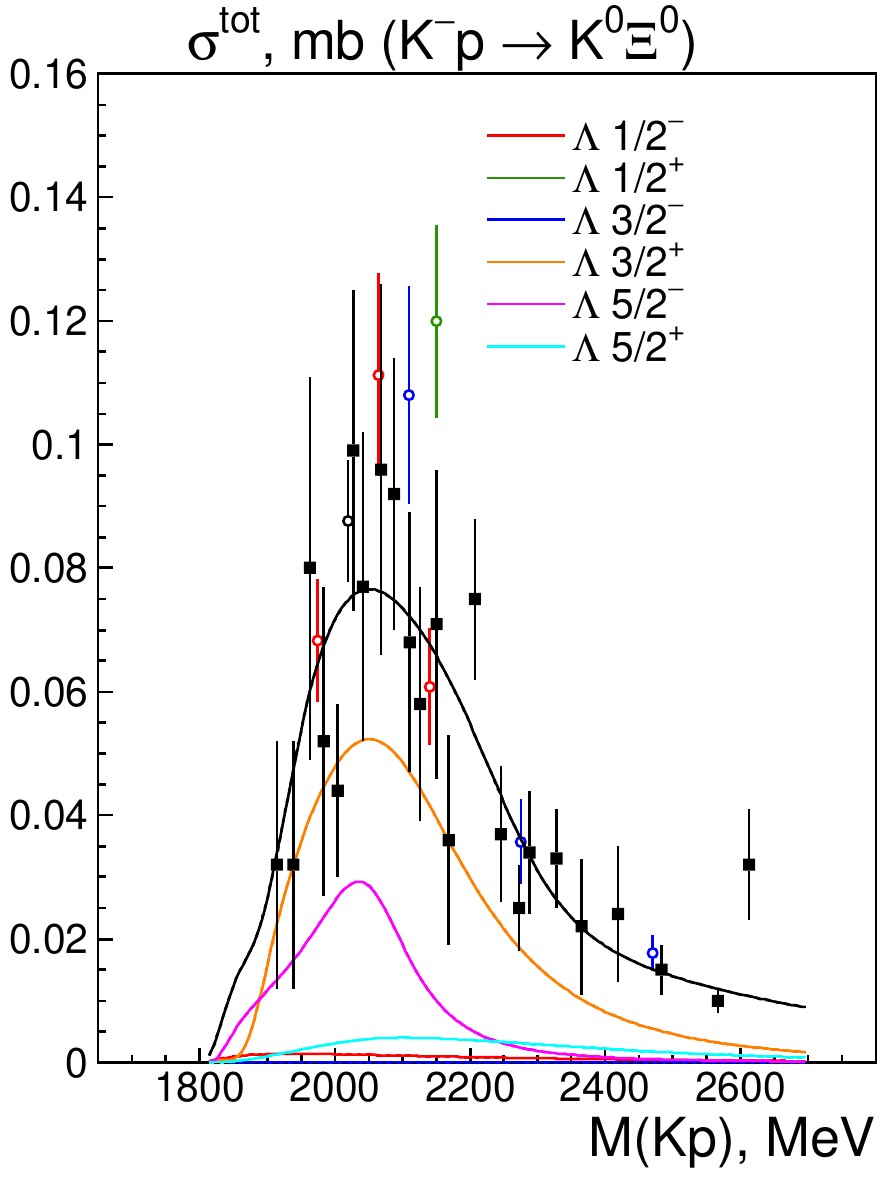}&
\hspace{-3mm}\includegraphics[width=0.25\textwidth,height=0.23\textheight]{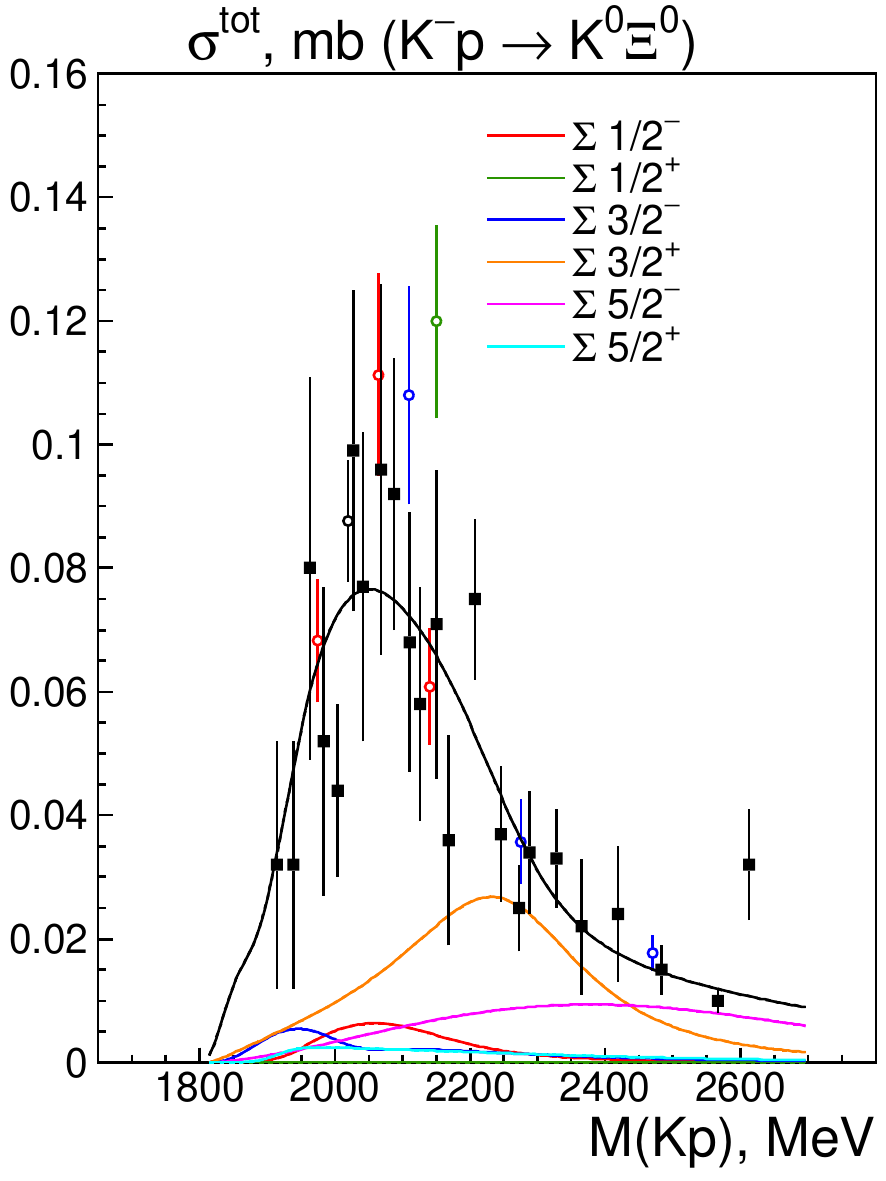}&
\hspace{-3mm}\includegraphics[width=0.25\textwidth,height=0.23\textheight]{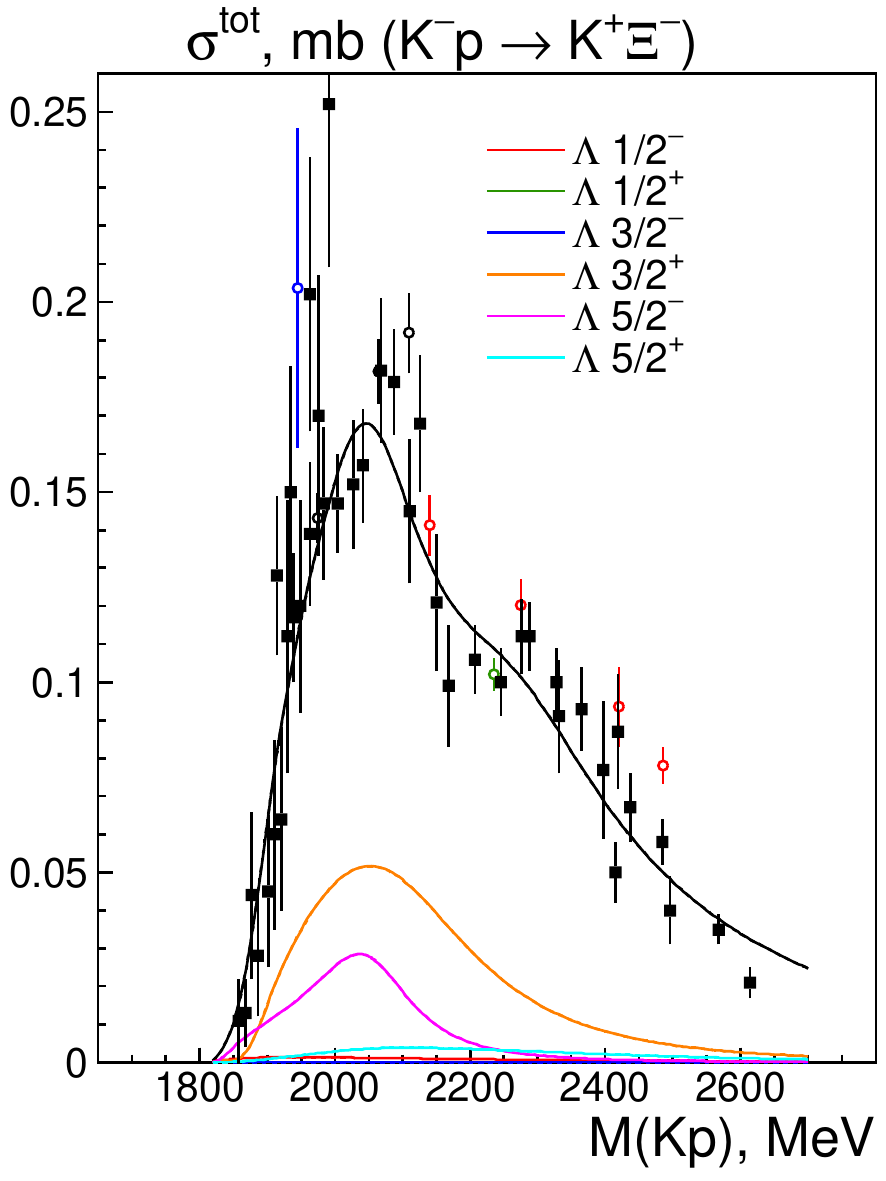}&
\hspace{-3mm}\includegraphics[width=0.25\textwidth,height=0.23\textheight]{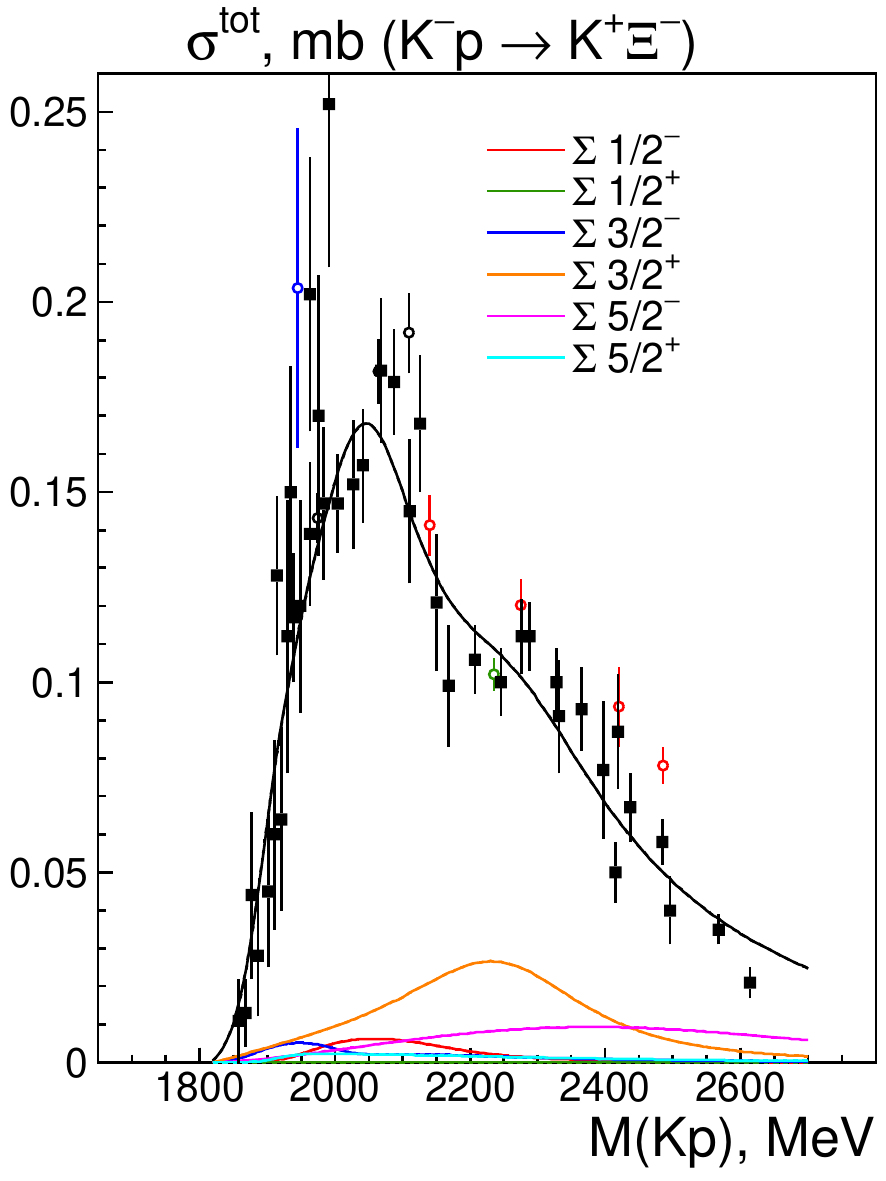}
\end{tabular}
 \ec
\caption{\label{fig:tot}Total cross sections for major $K^-$ induced reactions. The data on the
total cross sections are calculated from the corresponding differential cross sections. The black
solid curves represent the result of the final coupled-channel fit, the dotted line shows the result of the primary
Breit-Wigner fit. The partial wave contributions
for $\Lambda^*$'s and $\Sigma^*$'s are shown in separate subfigures. }
\end{figure*}
The cross sections for elastic scattering are significantly larger than those for charge exchange.
At $W=1500$\,MeV, the former is 2-4\,mb/sr while the latter is in the order of 1\,mb/sr. At
2000\,MeV the elastic cross section rises in forward direction and reaches 10\,mb/sr while charge
exchange leads to an oscillating cross section staying below 1\,mb/sr. Obviously, there are
significant $t$-channel contributions to the elastic channel -- likely due to Pomeron exchange --
while $t$-channel contributions to the charge-exchange reaction need $\rho$ exchange. The impact of
hyperon resonances is better seen in the charge-exchange data.

At the lowest mass -- 1465\,MeV -- the charge-exchange cross section falls off slightly in forward
direction; it is largely due to the dominant $S$-wave scattering with a small $P$-wave
contribution. $S$-wave or $P$-wave alone would both lead to a constant angular distribution. At
some masses, the reactions are dominated by one resonance. It may be helpful to compare the
experimental angular distributions with the theoretical distributions (see Fig.~\ref{simulation})
for a few mass intervals.

Figure~\ref{fig:CEX1} shows the angular distribution for $K^-p\to \bar K^0n$.
 At 1520\,MeV, there is a clear ($3\cos^2\theta + 1$) distribution
above a very small background. A comparison with Fig.~\ref{simulation} shows that $J=3/2$ is the
dominant wave: of course, it is the well-known $\Lambda(1520)$. With increasing mass, the $J=3/2$
contribution gets smaller and the minimum shifts without additional wiggles and without a strong
forward-backward asymmetry. This pattern signals additional contributions from $J^P=1/2^+$ and
$J^P=1/2^-$ waves.

At 1680\,MeV, a sharp minimum is seen in the total $K^-p\to\bar K^0n$ cross section
(see Fig.~\ref{fig:tot}) which is
assigned  to a sign change of the amplitude in the $N\bar K$ S-wave amplitude at about this mass.
$\Lambda(1670)1/2^-$ appears as a dip rather than as a peak (like $f_0(980)$ in $\pi\pi$
scattering). The effect is enhanced
by  the sudden rising of the $\Lambda(1690)3/2^-$ contribution.
Interestingly, the $K^-p\to \eta\Lambda$ cross section rises from threshold to a peak value of
above 0.1\,mb/sr at 1670\,MeV and has fallen below 0.03\,mb/sr in the highest-mass bin at
1696\,MeV.

In the subsequent energy bins of the $K^-p\to \bar K^0n$ differential cross sections, a stronger
forward-backward asymmetry develops which indicates the interference of odd and even partial waves.
Gradually, the angular distribution develops a strong w-shaped distribution, best recognized in the
1820 to 1850\,MeV mass bins. The comparison with Fig.~\ref{simulation} suggests significant $J=5/2$ contributions. The partial
wave analyses assigns this to a strong $\Lambda(1820)5/2^+$ production, the forward-backward
asymmetry to a smaller $\Sigma(1775)5/2^-$ amplitude and some smaller contributions from lower
partial waves. Above 2000\,MeV, the angular distributions are characterized by forward and backward
maxima with two additional maxima: In this mass region, $J^P=7/2^\pm$ are the most prominent
partial waves (see Fig.~\ref{simulation}). Similar observations can be made for the reaction $K^-p\to \pi^0\Lambda$.

The differential cross sections for the processes $K^-p\to \pi^0\Lambda$ and the $\Lambda$
polarization are shown in Figs.~\ref{fig:SpiL1}, \ref{fig:SpiL2}, and~\ref{fig:PpiL}, those for
$K^-p\to \pi^0\Sigma^0$ and the $\Sigma^0$ polarization in Figs.~\ref{fig:SpiS0} and
\ref{fig:PpiS0}. Only $\Sigma^*$ resonances contribute to $K^-p\to \pi^0\Lambda$, and only
$\Lambda^*$ resonances to $K^-p\to \pi^0\Sigma^0$.

Inspecting Figs.~\ref{fig:elast1}, \ref{fig:elast2}, \ref{fig:CEX1}, \ref{fig:CEX2} again, we
notice a few discrepancies between data and fit. Often, the discrepancies are enforced by data in
neighboring bins: the structure at $\cos\theta =0$ and 1935\,MeV in Fig.~\ref{fig:elast2} is, e.g.,
incompatible with the neighboring bins. The same statements holds for the data
from~\cite{Barber:1975rg} above 2200\,MeV which are incompatible with those
from~\cite{Daum:1968jey} and~\cite{deBellefon:1976qr}.

The polarization data are shown in Fig.~\ref{fig:elast}. Polarization data are available in the
full energy range considered here even though above 2\,GeV with limited statistical significance
only. For a single partial wave, the polarization vanishes. The complicated angular dependence of
$P$ indicates the presence of several partial waves which makes a direct interpretation difficult.
However, these data provide important constraints for the partial-wave analysis.

When the cross sections for $\pi^-\Sigma^+$ and $\pi^+\Sigma^-$ production in Figs.
\ref{fig:SpiSp1} and \ref{fig:SpiSp2} are compared, similar effects as in $\bar KN$ production are
seen. The cross sections for $\pi^-\Sigma^+$ are larger and the angular distributions at high
energies show forward peaking.  Note that for $t$-channel exchange, $\pi^+\Sigma^-$ production
requires an exotic particle to be exchanged. It is not included in the fits and not required by the
data. The contributing resonances are discussed using the data on $K^-p\to\pi^+\Sigma^-$.

The branching ratios for
$\Lambda(1520)\to \Lambda\pi\pi$, $\Sigma\pi\pi$ and $\Lambda\gamma$ of (12\er 1)\% are imposed as
missing width; the $\Lambda(1520)\to \pi\Sigma$ decay fraction is then fixed  fixed by unitarity
($\Gamma_{\rm tot}=\sum\Gamma_i$) when the data on elastic and charge exchange $K^-p$ scattering
are used; these data span the mass range down to $\sim$1465\,MeV.

The $\Lambda(1690)3/2^-$ can be recognized by the ($3\cos^2\theta+~1$) angular distribution. The
resonance makes a very significant contribution, jointly with $\Lambda(1670)1/2^-$. Both are even
partial waves, and the angular distribution remains approximately symmetric. At masses above
1800\,MeV, the w-shaped angular distribution turns up again, signalling $\Lambda(1820)5/2^+$ but
with a forward-backward asymmetry due an odd partial wave from $\Lambda(1830)5/2^-$.

Data on the reaction $K^-p\to \pi^0\Sigma^0$ were difficult to extract with the experimental
techniques of the 70ties of last century: The $\Sigma^0$ decays to $\Lambda\gamma$; with
$\pi^0\to\gamma\gamma$, there are three $\gamma$ in the final state. In \cite{London:1975zz}, a
bubble chamber was used that was filled with propane (C$_3$H$_8$) and freon (CF$_3$Br) in which
$\gamma$ ray have a high chance to convert. The $\gamma$ conversion probability depends on the
$\gamma$ energy which was difficult to simulate. In view of these difficulties, the agreement
between data and fit seems acceptable (see Fig.~\ref{fig:SpiS0}). The polarization $P$ for this
reaction (see Fig.~\ref{fig:PpiS0}) was determined using the Crystal Ball detector at
BNL~\cite{Prakhov:2008dc} with its excellent photon detection capability but with a limited energy
range.

The first mass bin in Fig.~\ref{fig:SpiS0} seems to suggest that there should be more
$\Lambda(1520)$ than the fit admits. A larger $\Lambda(1520)$ contribution would, however, worsen
the fit to the data on the $\pi^\mp\Sigma^\pm$ final states. In the 1650 to 1700\,MeV mass range, a
($3\cos^2\theta + 1$) contribution signals $\Lambda(1690)$ $3/2^-$ contributions above a small even
partial wave. The fit additionally identifies $\Lambda(1670)1/2^-$. Polarization data exist only
over a very limited range.

The low-energy region of the $K^-p\to \pi^0\Lambda$ differential cross section is dominated by the
interference of even and odd partial waves, of $S$- and $P$-waves. At higher energies similar
structures show up as we have seen them before: first a ($3\cos^2\theta + 1$), later a
$9/4\cdot$($5\cos^4\theta - 2\cos^2\theta +1)$ angular distribution distorted by contributions from
other waves. At masses above 2000\,MeV, more wiggles show up. The fit is sensitive to the
polarization data but their statistical value is limited again.

The differential cross sections for $K^-p\to \eta \Lambda$  are shown in Figs.~\ref{fig:etalam}.
They do not show striking structures.

The structure in the $K^-p\to K^+\Xi^-$ differential cross section (Fig.~\ref{fig:Xipm}) identifies
leading contributions in the $J^P=3/2^+$ and $3/2^-$ partial waves, the assignment to the $\Lambda$
or $\Sigma$ sector follows from the $K^-p\to K^0\Xi^0$ channel (Fig.~\ref{fig:Xi00}):
$K^0 \Xi^0\to\Sigma^0$ is forbidden. Description of the recoil asymmetry for both reactions is
shown in Fig.~\ref{fig:Xi_p}. The two reactions were studied in a single-channel
analysis~\cite{Landay:2018wgf}. Possibly contributing hyperon resonances were tested in a blindfold
identification process. Ten resonances were suggested to contribute to the reaction, among them
four 1* resonances for which we find no evidence in any final state. The strongest evidence is seen
for $\Sigma(2030)7/2^+$ which we do not observe in this decay mode.

The authors of Ref.~\cite{Horn:1972aj} reported measurements of the charge exchange reaction and
$\pi^0\Lambda$ formation. The results were expanded into associated Legendre polynomials:
\be
\label{Legendre}
 \frac{d\sigma}{d\cos\Theta}           =& \frac 12 \sum\limits_l A_l P^0_l(\cos\Theta)
\ee
The data were included in our fits. The data are shown in Fig.~\ref{fig:Legendre}. Significant
structures are seen, in particular for $K^-p\to \bar K^0n$ at 1800\,MeV where the interference of
$\Lambda(1800)1/2^-$, $\Lambda(1820)5/2^+$, $\Lambda(1830)5/2^-$ leads to a complicated pattern.
The largest contributions in $K^-p\to \pi^0\Lambda$ are at about 1700\,MeV which are strongly
influenced by $\Sigma(1620)$ $1/2^-$, $\Sigma(1660)1/2^+$, $\Sigma(1670)3/2^-$. With $S$, $P$, and
$D$-waves, the Legendre coefficients up to $A_4$ show traces of these resonances while $A_5$, $A_6$
and $A_7$ stay small.

\begin{figure*}[pt]
\bc
\begin{tabular}{cccc}
\includegraphics[width=0.240\textwidth]{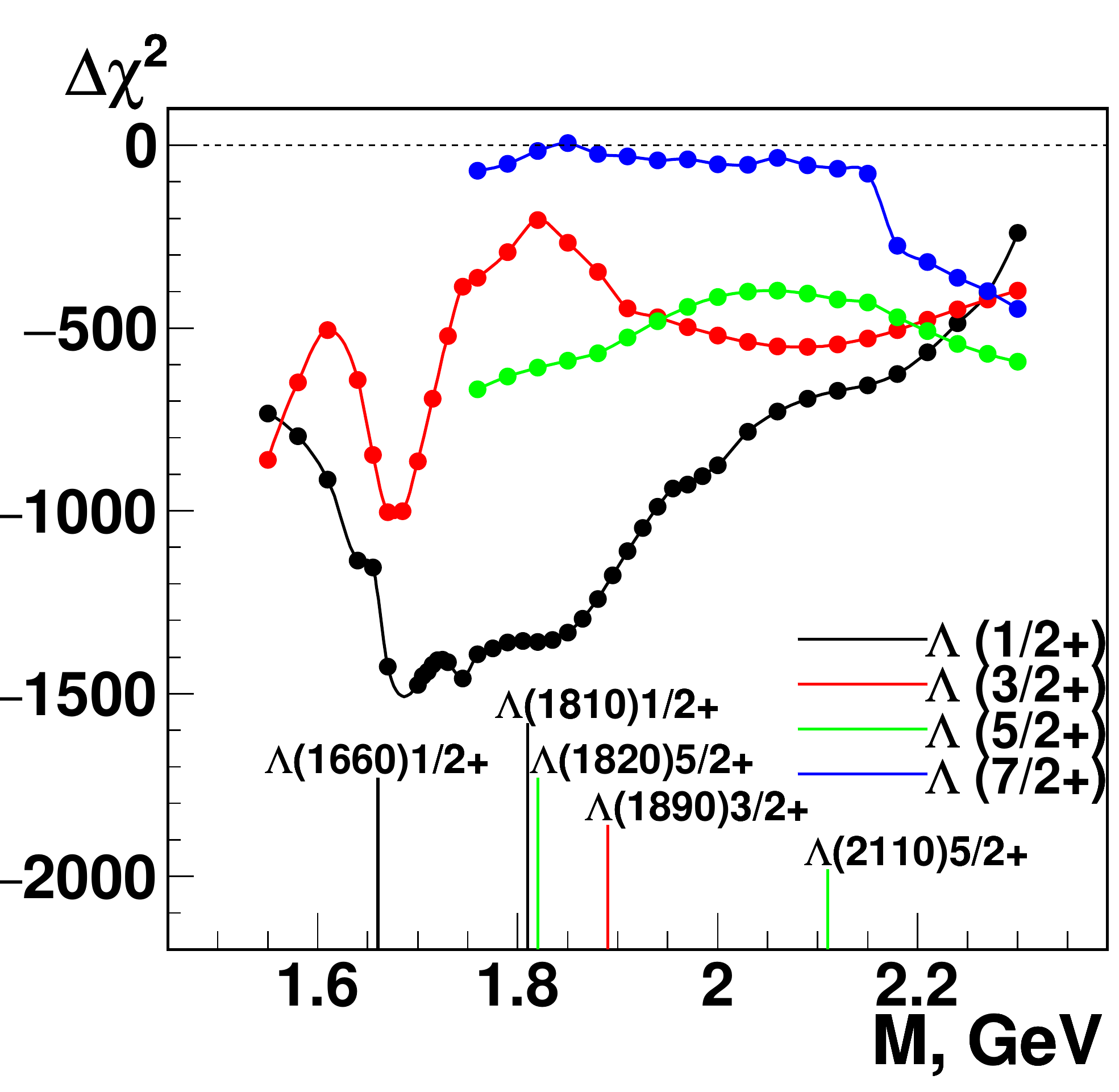}&
\hspace{-3mm}\includegraphics[width=0.240\textwidth]{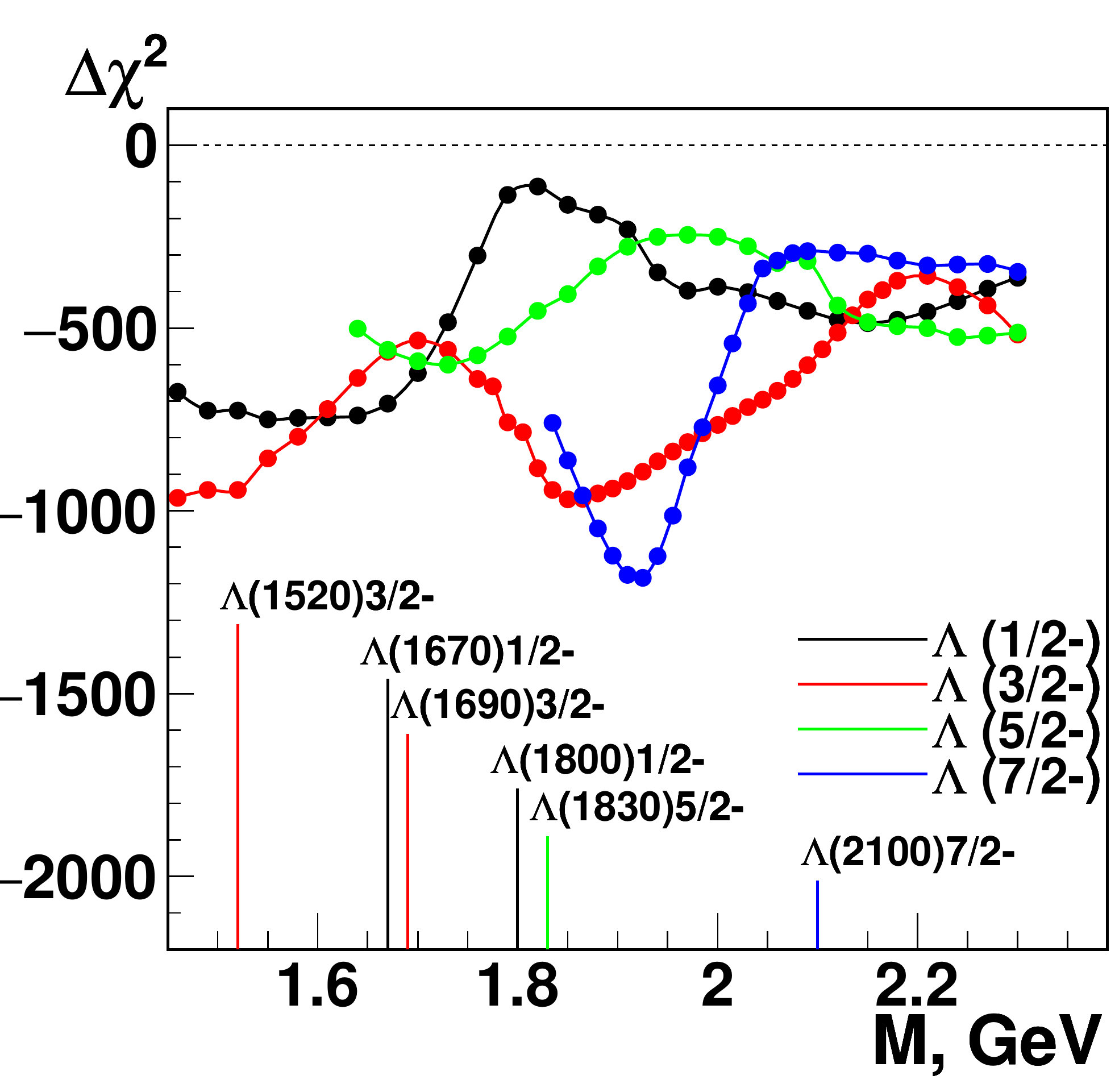}&
\hspace{-3mm}\includegraphics[width=0.240\textwidth]{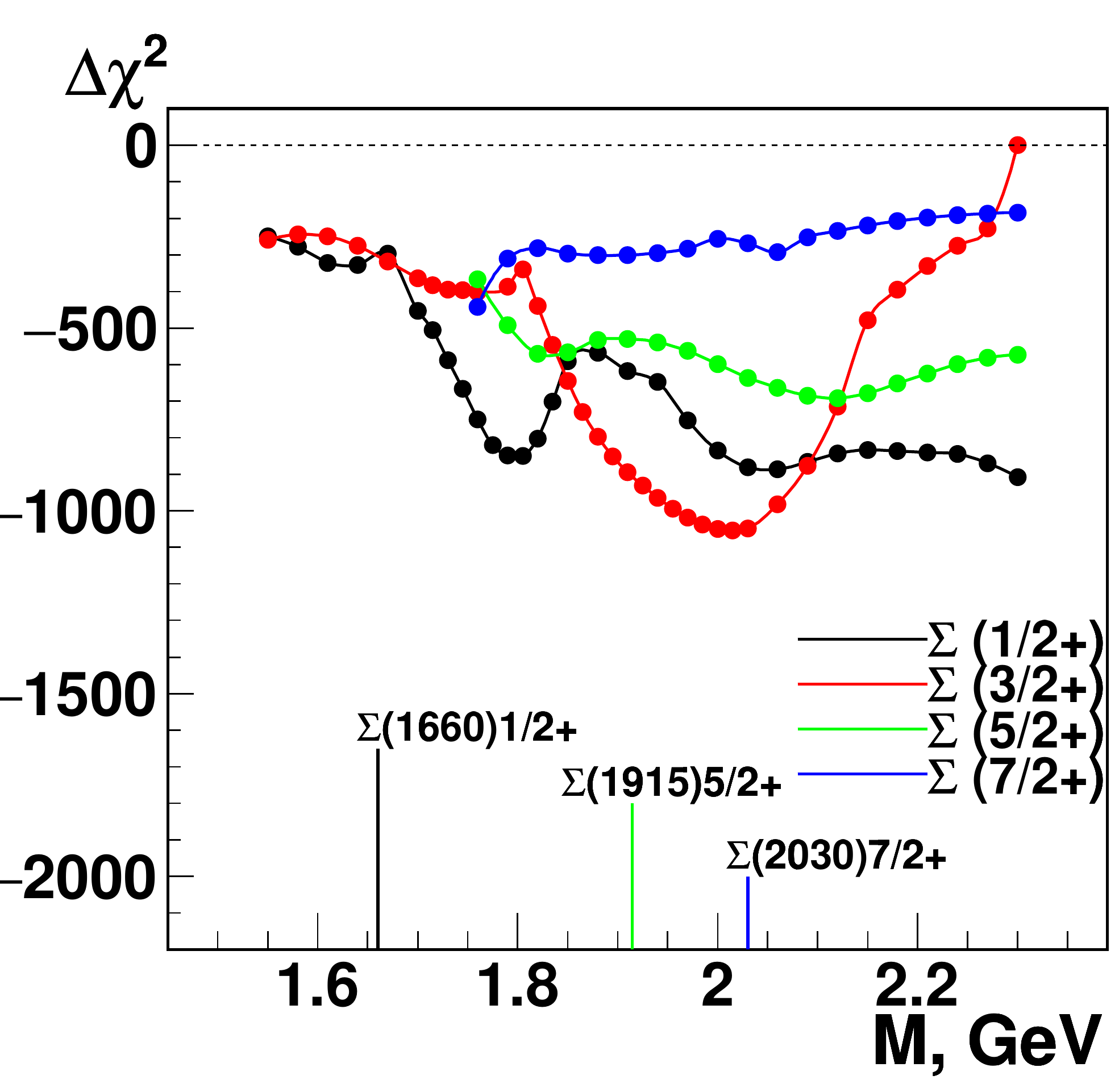}&
\hspace{-3mm}\includegraphics[width=0.240\textwidth]{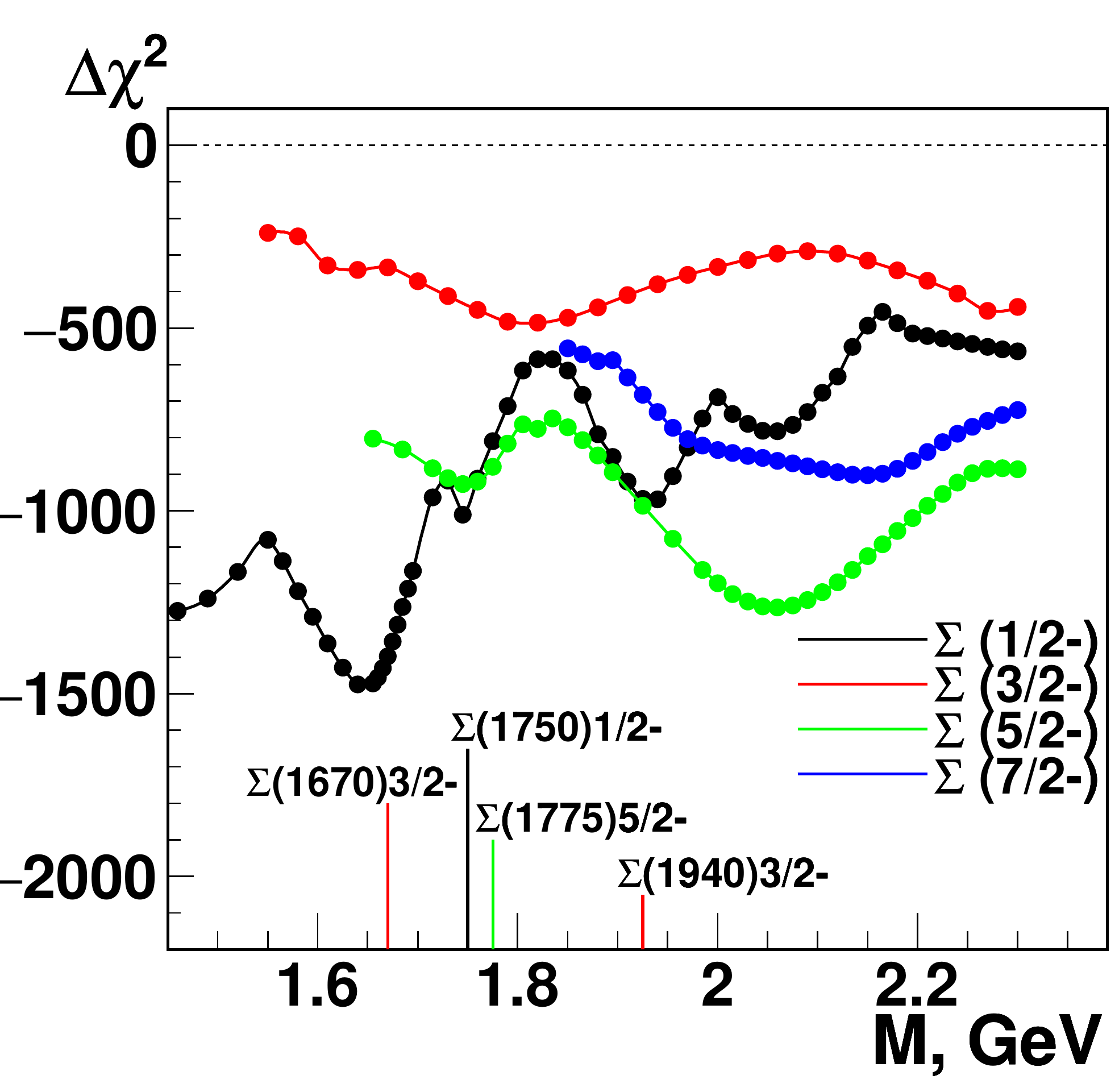}
\end{tabular}
\ec
\caption{\label{fig:explore}The scan of the primary fit for additional
resonances. The three vertical lines indicate the position of resonances which were assumed to
exist. The mass of one additional resonance is stepped through the mass range and the $\chi^2$ of
the fit is monitored. \vspace{-4mm}}
\bc
\begin{tabular}{cccc}
\includegraphics[width=0.24\textwidth]{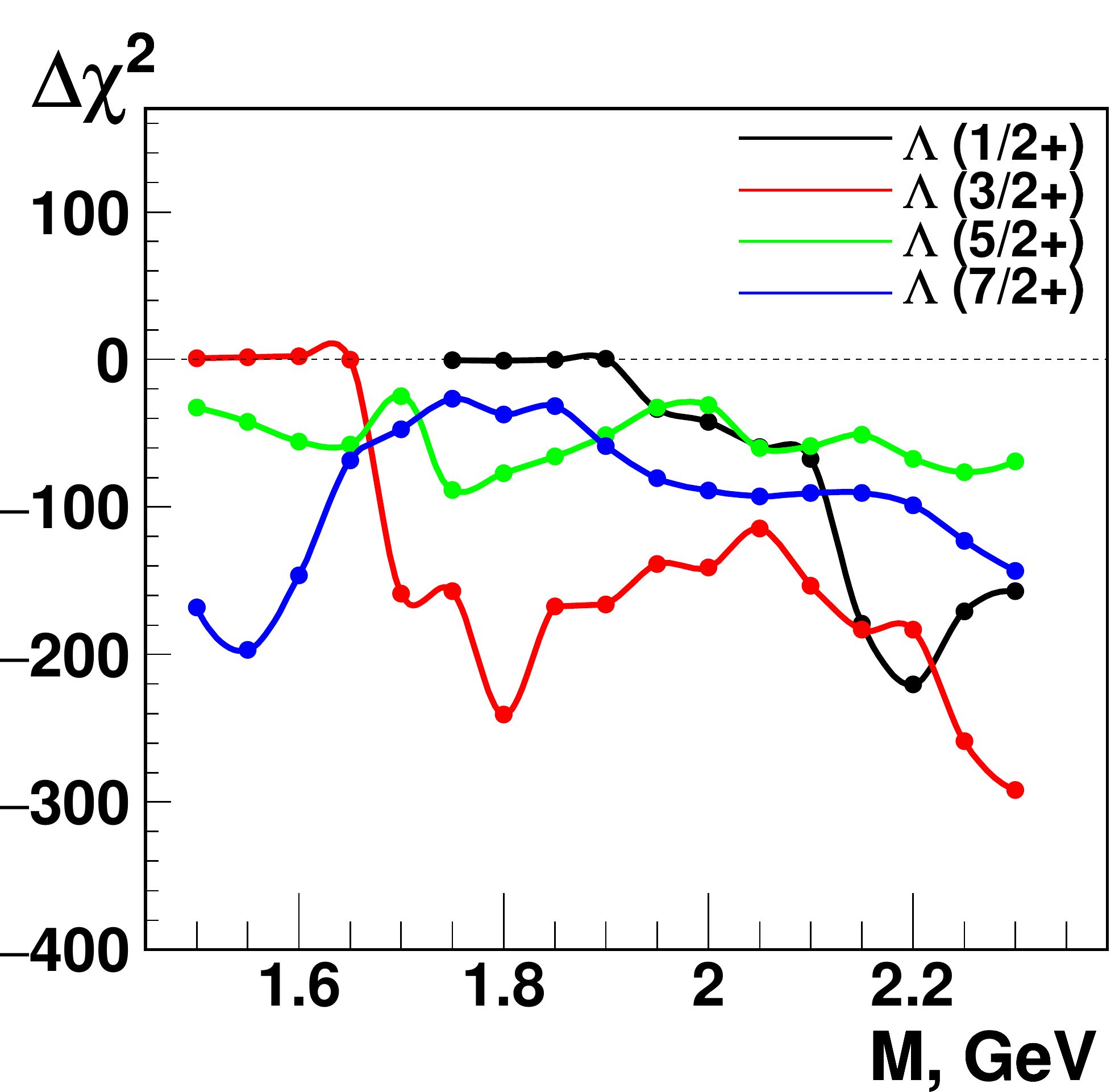}&
\hspace{-3mm}\includegraphics[width=0.24\textwidth]{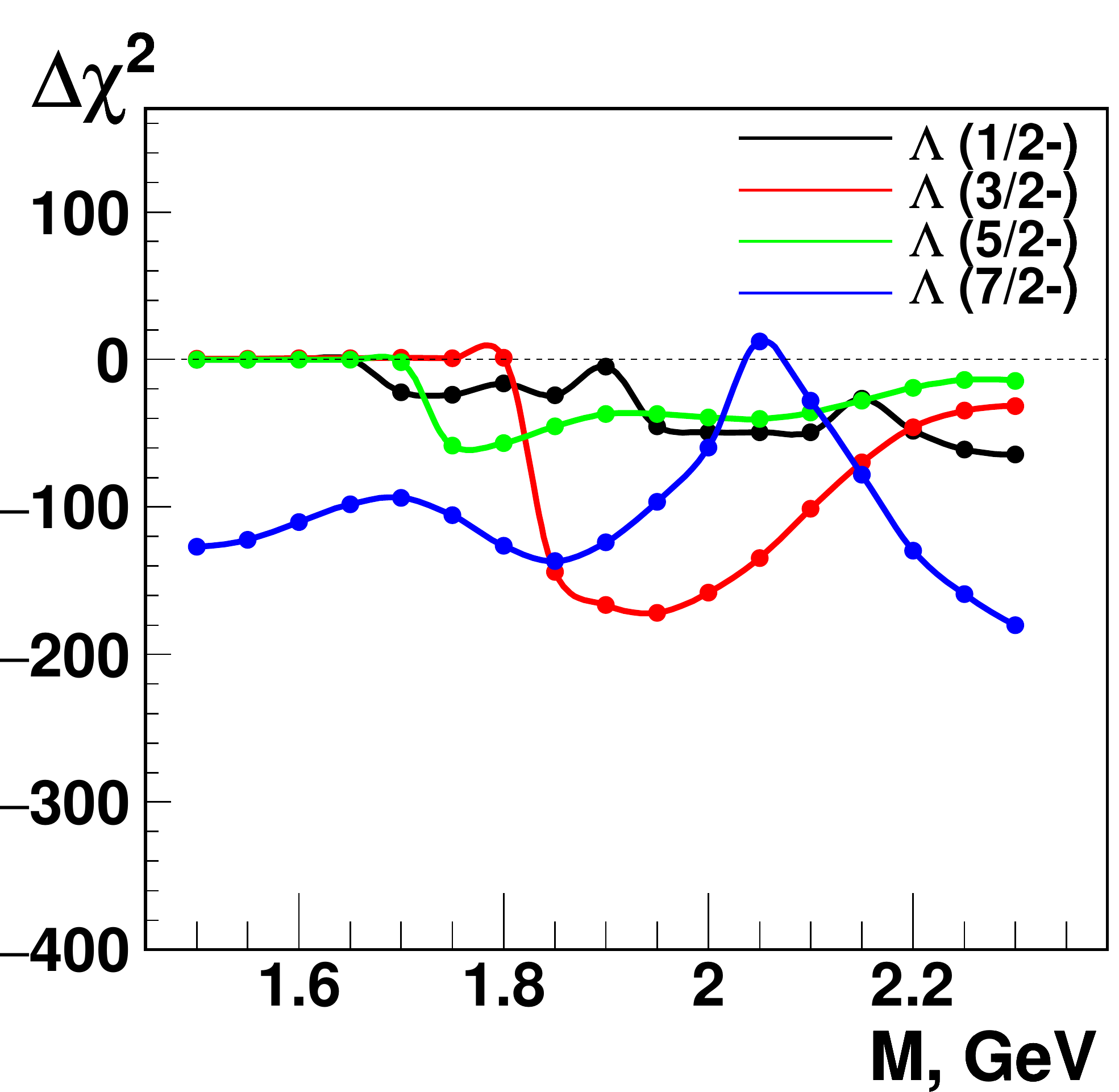}&
\hspace{-3mm}\includegraphics[width=0.24\textwidth]{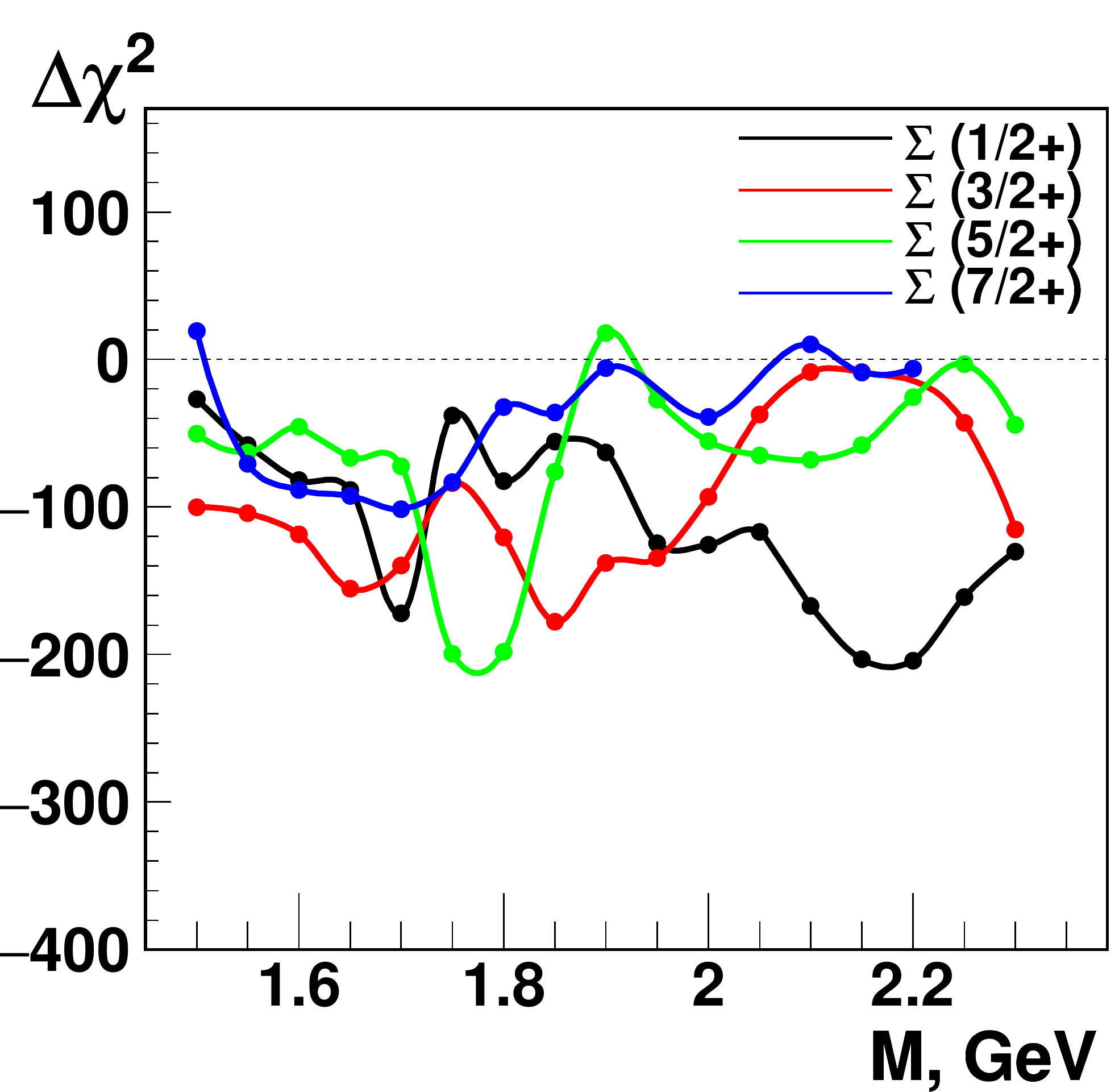}&
\hspace{-3mm}\includegraphics[width=0.24\textwidth]{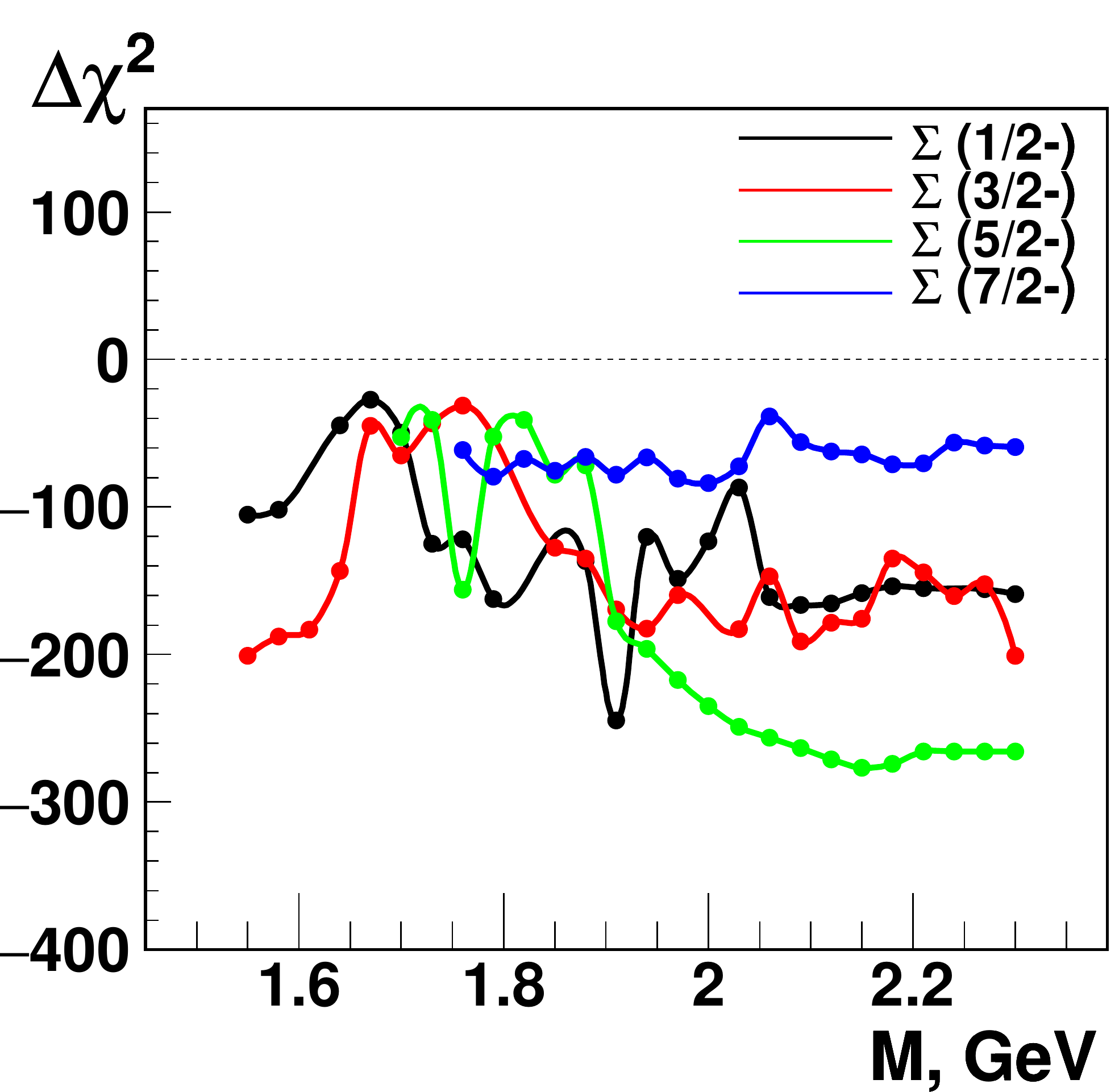}
\end{tabular}
\vspace{-2mm}
\ec
\caption{\label{fig:final_explore}Final scans for additional
resonances. The scale shows the improvement of $\Delta\chi^2$
calculated from the final solution. The final fit uses the formalism described
in section~\ref{pwa-bnga}, the resonances scanned are parametrized as Breit-Wigner resonances.}
\end{figure*}
\begin{figure}[pb]
\bc
\includegraphics[width=0.24\textwidth,height=0.43\textheight]{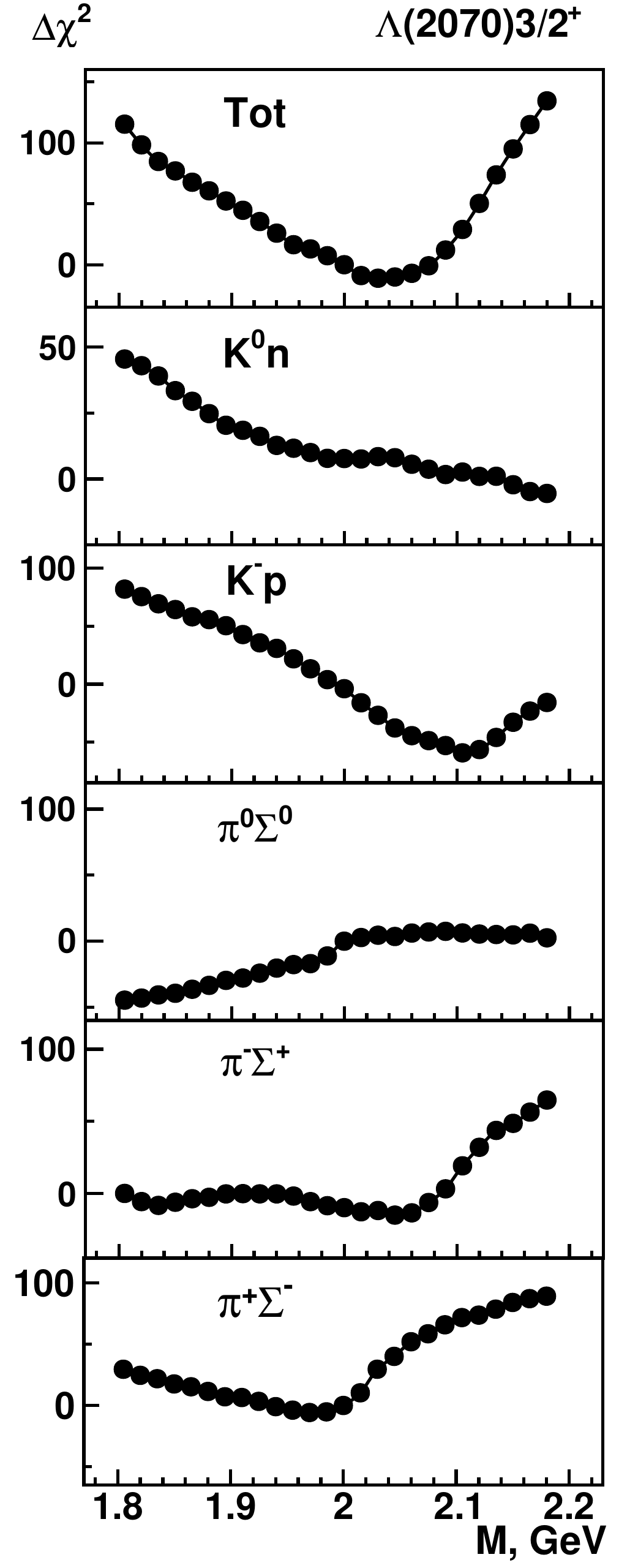}
\includegraphics[width=0.24\textwidth,height=0.43\textheight]{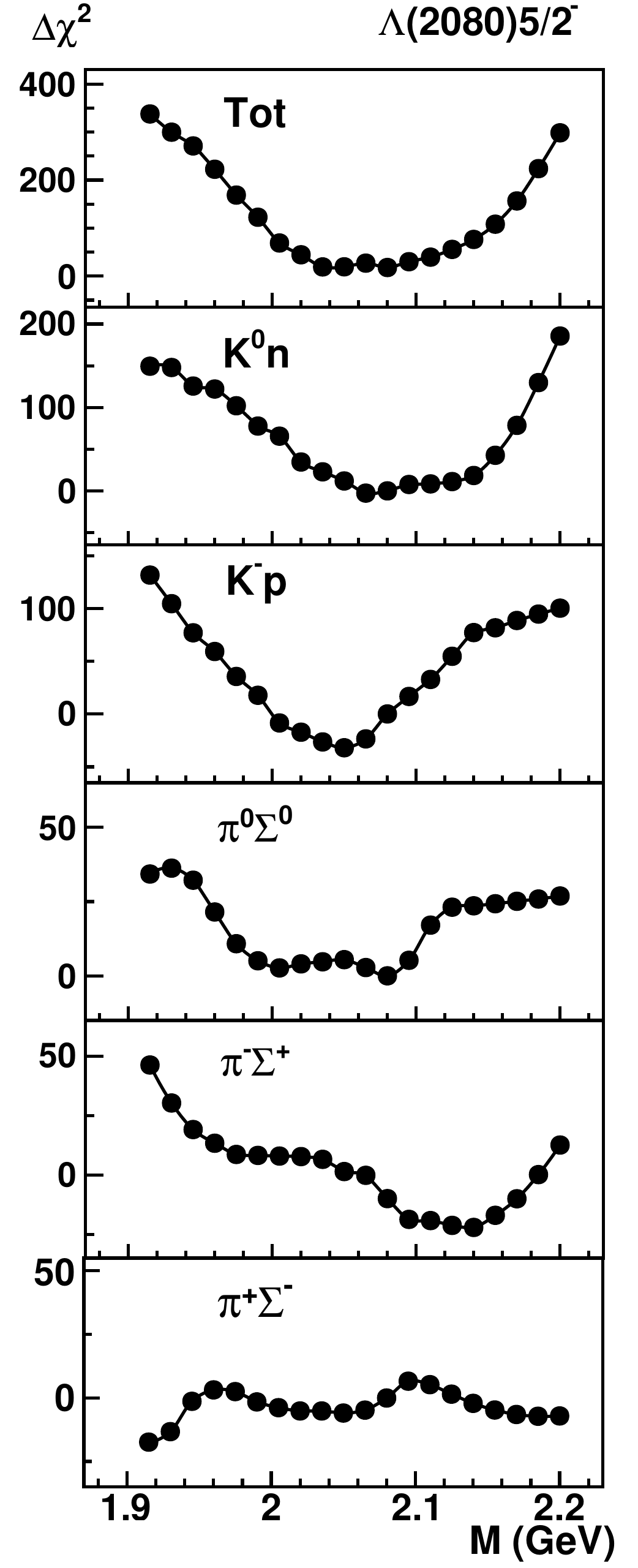}
\ec
\caption{\label{fig:dips}Scan of the $\Lambda(2070)3/2^+$ and
$\Lambda(2080)5/2^-$ states from the solution with all states
identified.}
\end{figure}
\section{\label{scan}Mass scans}

In the second step, we searched for new resonances and performed scans. In these {\it exploratory}
fits, the resonances were described by relativistic Breit-Wigner amplitudes.
States rated with
three and four stars in the RPP were assumed to exist and used in all fits. In the mass scans, the
masses of further states were scanned one by one in the corresponding mass regions.
Figure~\ref{fig:explore} shows the first series of scans. The {\it primary} fit contained five
positive and seven negative-parity $\Lambda$ resonances and four negative and four positive-parity
$\Sigma$ resonances in the $J^P=1/2^\pm, 3/2^\pm$, $5/2^\pm$ and $7/2^\pm$ partial waves. In the scan, we added one
resonance and varied its mass in steps. At each step, a full fit to the data was performed and the
$\chi^2$ of the fit recorded. The $\chi^2$'s as functions of the imposed mass vary from fit to fit;
sometimes a significant improvement is observed. The minimum defines a possible candidate for an
additional resonance.

In the case of scan for an additional positive-parity $\Lambda$ resonance, the deepest
minimum is seen in the scan of the $J^P=1/2^+$-wave. The minimum is rather broad and indicates that
more resonances could be required. In further fits, no significant structure in this partial waves
survives. The minimum in the $3/2^+$ wave is fake: the minimum is entirely due to the data on
$K^-p\to \pi^0\pi^0\Lambda$~\cite{Prakhov:2004ri} at the transition from one momentum to the
next momentum. In the scan for a  $J^P=7/2^-$ $\Lambda$ resonance, a narrow minimum is seen at about 1930\,MeV
which would be a rather low mass for a $7/2^-$-resonance. It is kept for further investigations, and is finally
not confirmed. Furthermore, there are indications for a $J^P=3/2^-$ state at
about 1870\,MeV. In the scans for positive-parity $\Sigma$ resonances, we keep a $1/2^+$ at
1780\,MeV and a $3/2^+$ at 2000\,MeV for further investigations. The scan of the  $J^P=1/2^-$-wave for
$\Sigma$ resonances, a very significant minimum at 1670\,MeV is seen even though a $\Sigma(1670)1/2^-$ resonance
is already included in the {\it primary} fit.  In the $5/2^-$-wave, a second
minimum is seen at 2080\,MeV.  However, this minimum faded away in further
studies.

These preliminary fits demonstrate that significant minima can be seen in fits but that the minima
do not necessarily correspond to true physical states. For the analysis presented here, we
performed several thousand fits with different hypotheses.
At the end, we had a set of resonances which improved the $\chi^2$ by more than 400 units. This is
called our {\it final} fit. These resonances will be discussed below (see
Table~\ref{list-of-baryons}).

A search for a further resonance -- beyond those listed in Table~\ref{list-of-baryons} -- led to
residual fluctuations of less than 300 in $\chi^2$. Figure~\ref{fig:final_explore} shows the scans
for positive and negative-parity $\Lambda$ and $\Sigma$ resonances. Some candidates could exist
which improve $\chi^2$ by about 200 or more: There might be one broad or two unresolved $\Lambda$
candidates with $3/2^-$ in the 1800 to 2100\,MeV range, or a $7/2^-$ and a $3/2^+$ $\Lambda$
candidate above 2300\,MeV. There are indications for a $\Lambda$ $3/2^+$ candidate at 1800\,MeV, a
$\Sigma(2180)1/2^+$ and a $\Sigma$ candidate with $5/2^+$ at 1775\,MeV
and a $5/2^-$ $\Sigma$ candidate above 2100\,MeV.  The $\chi^2$ minimum in
the $\Sigma$ $1/2^-$ scan at 1900\,MeV might indicate that the description of $\Sigma(1900)1/2^-$
is not perfect. We do not consider any of these residual minima as statistically significant.

Figure~\ref{fig:dips} shows the final scans of two new $\Lambda$ states after all resonances have
been identified.
The mass scan of $\Lambda(2080)5/2^-$, e.g., shows a
deeper minimum than the $\Lambda(2070)3/2^+$ state although the drop of the latter state has a more
significant effect for the data description. $\Lambda(2070)3/2^+$ is significantly broader and it
influences the via interference with other waves over a wider range. There is no clear minmum of
$\Lambda(2070)3/2^+$ in any reaction: the global minimum is due to a $\chi^2$ function decreasing
with mass for $K^-p\to \bar KN$ and increasing for $K^-p\to\pi\Sigma$. When this behavior is
observed, we assign at most 1* to the resonance. The 3* and 4* resonance show clear minima for
several final states.

\section{\label{results}The {\it final} fit}


{\small
\begin{table}[pt]
\caption{\label{Osaka}Comparison of the data base and the best $\chi^2$ for the ANL-Osaka fits in
their model A, their model B, and the final BnGa fit. The fits of the Legendre coefficients
extracted from differential cross section and density matrix elements are marked with (LC).}
\renewcommand{\arraystretch}{1.35}
\begin{tabular}{cccccc}
\hline\hline\\[-1ex]
              &\hspace{-2mm}ANL-Os.&BnGa& Model A/B&BnGa \\[0.2ex]
              & \multicolumn{2}{c}{$N_{\rm data}$}&\multicolumn{2}{c}{$\chi^2/N_{\rm data}$}\\[1ex]
\hline
$K^- p \to K^- p$ & \\
$d\sigma/d\Omega$ & 3962 & \gr 5170 & 3.07 / 2.98 & \gr 1.80 \\
$P$               & 510  & \gr 1180 & 2.04 / 2.08 &  \gr 1.41 \\
\hline
$K^- p \to \bar K^0 n$ &\\
$d\sigma/d\Omega$        & 2950 &\gr  3445 & 2.67 / 2.75 & \gr 1.55 \\
$d\sigma/d\Omega$ (LC)    &      &   134 &      /      &  1.86 \\
\hline
$K^- p \to \pi^- \Sigma^+$ &\\
 $d\sigma/d\Omega$ & 1792 &\gr  2455 & 3.37 / 3.49 &  \gr 1.45 \\
$P$               &  418  &\gr  593  & 1.30 / 1.28 &  \rd 2.09 \\
\hline
$K^- p \to \pi^0 \Sigma^0$ &\\
$d\sigma/d\Omega$ & 580 & 691 & 3.68 / 3.50 & \gr  1.96 \\
$P$               & 196 & \rd 124 & 6.39 / 5.80 & \gr 2.41 \\
\hline
$K^- p \to \pi^+ \Sigma^-$ &\\
$d\sigma/d\Omega$ & 1786 & \gr 2082 & 2.56 / 2.18 & \gr 1.59 \\
\hline
$K^- p \to \pi^0 \Lambda$ &\\
 $d\sigma/d\Omega$ & 2178 & \gr 2478 & 2.59 / 3.71 & \gr 1.66 \\
$P$               &  693 &  \gr 892 & 1.41 / 1.73 & \gr1.25 \\
\hline
$K^- p \to \eta \Lambda$ & \\
$d\sigma/d\Omega$ & 160& 160 & 2.69 / 2.03 & \gr 1.50 \\
 $P$               & 18 & --- & 0.94 / 3.83 & --- \\
\hline
$K^- p \to K^0 \Xi^0$ &\\
$d\sigma/d\Omega$ & 33 & \gr  67 & 1.24 / 1.61 & \gr0.89\\
$\sigma_{tot}$           &    & 16 &      /      & 1.00  \\
$P$                      &    & 11 &      /      & 1.70 \\
\hline
$K^- p \to K^+ \Xi^-$ &\\
$d\sigma/d\Omega$        & 92 & 193 & 2.05 / 1.74 & \gr 1.31 \\
$\sigma_{tot}$           &    & 29  &      /      &  1.57 \\
$P$                      &    & 18  &      /      &  0.93 \\
\hline
$K^- p \to \Lambda \omega$ &\\
 $d\sigma/d\Omega$ & -- & 300 & --- / --- & 1.03 \\
$\rho_{00},\rho_{10},\rho_{1-1}$(LC) & -- & 158  & --- / --- & 1.30 \\

\hline\hline\end{tabular}
\end{table}
}

While the primary fit uses a Breit-Wigner description for the contributing resonances,
the final fit uses the multi-channel K-/D-matrix-formalism described in Section~\ref{pwa-bnga}.
Figure~\ref{fig:tot} exhibits the total cross sections and the partial wave contributions
determined in the {\it primary} and the {\it final} fit. The total cross sections are shown twice:
the contributions from $\Lambda$ resonances are shown on the left figures, those from $\Sigma$
resonances on the right figures. Interferences between different partial waves -- which
play an important role in the analysis --  do not contribute to the total cross section.
Of relevance for the total cross section are on the other hand interferences of different isospin
contributions in the same partial wave, which are not shown.
Contributions from $t$- and $u$-channel exchanges are also not shown:
This is the reason why, e.g., the sum of the resonant contributions for $K^-p$ elastic scattering is much
less than the total cross section.
The $\chi^2$ contributions from the individual data sets are listed in Tables~\ref{data-diff1} to
\ref{data-three}. While the $\chi^2$'s for the {\it primary} fit are already acceptable, they are
considerably improved when additional resonances are included. Some data are perfectly described in
the final fit, other data contribute with a large $\chi^2$. One has, however, to have in mind that
the data often are not fully consistent. Thus a $\chi^2$ of one per degree of freedom cannot be
expected.

In Table~\ref{Osaka} we compare our $\chi^2$'s with those obtained by the ANL-Osaka group. We use
the same data but in several cases, our data set is slightly extended. Yet, for the polarization
observable in $K^-p\to \pi^0\Sigma^0$, we use only one of two existing data sets
\cite{Prakhov:2008dc,Manweiler:2008zz}. When both data sets were excluded from the analysis, the
predicted polarization was close to the values from \cite{Prakhov:2008dc} and disagreed with
\cite{Manweiler:2008zz}. Hence we decided not to use the latter data.

Due to a larger number of resonances, the BnGa fit achieves a better $\chi^2$ even though more data
are used. The Kent \cite{Zhang:2013sva} and Carnegie-Mellon \cite{Fernandez-Ramirez:2015tfa} groups
do not report the $\chi^2$'s achieved in their fits.


\begin{table*}[pt]
\caption{\label{list-of-baryons}Resonances found to contribute to $K^-$ induced reactions. The
columns give the Breit-Wigner mass and width, and the increase of $\chi^2$ when a hyperon is
removed from the fit, and our evaluation of the star rating. The small numbers give the PDG
entries. Some 4* resonances cannot be removed from the fit without destroying the fit;
$\delta\chi^2>$10\,000 is given for these resonances. A statistically significant very broad
enhancements in the $J^P$\,=\,$1/2^-$ wave are likely due to some background and not considered as
true resonances.  $^1$:  no estimate given in the RPP; our own estimate. }
\renewcommand{\arraystretch}{1.4}
\bc
\begin{tabular}{cc}
{\footnotesize
\begin{tabular}{lcccc}
\hline\hline\\[-2.0ex]
   &Mass&Width&  $\Delta\chi^2$ & Status\\[1.0ex]
\hline\hline\\[-2.0ex]
$\Lambda(1405)1/2^-$ & 1420\er 3 &  46\er 4  & $4070$  & ****  \\[-1ex]
                                  &\tiny 1405.1$^{+1.3}_{-1.0}$  &\tiny 50.5\er2.0  &  & \tiny  ****\\ \hline
$\Lambda(1670)1/2^-$ & 1677\er 2 &  33\er 4  & $3610$  & ****  \\[-1ex]
                                  &\tiny 1660 to 1680            & \tiny 25 to 50   &  & \tiny  ****\\ \hline
$\Lambda(1800)1/2^-$ & 1811\er10 & 209\er 18 & $ 1896$  & ***  \\[-1ex]
                                  &\tiny 1720 to 1850            & \tiny 200 to 400 &  & \tiny  ***\\ \hline
$\Lambda(1520)3/2^-$ & 1518.5\er 0.5 &  15.7\er 1.0  & $>$10\,000  & ****  \\[-1ex]  
                                  &\tiny 1519.5\er 1.0&\tiny 15.6\er 1.0            &  & \tiny  ****\\ \hline
$\Lambda(1690)3/2^-$ & 1689\er 3 &  75\er 5  & $>$10\,000 & ****  \\[-1ex] 
                                  &\tiny 1685 to 1695&\tiny 50 to 70                &  & \tiny  ****\\ \hline
$\Lambda(1830)5/2^-$ & 1821\er 3 &  64\er 7  & $ 1790$  & ***  \\[-1ex]
                                  &\tiny 1810 to 1830&\tiny 60  to 110              &  &\tiny  ****\\ \hline
$\Lambda(2080)5/2^-$ & 2082\er 13 & 181\er 29 & $ 770$  &  *  \\[-1ex]
                                  &\tiny -           &\tiny -                       &  & new \\ \hline
$\Lambda(2100)7/2^-$ & 2090\er 15 & 290\er 30 & $5412$  & ****  \\[-1ex]
                                  &\tiny 2090 to 2110&\tiny 100 to 250              &  &\tiny  ****\\ \hline
\hline\\[-2.0ex]
$\Lambda(1600)1/2^+$ & 1605\er 8  & 245\er 15 & $>$10\,000  & ****  \\[-1ex] 
                                  &\tiny 1560 to 1700&\tiny 50 to 250               &  &\tiny  ***\\ \hline
$\Lambda(1890)3/2^+$ & 1873\er 5  & 103\er 10  & $4480$  & ****  \\[-1ex]
                                  &\tiny 1850 to 1910&\tiny 60 to 200               &  &\tiny  ****\\ \hline
$\Lambda(2070)3/2^+$ & 2070\er 24 & 370\er 50   & $ 1144$  & *     \\[-1ex]
                                  &\tiny -           &\tiny -                       &  & new\\ \hline
$\Lambda(1820)5/2^+$ & 1822\er 4 & 80\er 8    & $>$10\,000  & ****  \\[-1ex] 
                                  &\tiny 1815 to 1825&\tiny 70 to 90                &  &\tiny  ****\\ \hline
$\Lambda(2110)5/2^+$ & 2086\er 12 & 274\er 25 & $ 1418 $  & **   \\[-1ex]
                                  &\tiny 2090 to 2140&\tiny 150 to 250              &  &\tiny  ***\\
\hline\hline
\end{tabular}
}
&
{\footnotesize
\begin{tabular}{lcccc}
\hline\hline\\[-2.0ex]
   &Mass&Width&  $\Delta\chi^2$ & Status\\[1.0ex]
\hline\hline\\[-2.0ex]
$\Sigma(1620)1/2^-$ & 1681\er 6  &  40\er 12  & $  386$  & (*)     \\[-1ex]
                                  &\tiny $\approx$1620&\tiny 10 to 400       &   &\tiny  *\\ \hline
$\Sigma(1750)1/2^-$ & 1692\er 11 & 208\er 18 & $  3032$   & **** \\[-1ex]
                                  &\tiny 1730 to 1800&\tiny 60 to 160        &   &\tiny  ***\\ \hline
$\Sigma(1900)1/2^-$ & 1938\er 12 & 155\er 30 & $  1500$  & **   \\[-1ex]
                                  &\tiny 1900\er 21&\tiny 191\er47           &   &\tiny  *\\ \hline
$\Sigma(2160)1/2^-$ & 2165\er 23 & 320$^{+300}_{-60}$ & 1612 & *\\[-1ex]
                    & &  &   & new\\ \hline
$\Sigma(1670)3/2^-$ & 1665\er 3  &  54\er 6 & $ 5894$  & ****  \\[-1ex]
                                  &\tiny 1665 to 1685&\tiny 40 to 80         &   &\tiny  ****\\ \hline
$\Sigma(1940)3/2^-$ & 1878\er 12 & 224\er 25 & $  1708$  &  ***  \\[-1ex]
                                 &\tiny 1900 to 1950&\tiny 150 to 300   &                &\tiny  ***\\ \hline
 $\Sigma(2000)3/2^-$ & 2005\er 14 & 178\er 23 & $  446$    & *   \\[-1ex]
                                  &-&-& & new\\ \hline
$\Sigma(1775)5/2^-$ & 1776\er 4 & 124\er 8 & $>$10\,000  & ****  \\[-1ex] 
                                  &\tiny 1770 to 1780&\tiny 105 to 135              &                &\tiny  ****\\ \hline
$\Sigma(2100)7/2^-$ & 2146\er 17 & 260\er 40  & 666  & *     \\[-1ex]
                                  &\tiny $\approx$2100&\tiny 50 to 150$^1$              &                &\tiny  *\\ \hline
\hline\\[-2.0ex]
$\Sigma(1660)1/2^+$ & 1665\er 20 & 300$^{+140}_{-40}$ & $ 1870$  & ***  \\[-1ex]
                                  &\tiny 1630 to 1690&\tiny 40 to 200              &                &\tiny  ***\\ \hline
$\Sigma(2230)3/2^+$ & 2240\er 27 & 345\er 50 & $1200$  &  *     \\[-1ex]
                                  &&     &                &new\\ \hline
$\Sigma(1915)5/2^+$ & 1918\er 6 & 102\er 12 & $ 2002$  & ****   \\[-1ex]
                                  &\tiny 1900 to 1935&\tiny 80 to 160              &                &\tiny  ****\\ \hline
$\Sigma(2030)7/2^+$ & 2032\er 6 & 177\er 12 & $ 2856$  & ****  \\[-1ex]
                                  &\tiny 2025 to 2040&\tiny 150 to 200              &                &\tiny  ****\\
\hline\hline
\end{tabular}
}
\end{tabular}
\ec
\end{table*}

Finally, we give in Table~\ref{list-of-baryons} a list of the hyperon resonances which we use in the
fits. One $\Lambda$ resonance is not listed. We find a very broad $\Lambda$ resonance
with $J^P=1/2^-$ at 2230\,MeV
with a width of $\approx 450$\,MeV. It is statistically significant; according to the criteria
described in Section~\ref{star}, it would be listed with three stars. Due to its large width, we
think the resonance may hide a number of resonances with different spin and parities. Hence we do
not include it in Table~\ref{list-of-baryons}. A similar resonance shows up in the $\Sigma$ sector.
The $\Sigma(2160)1/2^-$ may have a width of up to 600\,MeV and may
also cover the contribution of several resonances. It could, however, also have a width of
260\,MeV; that is the reason that we include this resonance in the listings. But due to this
uncertainty, we assign only one star to it even this its statistical evidence would justify three
stars.

There are several one-star resonances for which we see no evidence underlying their weakness in kaon-induced
reactions. Most four and three-star resonances are confirmed.

Compared to our expectation of seven $\Lambda^*$ and seven $\Sigma^*$ resonances with negative
parity and below $\approx1900$\,MeV, we were missing one resonance in both sectors. The expected
$3/2^-$ companion of $\Lambda(1800)1/2^-$ and $\Lambda(1830)5/2^-$ is definitely not seen, see the
discussion in Section~\ref{scan}.  In the $\Sigma^*$ sector, a low-mass partner of
$\Sigma(1670)3/2^-$ with $J^P=1/2^-$ is missing. There is a 1* candidate $\Sigma(1620)1/2^-$. If we
include this in the fit, the $\chi^2$ gain is 386, just below our limit at 400 (see below). Thus we
keep it as 1* resonance. Its mass of 1681\,MeV is unexpectedly close to $\Sigma(1750)1/2^-$ which
we find at 1692\,MeV. We notice that the K-matrix poles of the three low-mass $\Sigma^*$ resonances
with $J^P=1/2^-$ are 1610, 1695, and 1900\,MeV. In the nucleon sector, it had turned out that the
helicity amplitudes for photoexcitation of $N(1535)1/2^-$ and $N(1650)1/2^-$ off protons and
neutrons determined at the K-matrix pole~\cite{Anisovich:2015tla} agree very well with expectations
of the Single-Quark-Transition-Model~\cite{Burkert:2002zz} and are at variance with SU(3) relations
at the T-matrix poles~\cite{Boika:2014aha}.

\section{\label{star}Hyperon resonances and their star rating}

The final fit converged with a $\chi^2$ minimum at $\chi^2=40615$. Table~\ref{list-of-baryons}
lists the resonances used in the fit, their masses and widths. From the full list of resonances
used to fit the data we removed individual resonances one by one. The new fit with readjusted
parameters deteriorated, the increase in $\chi^2$ is used to estimate the significance of the
resonance. When some dominant 4* resonances were removed, the fit became very bad. In these cases
we did not try to improve the fit but just left these resonances with their 4* rating.

In the RPP - not counting $\Lambda$(1116), $\Sigma$(1193), and  $\Sigma$(1385),
there are at present 12 $\Lambda$ and $\Sigma$ resonances with 4*'s. We use these
resonances to define criteria to estimate the star rating of resonances. In our analysis, the least
significant 4* resonances are $\Lambda(1830)5/2^-$ and $\Sigma(1915)5/2^+$; when removed they led to an
increase in $\chi^2$ of 1790 or 2002, respectively. Since we do not wish to drastically alter the
criteria for the star rating, we assign a 4* rating to resonances for which a $\chi^2$ change of
more than 2000 is observed. Thus 11 of the 12 4*-star resonances kept their star rating.
We defined 400 as the minimum $\chi^2$ change to accept a resonance with 1*.
The ratings are thus defined by
\bea
 1*:  400<\delta\chi^2<1000; &\qquad 2*:&  1000<\delta\chi^2<1500;\nonumber\\
 3*:  1500<\delta\chi^2<2000; &\qquad 4*:&  2000<\delta\chi^2.
\eea

\begin{figure*}
\bc
\begin{tabular}{ccc}
\hspace{-2mm}\includegraphics[width=0.5\textwidth]{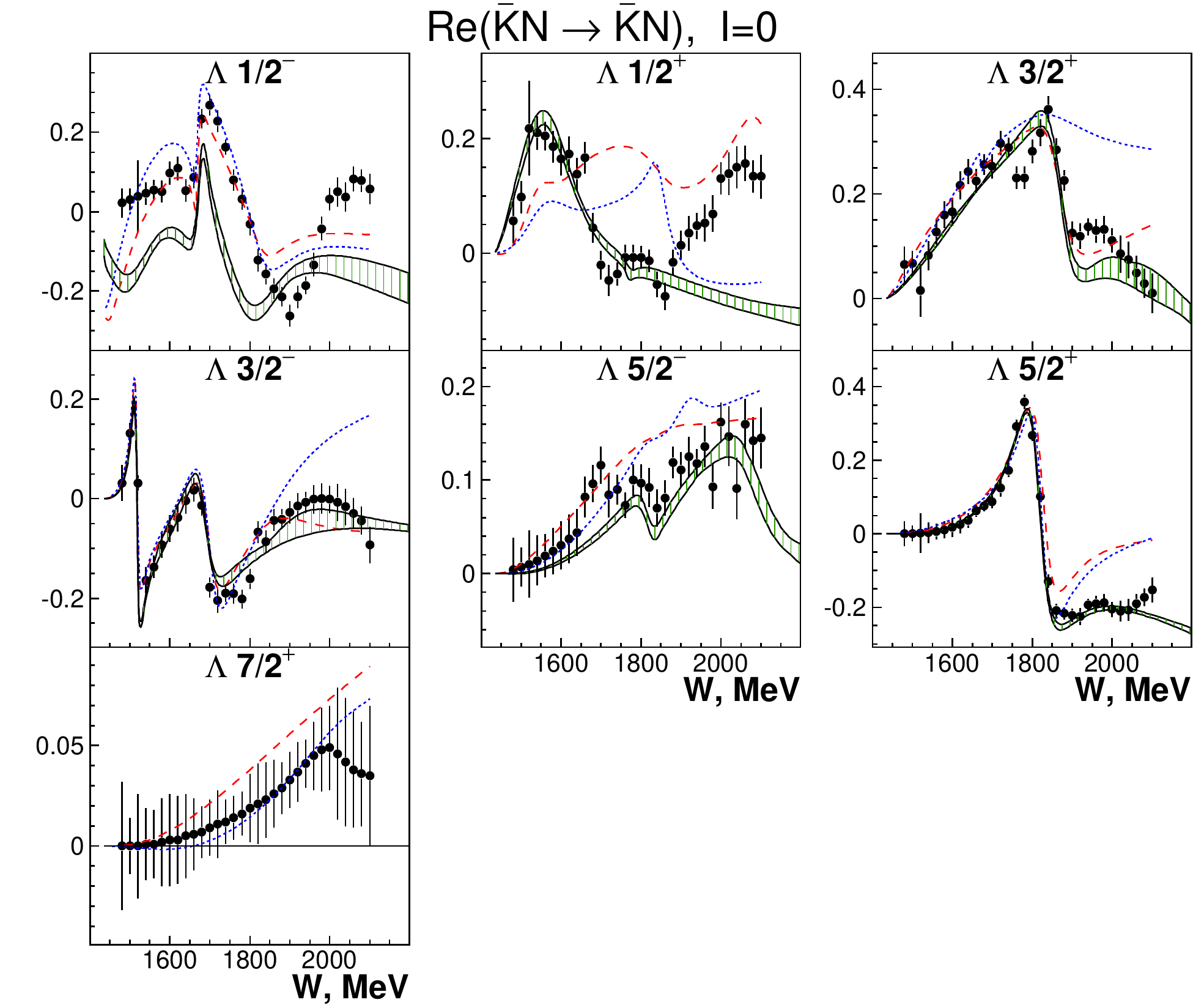}&
\hspace{-4mm}\includegraphics[width=0.5\textwidth]{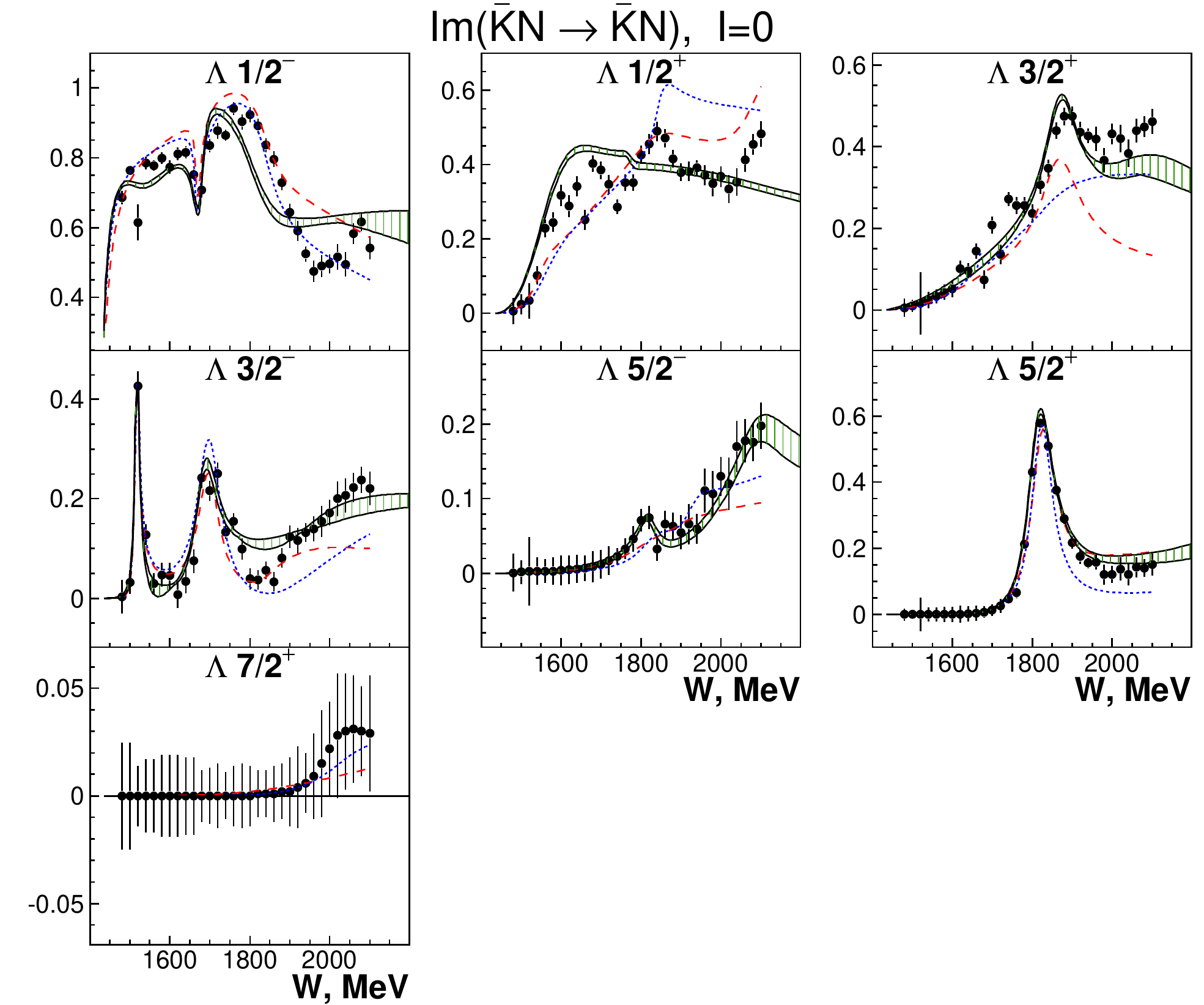}\\
\hspace{-2mm}\includegraphics[width=0.5\textwidth]{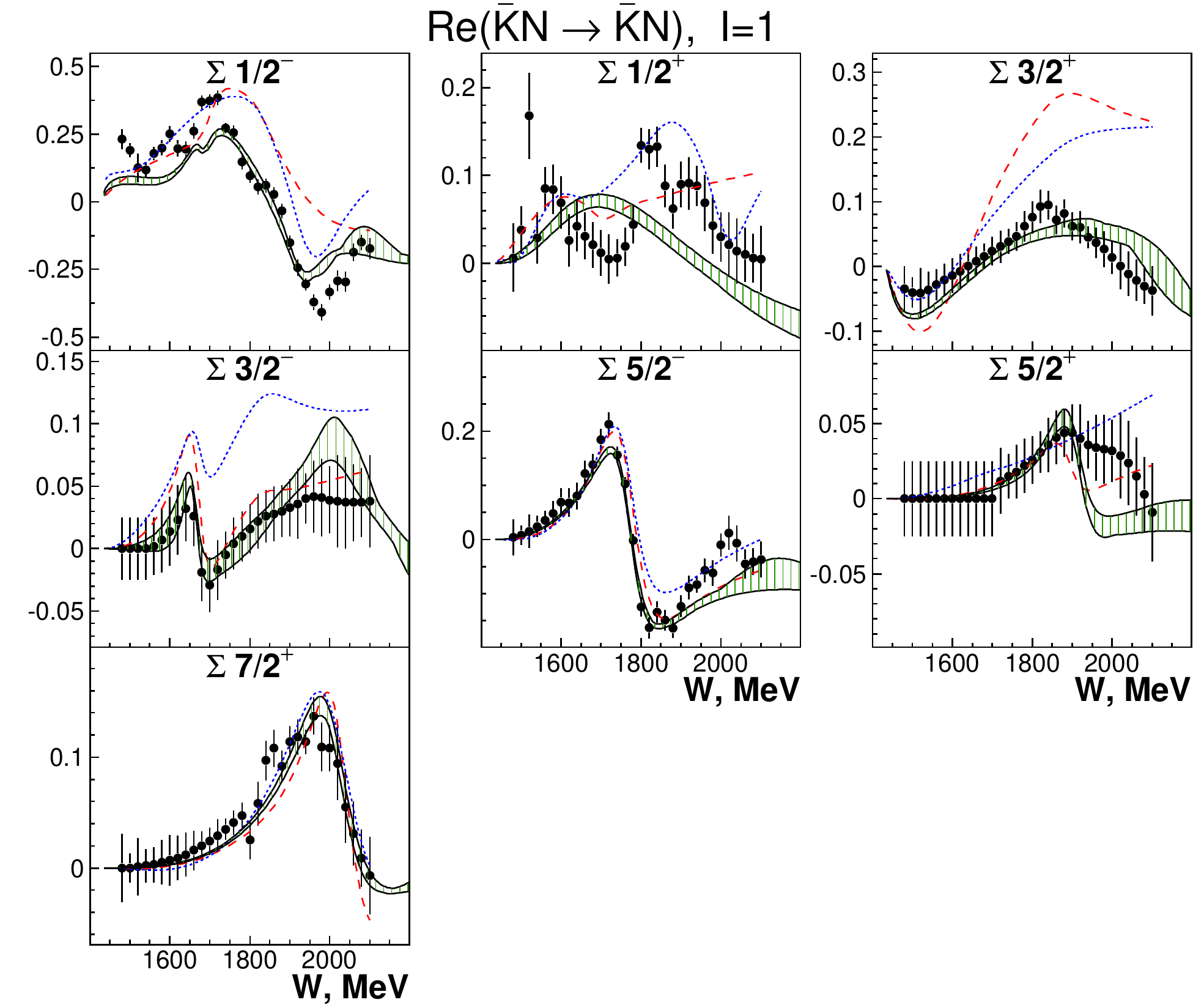}&
\hspace{-4mm}\includegraphics[width=0.5\textwidth]{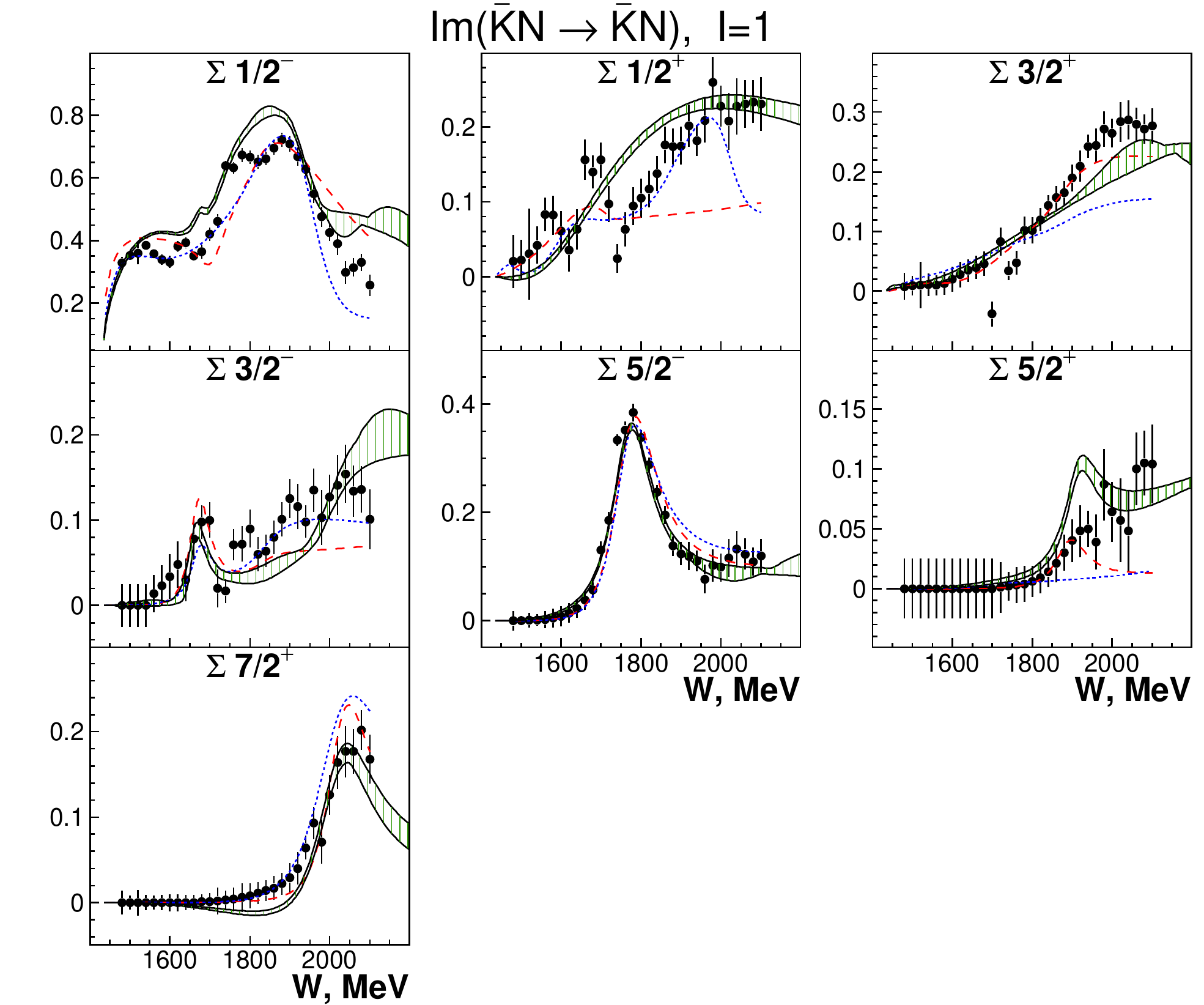}
\end{tabular}\vspace{-4mm}
 \ec
\caption{\label{fig:waves-1}Real and imaginary parts of $K^-p$ induced scattering amplitudes. The
solid points with error bars are the energy independent amplitudes derived in~\cite{Zhang:2013cua}.
The larger part of the data used in~\cite{Zhang:2013cua} were fitted by the authors of
Ref.~\cite{Kamano:2014zba}. Their amplitudes for solution A and B are given as short-dashed blue
and long-dashed red curves. The green-shaded area represents the spread of results from our main
solutions and from solution  weak resonances turned off.}
\end{figure*}

\begin{figure*}
\bc
\begin{tabular}{ccc}
\hspace{-2mm}\includegraphics[width=0.5\textwidth]{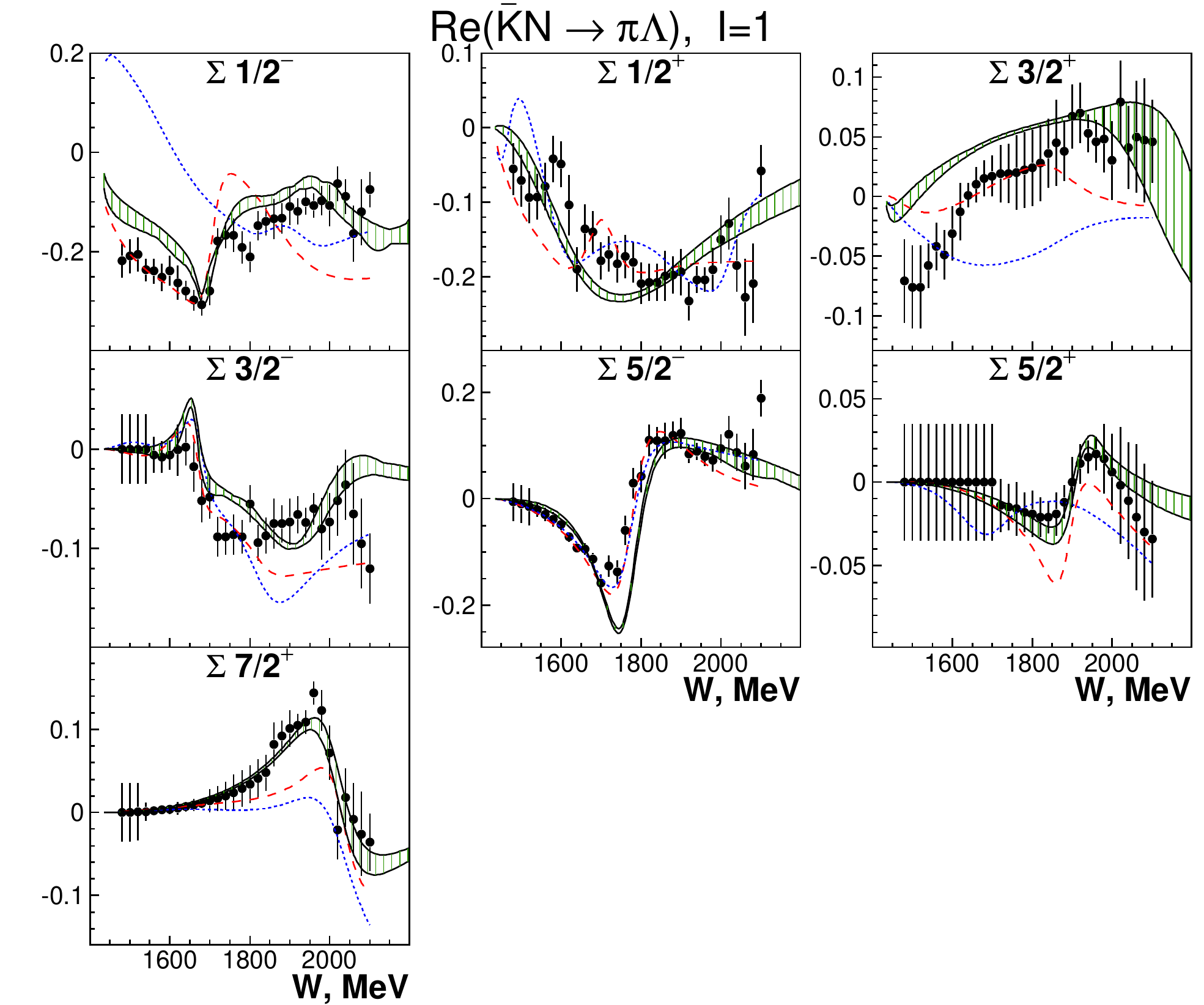}&
\hspace{-4mm}\includegraphics[width=0.5\textwidth]{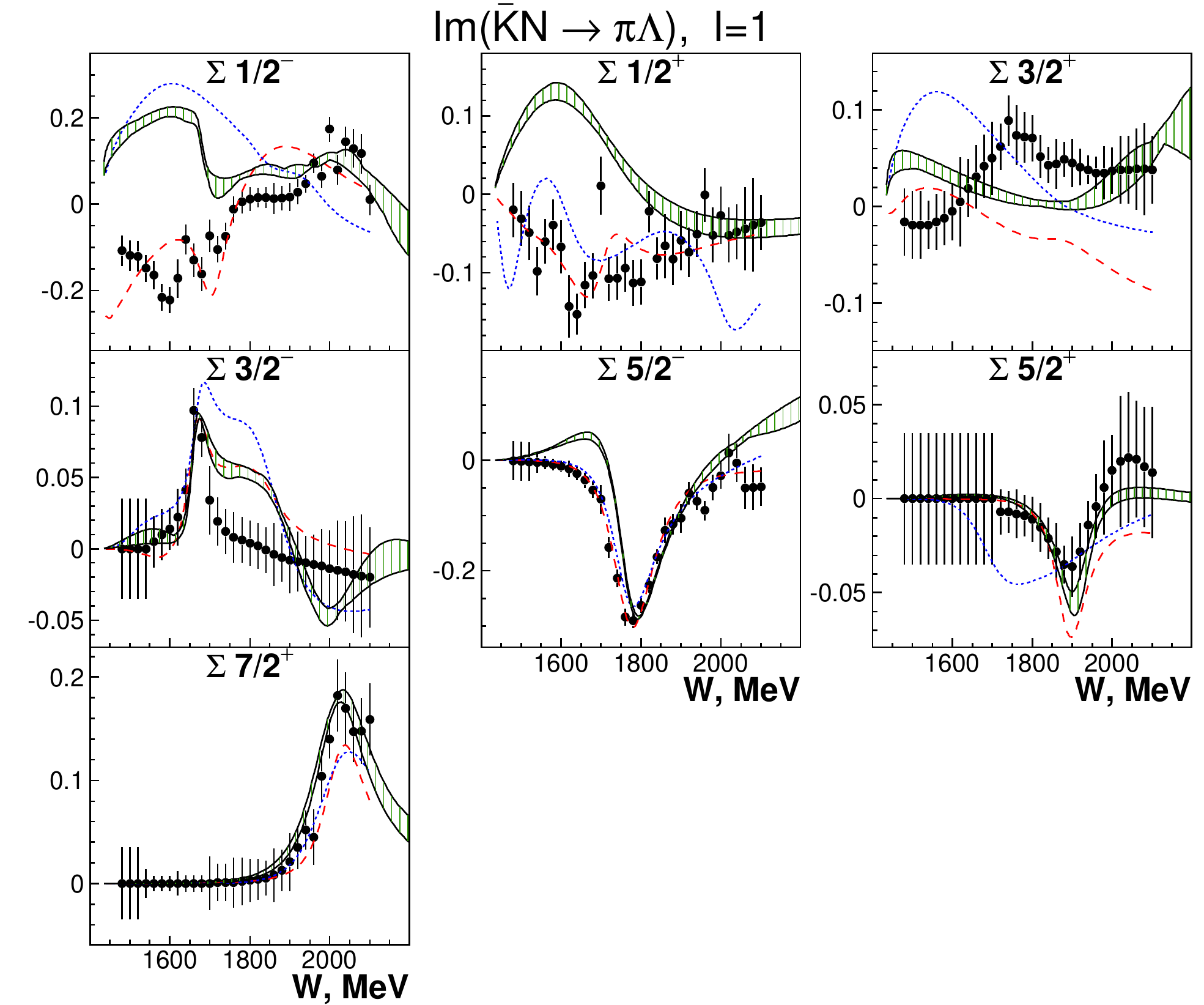}\\
\hspace{-2mm}\includegraphics[width=0.5\textwidth]{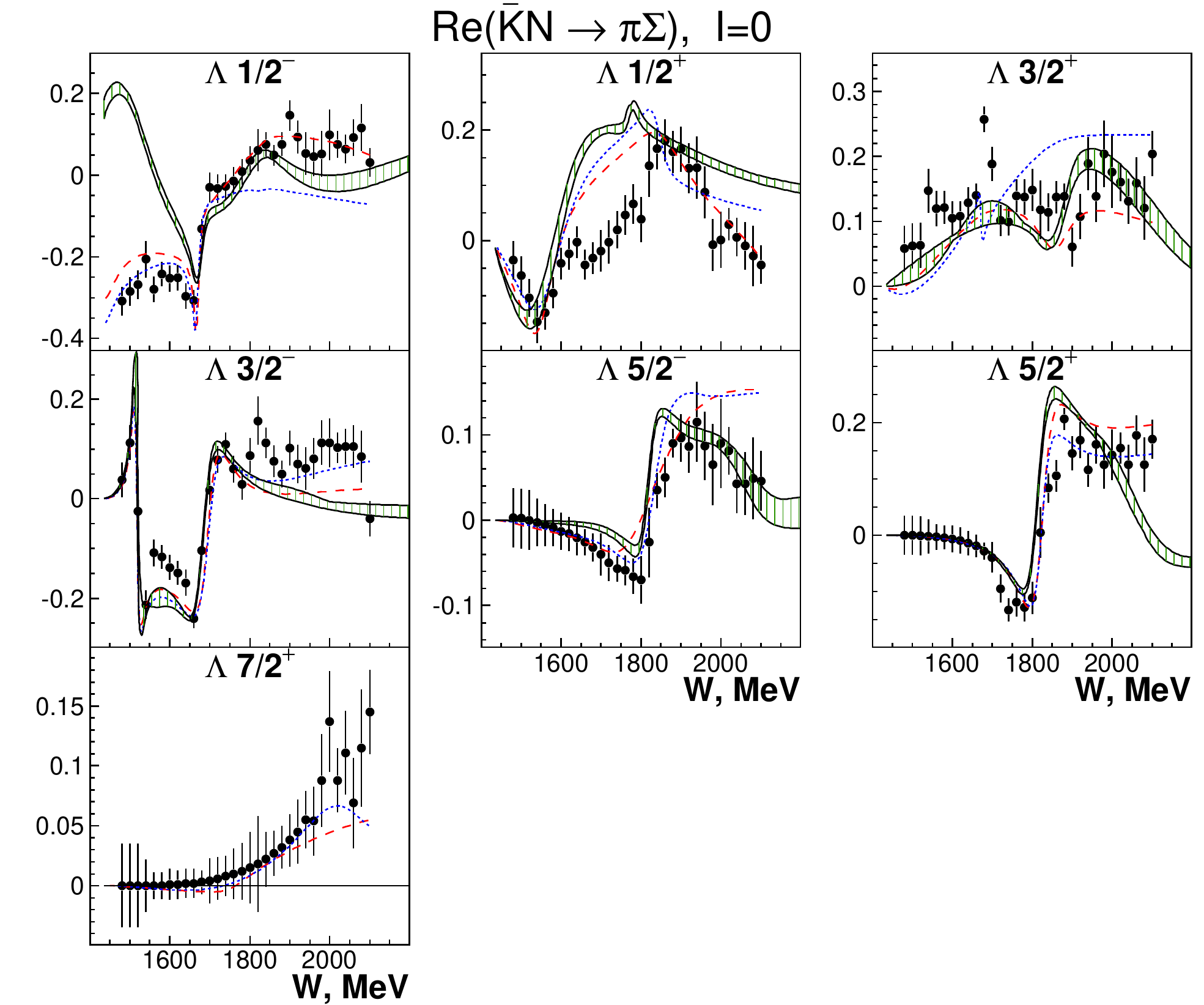}&
\hspace{-4mm}\includegraphics[width=0.5\textwidth]{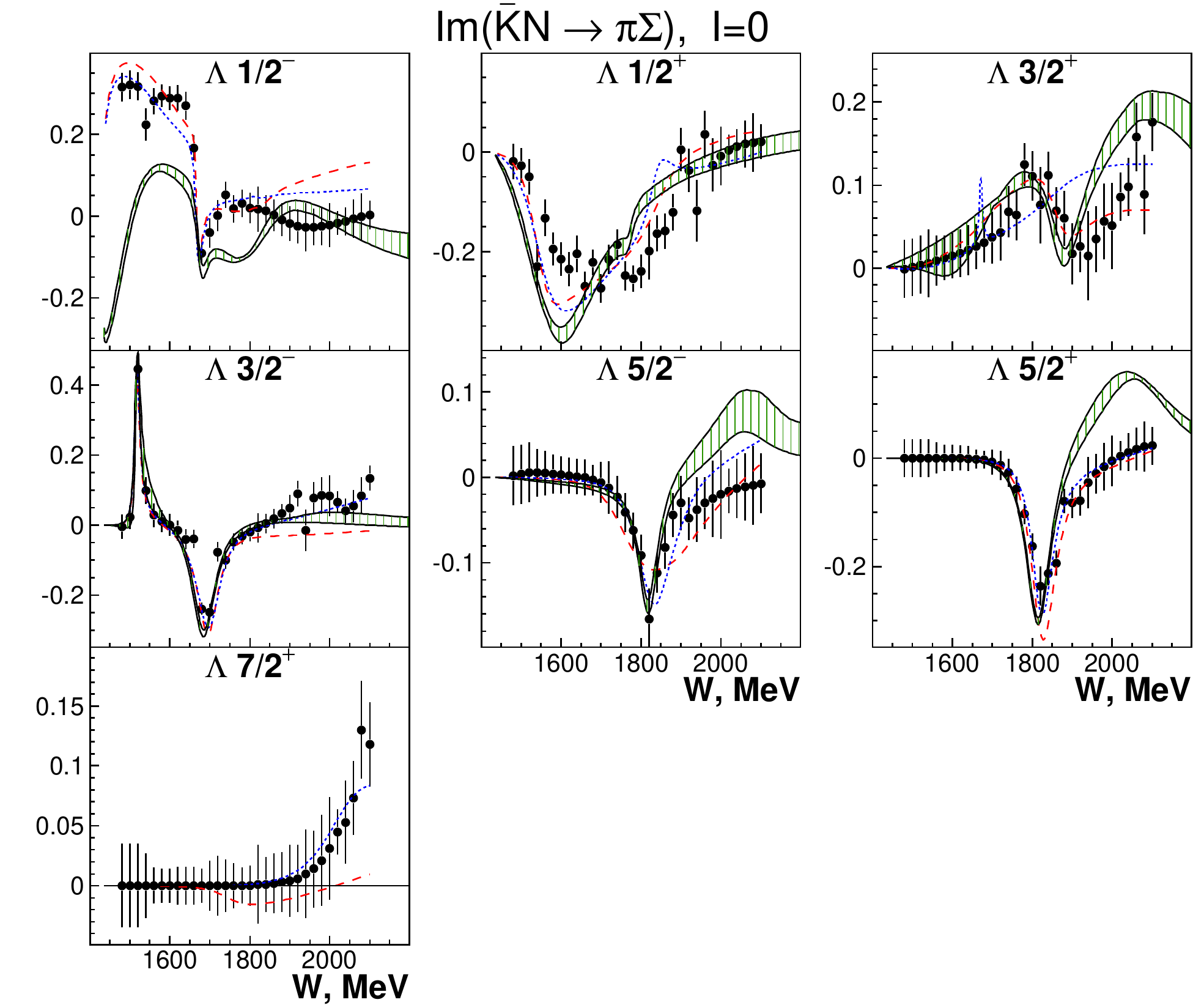}\\
\hspace{-2mm}\includegraphics[width=0.5\textwidth]{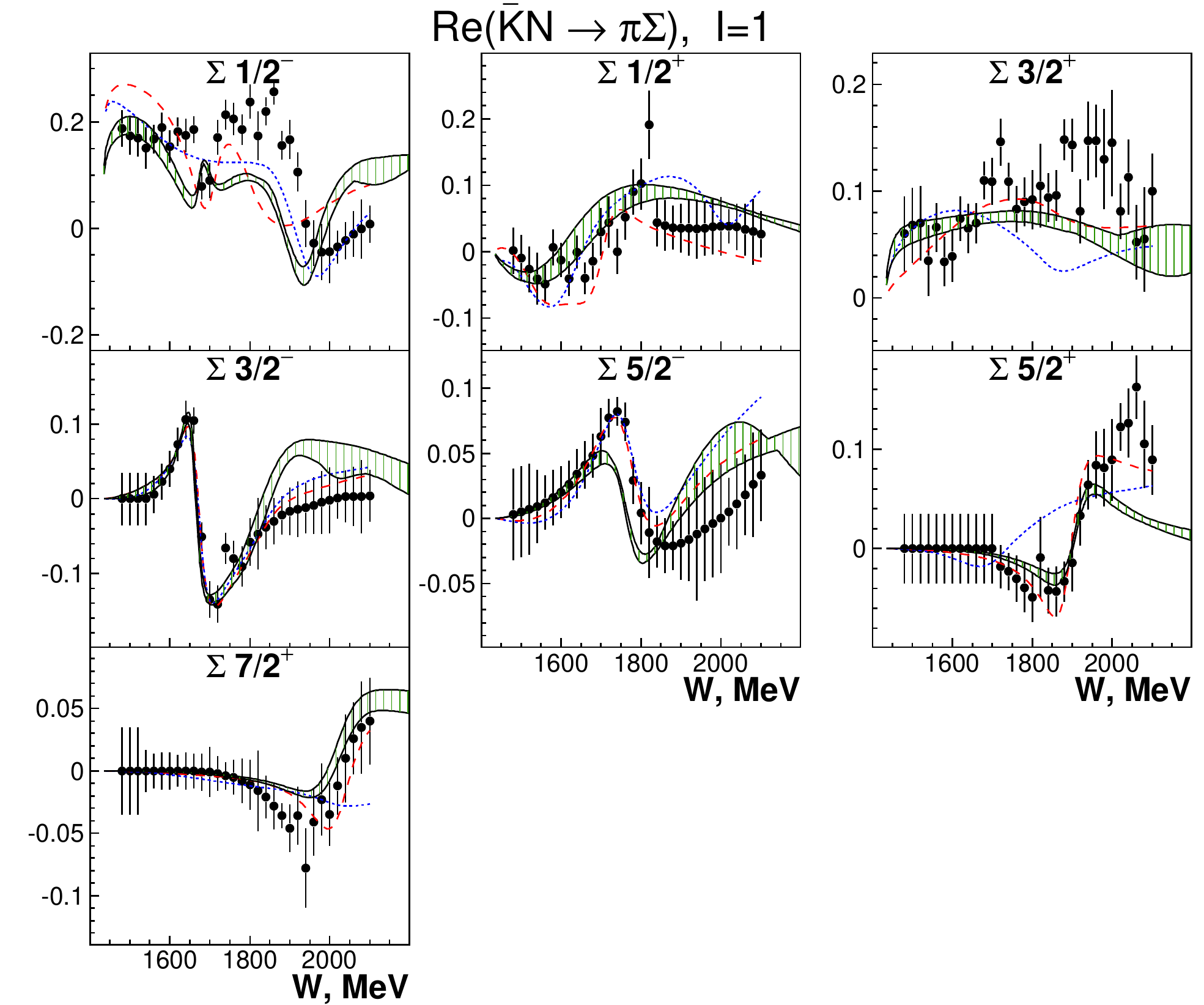}&
\hspace{-4mm}\includegraphics[width=0.5\textwidth]{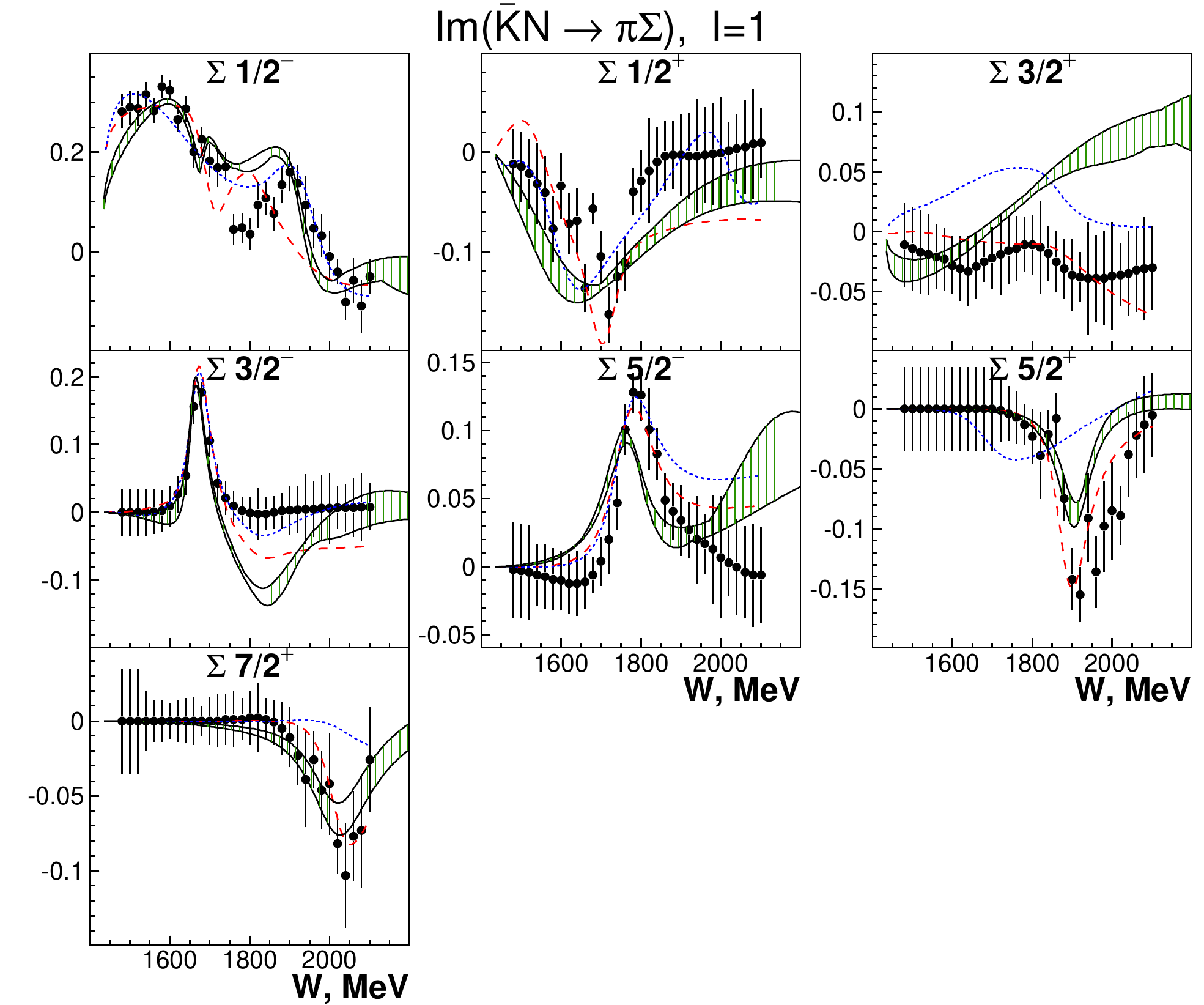}
\end{tabular}\vspace{-4mm}
 \ec
\caption{\label{fig:waves-2}Real and imaginary parts of $K^-p$ induced scattering amplitudes. The
solid points with error bars are the energy independent amplitudes derived in~\cite{Zhang:2013cua}.
The larger part of the data used in~\cite{Zhang:2013cua} were fitted by the authors of
Ref.~\cite{Kamano:2014zba}. Their amplitudes for solution A and B are given as short-dashed blue
and long-dashed red curves. The green-shaded area represents the spread of results from our main
solutions and from solution with weak resonances turned off.}
\end{figure*}

We do not observe any evidence for the low-mass $\Sigma$ ``bumps'' at 1480, 1560, 1620, 1670, or
1690\,MeV, observed in production experiments, and no evidence for the 1* $\Sigma(1580)3/2^-$
in agreement with the authors of
Ref.~\cite{Olmsted:2003is}. Below 2200\,MeV, we find no evidence for the 1* resonances $\Sigma(1620)$ $1/2^-$,
$\Sigma(1730)3/2^+$, $\Sigma(1770)1/2^+$, $\Sigma(1840)3/2^+$, $\Sigma(1940)3/2^+$, $\Sigma(2000)1/2^-$,
$\Sigma(2070)5/2^+$, and no evidence for the 2* $\Sigma(1880)1/2^+$ and $\Sigma(2080)3/2^+$. We do
not observe the 1* resonances $\Lambda(1710)1/2^+$, $\Lambda(2000)$, $\Lambda(2020)7/2^+$, and
$\Lambda(2050)3/2^-$. $\Lambda(1810)1/2^+$ is the only 3* resonance for which we find no evidence.
If we include it, the $\chi^2$ improves by 106 units only, and mass and width are fitted to
$M$=1773\er 7; $\Gamma$= 39\er 15\,MeV. Two resonances above 2230\,MeV are included in the
analysis but are not included in Table~\ref{list-of-baryons}.

The changes in star rating suggested here are collected in Table~\ref{sr}.
\begin{table}[pt]
\caption{\label{sr}Suggested changes in star rating (the octet and decuplet
ground states are not included in the counting). Two states, $\Sigma(1620)1/2^-$ and
$\Lambda(1810)1/2^+$, we keep as one star resonances, even though they cannot be
confirmed by our fits (see text and \cite{Hyperon-III}).
}
\bc
\renewcommand{\arraystretch}{1.1}
\begin{tabular}{lccccc}
\hline\hline
            & 0* &  1* &  2* & 3* &  4* \\\hline
4*   to   &   -     &   -    &   -     &  1      & 11 \\
3*   to   &   -     &   1    &   1    &  3      & 2 \\
2*   to   &   4     &   -    &   -     &  -      & - \\
1*   to   &   12    &   2    &   2     &   -      & - \\  
new       &   -     &   5    &   -    &   -     & - \\
\hline\hline
\end{tabular}
\renewcommand{\arraystretch}{1.0}
\ec
\end{table}
The most significant changes are suggested for resonances which so far had a 1* or 2*
rating.  Most of them are not seen here. The changes suggested for  3*
and 4* resonances are moderate. Five new 1*-resonances are suggested.

In most cases, the fits finds a minimum within the mass-boundaries of the RPP, except for
four exceptions: We find $\Sigma(1620)1/2^-$ at $1681\pm 6$\,MeV and $\Sigma(1750)1/2^-$ at
$1692\pm 11$\,MeV, these two resonances are very close in mass and the evidence for the lower
mass state is weak as discussed in the previous section. $\Sigma(1940)3/2^-$ is found at
$1878\pm12$\,MeV. In the $3/2^-$ partial waves, a new resonance is found, and it is not surprising that
the masses of known resonances are shifted. In addition the $\Lambda(1405)$ is found at $1420\pm 3$\,MeV.
This state is discussed in detail in \cite{Hyperon-I}.

Finally, we made {\it error defining fits}. We chose solutions with an additional resonance and a
local minimum as shown in Figs.~\ref{fig:final_explore}. From the spread of results we estimated
the errors given to masses, widths and other properties (see~\cite{Hyperon-III}).

\section{\label{amplitudes}Partial wave amplitudes}
Our partial wave amplitudes are not derived from energy-independent fits (i.e. from fits in slices
of energy). Measurements of the spin-rotation parameters -- needed for a truly energy-independent
analysis -- do not exist. The Kent group succeeded nevertheless to construct the amplitudes by
first determining the leading waves and then defining the smaller ones. Our amplitudes are
determined from energy-dependent fits to the differential cross sections and polarization data. As
discussed above, we have made numerous fits, in particular also a large number of fits with
different resonance contents. Some additional poles led to a small $\chi^2$ reduction, of less than
400. For these fits, all partial wave amplitudes were determined as well. Thus, we derived a set of
partial wave amplitudes which all are about consistent with the data. The spread of these results
were used to determine a band for each partial wave amplitude.

Figures~\ref{fig:waves-1} and \ref{fig:waves-2} show the real and imaginary parts of the partial
wave amplitudes for $\bar K N$ elastic scattering in the two isospin channels, for the isospin 1
$\bar K N\to \pi\Lambda$ scattering and for the two isospins in $\bar K N\to \pi\Sigma$ determined
in this analysis and in Refs.~\cite{Zhang:2013cua} and~\cite{Kamano:2014zba}. Giving the
limitations of the data, the comparison shows reasonable consistency, at least for the leading
contributions. The partial waves are very similar in all four solutions where strong resonances
like
 $\Lambda(1520)3/2^-$, $\Lambda(1820)5/2^+$, $\Lambda(1830)5/2^-$, $\Sigma(1670)3/2^-$, $\Sigma(1775)5/2^-$,
$\Sigma(1915)5/2^+$, or $\Sigma(2030)7/2^+$  dominate the partial waves. For $\Lambda(1600)$ $1/2^+$,
the imaginary part is similar in all four analyses while the real part of
analysis~\cite{Zhang:2013cua} deviates. The $\Lambda$\,$1/2^-$ low-mass amplitudes for $\bar K N\to
\pi\Sigma$ from~\cite{Zhang:2013cua} and~\cite{Kamano:2014zba} are consistent but inconsistent with
our findings. However, a significant $\Lambda(1670)1/2^-$ structure is seen in all analyses.

Above the lowest-mass resonance, the structure of the amplitudes shows significant differences.
This is to be expected since the resonance content of the four analyses is different.
To resolve these discrepancies, new data are likely mandatory.

\section{\label{summary}Summary}
We have collected existing data on hyperon formation in $K^-p$ elastic and inelastic scattering.
The data were fitted in a coupled-channel analysis within the BnGa framework. We looked
systematically for contributing resonances in a large number of fits and mass scans. The
statistical significance of all resonances was evaluated. We find five new resonances; some
resonances are suggested to be upgraded others to be downgraded. For eighteen resonances -- mostly
listed as 1* or 2* resonances -- we did not find any signature. The partial-wave amplitudes derived
in our fits are compared to those from other analyses.

\section*{Acknowledgements}
We are very grateful to Sergey Prakhov for providing to us the Crystal Ball data on
the $K^-p\to \pi^0\pi^0\Lambda$ and $K^-p\to \pi^0\pi^0\Sigma$ reactions. Comments
by D. M. Manley and H. Kamano are kindly acknowledged.
This work was supported by the \textit{Deutsche Forschungsgemeinschaft} (SFB/TR110)
and the \textit{Russian Science Foundation} (RSF 16-12-10267).

\bibliographystyle{apsrev}

\end{document}